\documentclass[11pt,draftcls,onecolumn]{IEEEtran}
\usepackage{amssymb, amsmath, graphicx, cite, color, epsfig}

\newcommand{\complex}{\mathbb{C}}

\newcommand{\Fig}   {\mbox{Fig.} }
\newcommand{\Figs}  {\mbox{Figs.} }

\newcommand{\eqdef}{ := }
\newcommand{\twolines}[2]{\genfrac{}{}{0pt}{}{#1}{#2}}

\newcommand{\thmend}{\hspace*{\fill}~\QEDopen\par\endtrivlist\unskip}

\newcommand{\na} { {\sf a} }
\newcommand{\nb} { {\sf b} }
\newcommand{\nr} { {\sf r} }

\newcommand{\pa} { {(1)} }
\newcommand{\pb} { {(2)} }
\newcommand{\pc} { {(3)} }
\newcommand{\pd} { {(4)} }

\newtheorem{theorem}{Theorem}

\newtheorem{remark}[theorem]{Remark}

\allowdisplaybreaks[1]

\begin{document}
\title{Achievable rate regions for bi-directional relaying}
\author{Sang Joon Kim,
Natasha Devroye,
Patrick Mitran,
and Vahid Tarokh%
\thanks{ Sang Joon Kim and Vahid Tarokh are with the School
of Engineering and Applied Sciences, Harvard University, Cambridge,
MA 02138. Emails:~sangkim@fas.harvard.edu, vahid@deas.harvard.edu.
Natasha Devroye is with the Department of Electrical and Computer
Engineering, University of Illinois at Chicago, Chicage, IL 60607.
Email:~devroye@ece.uic.edu.
Patrick Mitran is with the Department of Electrical and Computer Engineering,
University of Waterloo, Waterloo, Canada. Email:~pmitran@ecemail.uwaterloo.ca. This research is supported in part by NSF grant number ACI-0330244 and ARO MURI grant number W911NF-07-1-0376. This work was supported in part by the Army Research Office,under the MURI award N0. N00014-01-1-0859.  The views expressed in this paper are
those of the author alone and not of the sponsor.
} }

\maketitle

\begin{abstract}
In a bi-directional relay channel, two nodes wish to exchange independent messages over a shared wireless half-duplex channel with the help of a relay. In this paper, we derive achievable rate regions for four new half-duplex protocols and compare these to four existing half-duplex protocols and outer bounds.  In time, our protocols consist of either two or three phases. In the two phase protocols, both users simultaneously transmit during the first phase and the relay alone transmits during the second phase, while in the three phase protocol the two users sequentially transmit followed by a transmission from the relay. The relay may forward information in one of four manners;  we outline existing Amplify and Forward (AF), Decode and Forward (DF) and Compress and Forward (CF) relaying schemes and introduce the novel Mixed Forward scheme. The latter is a combination of CF in one direction and DF in the other. We derive achievable rate regions for the CF and Mixed relaying schemes for the two and three phase protocols.  In the last part of this work we provide a comprehensive treatment of 8 possible half-duplex bi-directional relaying protocols in Gaussian noise, obtaining their respective achievable rate regions, outer bounds, and their relative performance under different SNR and relay geometries.
\end{abstract}



\section{Introduction}
\label{sec:intro}

Bi-directional relay channels, or wireless channels in which two nodes  ($\na$ and $\nb$)\footnote{We call the nodes $\na $ and $\nb$ \emph{terminal} and \emph{source} nodes interchangeably.}  wish to exchange independent messages with the help of a third relay node $\nr$, are both of fundamental and practical interest. Such channels may be relevant to ad hoc networks as well as to networks with a centralized controller through which all messages must pass.  From an information theoretic perspective, an understanding of these fundamental bi-directional channels would bring us closer to a coherent picture of multi-user information theory. To this end, we study bi-directional relay channels with the goal of determining spectrally efficient achievable rate regions and tight outer bounds to the capacity region.  

 This two-way channel \cite{Cover:2006} was first considered in \cite{Shannon:1961} where an achievable rate region and an outer bound for the case in which nodes operate in full-duplex were obtained.  Non-orthogonal\footnote{By non-orthogonal we mean that no additional space, time, frequency, or coding dimensions are used to separate sent and receive signals.}  full-duplex operation requires nodes to transmit and receive on the same antenna and frequency simultaneously. However,  it may not be practically feasible to do so since the intensity of the near field of the transmitted signal is
much higher than that of the far field of the received signal. In this work, we thus consider half-duplex communication in which a node may either transmit or receive at some time, but not both. Our goal is to determine spectrally efficient (measured in bits per channel use) transmission schemes and outer bounds for the half-duplex bi-directional relay channel and to compare their performance in a number of scenarios. These scenarios highlight the fact that different protocols may be optimal under different channel conditions.

An obvious half-duplex bi-directional relay protocol is the four
phase protocol, $\na \rightarrow \nr$, $\nr \rightarrow \nb$, $\nb
\rightarrow \nr$ and $\nr \rightarrow \na$, where the phases are
listed chronologically. However,  this protocol is spectrally
inefficient and does not take full advantage of the broadcast nature
of the wireless channel. One way to take advantage of the shared
wireless medium would be to combine the second and the fourth phases
into a single broadcast transmission by using, for example,  network
coding \cite{Ahlswede:2000}. That is, if the relay $\nr$ can decode
the messages $w_\na$ and $w_\nb$ from nodes $\na$ and $\nb$
respectively, it is sufficient for the relay $\nr$ to broadcast
$w_\na \oplus w_\nb$ to both $\na$ and $\nb$.

In this paper we consider two possible bi-directional relay protocols which differ in their number of phases. Throughout this work, \emph{phases} will denote temporal phases, or durations.  The three phase protocol is called the {\it Time Division Broadcast} (TDBC) protocol, while the two phase protocol is called the {\it Multiple Access Broadcast} (MABC) protocol.  One of the main conceptual differences between these two protocols is the possibility of \emph{side-information} in the TDBC protocol but not in the MABC protocol. By side-information we mean information obtained from the wireless channel in a particular phase which may be combined with information obtained in different stages to potentially improve decoding or increase transmission rates.  The two considered protocols may be described as:
\begin{enumerate}
\item  TDBC protocol: this consists of the three phases $\na \rightarrow \nr$, $\nb \rightarrow \nr$ and $\na \leftarrow \nr \rightarrow \nb$. In this protocol, only a single node is transmitting at any given point in time. Therefore, by the broadcast nature of the wireless channel, the non-transmitting nodes may listen in and obtain  ``side information'' about the transmissions of the other nodes. This may be used for more efficient decoding, i.e. improved rates.
\item  MABC protocol: this protocol combines the first two phases of the TDBC protocol and consists of the two phases $\na \rightarrow \nr \leftarrow \nb$ and $\na \leftarrow \nr \rightarrow \nb$. Due to the half-duplex assumption, during phase 1 both source nodes are transmitting and thus cannot obtain any  ``side information'' regarding the other nodes' transmission. It may nonetheless be spectrally efficient since it has less phases than the TDBC protocol and may take advantage of the multiple-access channel in phase 1.
\end{enumerate}
 We consider restricted protocols in the sense that the receivers must decode their messages at the end of the third phase (TDBC) or second phase (MABC) and collaboration accross multiple successive runs of the protocols are not possible.
For each of the MABC and TDBC protocols, the relay may process and
forward the received signals differently. These different forwarding
schemes are motivated by different {\em relaying capabilities or
assumptions} (about the required complexity or knowledge). Combining
the relaying schemes with the temporal protocols, we can obtain
various protocols whose rate regions are not in general subsets of
one another. The relative benefits and merits of the two protocols
and four relaying schemes are summarized in Tables
\ref{table:protocol} and  \ref{table:relaying}.  The four relaying
schemes we consider are:
\begin{enumerate}
\item {\it Amplify and Forward} (AF): the relay $\nr$ constructs its symbol by symbol replication of the received symbol. The AF scheme does not require any computation for relaying, and carries the noise incurred in the first stage(s) forward during the latter relaying stage.
\item {\it Decode and Forward} (DF): the relay decodes both messages from nodes $\na$ and $\nb$ before re-encoding them for transmission. The DF scheme requires  the full codebooks of both $\na$ and $\nb$ and a large amount of computation at the relay $\nr$.
\item {\it Compress and Forward} (CF):
the relay does not decode the messages of $\na$ and $\nb$, nor does
it simply amplify the received signal, but it performs something in
between these two extremes. It \emph{compresses} the received
signal, which it then transmits. To do so, the relay does not
require the codebooks of the source nodes, but it does require the
channel output distribution $p(y_\nr)$
at the relay.
\item {\it Mixed Forward}:  the relay decodes and forwards (DF) the data traveling in one direction (from $\na \rightarrow \nb$), while it compresses and forwards (CF) the data traveling in the opposite direction (from $\na \leftarrow \nb$). For the mixed scheme, one of the codebooks and the channel output distribution are needed at the relay.
\end{enumerate}
 In the CF scheme the relay searches the compression codebook to find an appropriate codeword. While the search operation is similar to the decoding operation in the DF scheme, the CF scheme may be less complex since the relay can choose a codebook for compression whose search space is smaller than the DF codebook.

Some of these protocols and relaying schemes have been considered in
the past. In \cite{Larsson:2006}, the DF TDBC protocol is
considered. There, network coding in $\mathbb{Z}_2^k$ is used to
encode the message of relay $\nr$ from the estimated messages
$\tilde{w}_\na$ and $\tilde{w}_\nb$. The works of
\cite{Popovski:2006b} and \cite{Popovski:2006a} consider the MABC
protocol, where an amplification and denoising relaying scheme are
introduced. In \cite{Narayanan:2007} a lattice code is used for the Gaussian channel in the MABC protocol. The capacity region of the broadcast phase in the MABC protocol assuming the relay has both messages $w_\na$ and $w_\nb$ is found in  \cite{Oechtering:2007}. In \cite{Tuncel:2006}, \cite{Nayak:2008}  Slepian-Wolf coding is extended to lossy broadcast channels with side information at the receivers. In \cite{SKim:2007}, achievable rate regions and outer
bounds of the MABC protocol and the TDBC protocol with the DF
relaying scheme are derived. There, network coding and random
binning are the techniques employed to determine achievable rate
regions.
Uni-directional CF relaying in the full-duplex channel is first
introduced in \cite{cover:1979}. An achievable region in the CF MABC protocol is derived in \cite{Schnurr:2007}. In \cite{Gunduz:2008} a comparison between DF and CF schemes in full-duplex channels is performed, while in \cite{Rankov:2005} a comparison of AF and DF schemes with two relays in the MABC protocol is performed.

In this paper, we derive achievable regions for new CF and mixed
relaying schemes in both the TDBC and MABC half-duplex protocols. We also obtain
outer bounds for the TDBC and MABC protocols based on cut-set
bounds. We compare the achievable rate regions of these four novel
schemes with the regions and outer bounds derived in\cite{SKim:2007}
as well as a simple AF scheme in Gaussian noise.  We thus present a
comprehensive overview of the bi-directional relay channel which
highlights the relative performance and tradeoffs of the different
schemes under different channel conditions and relay processing
capabilities. Notably, we find that under some channel conditions the mixed TDBC protocol outperforms the other protocols and similarly, there are channel conditions for which the CF TDBC protocol has the best performance.

This paper is structured as follows: in Section \ref{sec:prelim}, we introduce our notation, review previously determined achievable rate regions and outer bounds and define the protocols that we will consider. In Section \ref{sec:bounds} we derive achievable rate regions for the CF and mixed relaying schemes. In  Section \ref{sec:gaussian} we obtain explicit expressions for  these, and previous rate regions and outer bounds in Gaussian noise. In Section \ref{sec:regions}, we numerically compute these bounds in the Gaussian noise channel and compare the results for different powers and channel conditions.

\begin{table}
\caption{Comparison between two protocols }
\label{table:protocol}
\centering
\begin{tabular}{c||c|c|c}
  \hline
  Protocol & Side information & Number of phases & Interference\\
  \hline
  MABC & not present & 2 & present  \\
  TDBC & present & 3 & not present \\
  \hline
\end{tabular}
\end{table}

\begin{table}
\caption{Comparison between four relaying schemes }
\label{table:relaying}
\centering
\begin{tabular}{c||c|c|c}
  \hline
  Relaying & Complexity & Noise at relay & Relay needs \\
  \hline
  AF & very low & carried plus noise at rx & nothing \\
  DF & high & perfectly eliminated & full codebooks \\
  CF & low & carried plus distortion & $p(y_\nr)$ \\
  Mixed & moderate & partially carried & one codebook, $p(y_\nr)$ \\
  \hline
\end{tabular}
\end{table}


\section{Preliminaries}

\label{sec:prelim}

In this work, we will determine and compare the rate regions of eight  bi-directional relay protocols. We consider the 2 phase Multiple Access and Broadcast (MABC) and the 3 phase Time Division Broadcast (TDBC) protocol versions of Amplify and Forward (AF), Decode and Forward (DF), Compress and Forward (CF) as well as a Mixed scheme which combines Decode and Forward in one direction with Compress and Forward in the other. The AF, DF protocol regions and the CF MABC protocol region have been derived in prior work \cite{Popovski:2006a, SKim:2007, Schnurr:2007} while the CF and Mixed protocol regions described in Section \ref{sec:bounds} are determined here. Also we slightly improve upon the CF MABC protocol region in \cite{Schnurr:2007}. We formally define our notation and problem next.

\subsection{Notation and Definitions}
 We consider two terminal nodes $\na$ and $\nb$, and one relay node $\nr$. Terminal
 node $\na$ (resp. $\nb$) has its own message that it wishes
 to send to the opposite terminal node, node $\nb$ (resp. $\na$).
The relay node $\nr$ may assist in the bi-directional endeavor.
This paper will determine achievable rate regions for bi-directional relay protocols over half-duplex, discrete-time memoryless channels. The half-duplex constraint implies that a node cannot simultaneously transmit and receive data.

 We first start with a somewhat more general formulation of the problem that simplifies the application of cut-set outer bounds and then apply it the MABC and TDBC protocols considered here. We consider an $m$ node
set, denoted as ${\cal M }\eqdef \{1, 2, \cdots,m\}$ (where $\eqdef$ means defined as).
We use $R_{i,j}$ to denote the transmitted data rate of message $W_{i,j}$ from node $i \in {\cal M}$ to node $j\in {\cal M}$, i.e., $W_{i,j} \in \{0, \ldots, \lfloor 2^{nR_{i,j}} \rfloor - 1\} \eqdef {\cal S}_{i,j}$. The protocols considered have either $L=2$ (MABC) or $L=3$ (TDBC) \emph{phases.} We denote by $\Delta_\ell \geq 0$ the relative time duration of the $\ell^{th}$ phase, where $\sum_\ell \Delta_\ell = 1$. For a given block size $n$, $\Delta_{\ell,n}$ denotes  the duration of the $\ell^{th}$ phase. Obviously, $\Delta_{\ell,n} \rightarrow \Delta_{\ell}$ as $n \rightarrow \infty$.

For notational convenience, we define the messages $W_\na \eqdef W_{\na,\nb}$, $W_\nb \eqdef W_{\nb,\na}$ and the corresponding rates $R_\na \eqdef R_{\na,\nb}$ and $R_\nb \eqdef R_{\nb,\na}$. The two distinct messages $W_\na$ and $W_\nb$ are taken to be independent and uniformly distributed in the set of $\{0,\ldots,\lfloor 2^{nR_\na}\rfloor  -1\}\eqdef {\cal S}_\na$ and $\{0,\ldots,\lfloor 2^{nR_\nb} \rfloor -1\}\eqdef {\cal S}_\nb$, respectively.

We use channel input alphabet ${\cal X}_i$ and channel output
alphabet ${\cal Y}_i$ for node $i$. We will be
constructing Compress and Forward schemes in which received signals
are compressed or quantized before being re-transmitted. We let
$\hat{Y}_i$ denote the compressed representation of the received signal at node $i$,
which lies in the corresponding compression alphabet $\hat{\cal
Y}_i$ for node $i$. $\hat{\cal Y}_i$ is not necessarily equal to
${\cal Y}_i$. We summarize the input/output alphabets of the MABC and TDBC protocols in Table \ref{table:alphabet}. In Section \ref{sec:gaussian} and \ref{sec:regions}, we
consider the case ${\cal X}_i = {\cal Y}_i = \hat{\cal Y}_i =\complex$, $\forall i$.

\begin{table}
\caption{Input and output alphabets }
\label{table:alphabet}
\centering
\begin{tabular}{c||c|c}
  \hline
   & MABC & TDBC \\
  \hline
  Phase 1 & ${\cal X}_\na, {\cal X}_\nb, {\cal Y}_\nr, {\hat{\cal Y}}_\nr$ & ${\cal X}_\na,{\cal Y}_\nb,{\cal Y}_\nr, {\hat{\cal Y}}_\nr$ \footnotemark\\
  Phase 2 & ${\cal X}_\nr, {\cal Y}_\na, {\cal Y}_\nb$ & ${\cal X}_\nb,{\cal Y}_\na,{\cal Y}_\nr, {\hat{\cal Y}}_\nr $  \\
  Phase 3 & N/A & ${\cal X}_\nr, {\cal Y}_\na, {\cal Y}_\nb$ \\
  \hline
\end{tabular}
\end{table}

\footnotetext{${\hat{\cal Y}}_\nr$ is used in the CF TDBC protocol only.}

 For a given block length $n$, it will be
convenient to denote the transmission at time $1\leq k\leq n$ at
node $i$ by $X_i^k$, the reception at node $i$ at time $k$ by
$Y_i^k$. Note that the distributions of $X_i^k$ and $Y_i^k$ depend on
the value of $k$, e.g. for $k\leq \Delta_{1,n} \cdot n$ we are in phase 1, for
$\Delta_{1,n} n < k \leq (\Delta_{1,n}+\Delta_{2,n}) n$ we are in phase 2 and for
$(\Delta_{1,n} +\Delta_{2,n}) n < k \leq n$ we are in phase 3 (in TDBC
protocols only). During phase $\ell$ we use $X_i^{(\ell)}$ to denote the random variable with alphabet ${\cal X}_i$ and input distribution $p^{(\ell)}(x_i)$.
It is also convenient to denote by $X_S^k \eqdef \{X_i^k | i\in S\}$, the set of transmissions by all nodes in the set $S$ at time $k$, and by $X_{S}^{(\ell)} \eqdef \{X_i^{(\ell)}|i\in S\}$, a set of random variables with channel input distribution $p^{(\ell)}(x_S)$ for phase $\ell$, where $x_S \eqdef \{x_i | i\in S\}$. Lower case letters $x_i$ will denote instances of the upper case $X_i$ which lie in the calligraphic alphabets ${\cal X}_i$. Boldface ${\bf x}_i$ represents a vector indexed by time at node $i$. Finally, we denote ${\bf x}_S \eqdef \{{\bf x}_i | i\in S\}$ as a set of vectors indexed by time. In this work, $Q$ will denote a discrete time-sharing random variable with distribution $p(q)$.

In order to define bi-directional communication rates we must define the encoders, decoders and associated probability of errors. We define $W_{S,T} \eqdef \{W_{i,j} | i\in S ,~ j\in T,~ S,T\subset {\cal M}\}$. For a block length $n$, let $\overleftarrow{{\bf y}}_i^{(\ell)}$ denote the set of received signals at node $i$ up until the end of phase $\ell$, i.e., until the end of time $n\cdot\sum_{m=1}^\ell \Delta_{m,n}$. Encoders and decoders are functions $X_i^{(\ell)}(W_{\{i\},{\cal M}},\overleftarrow{{\bf y}}_i^{(\ell)}) \in {\cal X}_i$ and $\tilde{W}_{j,i}(\overleftarrow{{\bf y}}_i^{(L)}, W_{\{i\},{\cal M}})$ respectively for $\ell \in \{1,2,\cdots,L\}$ and for our proposed schemes, we will obtain single letter bounds. We define error events $E_{i,j} \eqdef \{W_{i,j} \neq \tilde{W}_{i,j}(.)\}$ for decoding the message $W_{i,j}$
at node $j$ at the end of the block of length $n$, and $E_{i,j}^{(\ell)}$ as the error event at node $j$ in which node $j$ attempts to decode $w_i$ at the end of phase $\ell$ using a joint typicality decoder.

Let $A^{(\ell)}(UV)$ represent the set of $\epsilon$-typical
$({\bf u}^{(\ell)},{\bf v}^{(\ell)})$ sequences of length $n \cdot \Delta_{\ell,n}$
according to the distributions $U$ and $V$ in phase $\ell$. The events $D^{(\ell)}({\bf u},{\bf v})\eqdef
\{({\bf u}^{(\ell)},{\bf v}^{(\ell)}) \in A^{(\ell)}(UV)\}$.
In general, joint typicality is
non-transitive. However, by using strong joint-typicality, and the fact that for the
distributions of interest $x \rightarrow y \rightarrow \hat{y}$, we will be able to argue
joint typicality between ${\bf x}$ and $\bf {\hat{y}}$ by the {\it Markov lemma} of Lemma 4.1
in \cite{berger:1977} and the extended Markov lemma (Lemma 3 of \cite{Oohama:1997}, Remark 30 of \cite{Kramer:2005}).

A set of rates $R_{i,j}$ is said to be achievable for a protocol with phase durations $\{\Delta_{\ell}\}$ if there exist  encoders/decoders of block length $n = 1, 2, \ldots$   with both $P[E_{i,j}]\rightarrow 0$ and $\Delta_{\ell , n}\rightarrow \Delta_{\ell}$ as $n\rightarrow \infty$ for all $\ell$. An achievable rate region (resp. capacity region) is the closure of a set of (resp. all) achievable rate
tuples for fixed $\{\Delta_{\ell}\}$.
\subsection{Previous results}
We use the following outer bounds and achievable rate regions of decode and forward protocols, derived in \cite{SKim:2007} for comparison purposes in Sections \ref{sec:gaussian} and \ref{sec:regions}. We simply state the results here for completeness.

\begin{theorem}
\label{theorem:MABC:out}
(Outer bound)  The capacity region of the bi-directional
relay channel constrained to the MABC protocol is outer bounded by the union
of
\begin{align}
R_{\na} &\leq \min \{ \Delta_1 I(X_{\na}^\pa ; Y_{\nr}^\pa |X_{\nb}^\pa,Q),
            \Delta_2 I(X_{\nr}^\pb ; Y_{\nb}^\pb |Q)\} \\
R_{\nb} &\leq \min \{ \Delta_1 I(X_{\nb}^\pa ; Y_{\nr}^\pa | X_{\na}^\pa,Q),
            \Delta_2 I(X_{\nr}^\pb ; Y_{\na}^\pb |Q)\}
\end{align}
over all joint distributions
$p(q)p^\pa(x_{\na}|q)p^\pa(x_{\nb}|q)p^\pb(x_{\nr}|q)$ with $|{\cal
Q}| \leq 4$ over the alphabet ${\cal X}_\na \times {\cal X}_\nb
\times {\cal X}_\nr $.
\thmend
\end{theorem}

\begin{theorem}
\label{theorem:TDBC:out}
(Outer bound) The capacity region of the half-duplex bi-directional relay channel
constrained to the TDBC protocol is outer bounded by
\begin{align}
R_\na &\leq \min\{{\Delta}_1 I(X_\na^\pa;Y_\nr^\pa,Y_\nb^\pa|Q),{\Delta}_1 I(X_\na^\pa;Y_\nb^\pa|Q) +
          {\Delta}_3 I(X_\nr^\pc;Y_\nb^\pc|Q) \} \\
R_\nb &\leq \min\{{\Delta}_2 I(X_\nb^\pb;Y_\nr^\pb,Y_\na^\pb|Q),{\Delta}_2 I(X_\nb^\pb;Y_\na^\pb|Q) +
           {\Delta}_3 I(X_\nr^\pc;Y_\na^\pc|Q)\} \\
R_\na + R_\nb &\leq \Delta_1 I(X_\na^\pa; Y_\nr^\pa |Q) + \Delta_2 I(X_\nb^\pb; Y_\nr^\pb|Q)
\end{align}
over all joint distributions $p(q)p^\pa(x_{\na}|q)p^\pb(x_{\nb}|q)$
$p^\pc(x_{\nr}|q)$ with $|{\cal Q}| \leq 5$ over the alphabet ${\cal
X}_\na \times {\cal X}_\nb \times {\cal X}_\nr$. \thmend
\end{theorem}

\begin{theorem}
\label{theorem:MABC:DF}
An achievable rate region for the half-duplex bi-directional relay channel with the MABC protocol is
the closure of the set of all points $(R_\na,R_\nb)$ satisfying
\begin{align}
R_{\na} &< \min \left\{ \Delta_1 I(X_{\na}^\pa ; Y_{\nr}^\pa | X_{\nb}^\pa,Q),
            \Delta_2 I(X_{\nr}^\pb ; Y_{\nb}^\pb | Q)\right\} \\
R_{\nb} &< \min \left\{ \Delta_1 I(X_{\nb}^\pa ; Y_{\nr}^\pa | X_{\na}^\pa,Q),
            \Delta_2 I(X_{\nr}^\pb ; Y_{\na}^\pb | Q)\right\}\\
R_{\na} + R_{\nb} &<    \Delta_1 I(X_{\na}^\pa , X_{\nb}^\pa ; Y_{\nr}^\pa|Q)
\end{align}
over all joint distributions
$p(q)p^\pa(x_{\na}|q)p^\pa(x_{\nb}|q)p^\pb(x_{\nr}|q)$ with $|{\cal
Q}| \leq 5$ over the alphabet ${\cal X}_\na \times {\cal X}_\nb
\times {\cal X}_\nr $.
\thmend
\end{theorem}

\begin{theorem}\label{theorem:TDBC:DF}
An achievable rate region for the half-duplex bi-directional relay channel with the TDBC protocol is
the closure of the set of all points $(R_\na,R_\nb)$ satisfying
\begin{align}
R_\na &< \min\big\{{\Delta}_1 I(X_\na^\pa;Y_\nr^\pa|Q),{\Delta}_1 I(X_\na^\pa;Y_\nb^\pa|Q) + {\Delta}_3 I(X_\nr^\pc;Y_\nb^\pc|Q)\big\}\\
R_\nb &< \min\big\{{\Delta}_2 I(X_\nb^\pb;Y_\nr^\pb|Q),{\Delta}_2 I(X_\nb^\pb;Y_\na^\pb|Q) + {\Delta}_3 I(X_\nr^\pc;Y_\na^\pc|Q)\big\}
\end{align}
over all joint distributions
$p(q)p^\pa(x_{\na}|q)p^\pb(x_{\nb}|q)p^\pc(x_{\nr}|q)$ with $|{\cal
Q}| \leq 4$ over the alphabet ${\cal X}_\na \times {\cal X}_\nb
\times {\cal X}_\nr $.
\thmend
\end{theorem}

\subsection{Compress and Forward using two joint typicality decoders}
In Compress and Forward protocols, unlike in Decode and Forward protocols,  the relay node $\nr$ does not decode the message $w_\na$ or $w_\nb$. Thus, network coding techniques such as the algebraic group operation $w_\na \oplus w_\nb$ used in \cite{SKim:2007} cannot be used to generate $w_\nr$ for the current CF schemes.  Instead, two jointly typical decoders at each node are used to decode  $w_\nr$.

\begin{figure}[th]
\begin{center}
\epsfig{keepaspectratio = true, width = 11cm, figure = 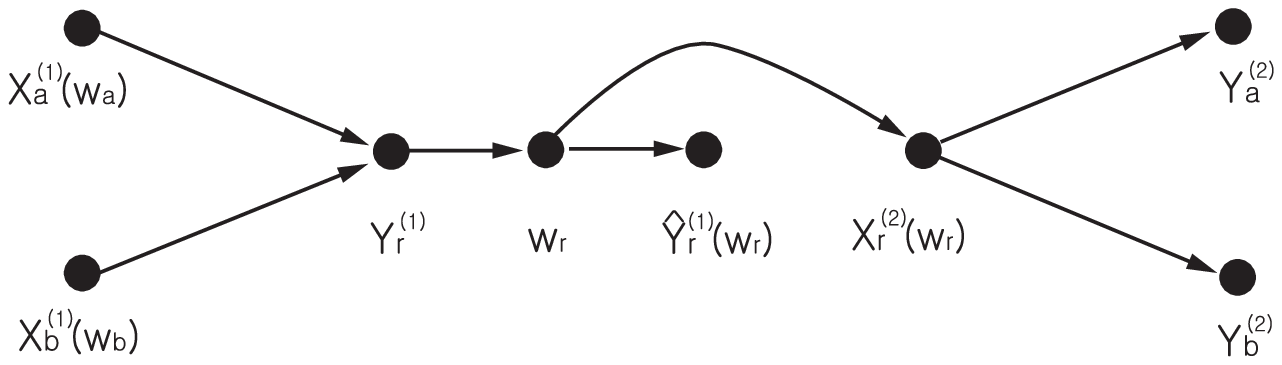}
\end{center}
\caption{The data flow in the compress and forward MABC protocol }
\label{fig:decoding}
\end{figure}

To illustrate the decoding scheme, consider the decoder at node $\na$ which wishes to decode the relay message $w_\nr$ in order to ultimately decode the desired message from node $\nb$, $w_\nb$.  After phase 2, node $\na$ has the known sequences ${\bf x}^\pa_\na(w_\na)$ and ${\bf y}_\na^\pb$. Node $\na$ then finds the sets of all $\hat{\bf y}^\pa_\nr(w_\nr)$ and ${\bf x}^\pb_\nr(w_\nr)$ such that  $({\bf x}^\pa_\na(w_\na),\hat{\bf y}^\pa_\nr(w_\nr))$ and $({\bf x}^\pb_\nr(w_\nr),{\bf y}_\na^\pb)$ are  two pairs of jointly typical sequences, as shown in \Fig \ref{fig:decoding}.  Then node $\na$ decodes $w_\nr$ correctly if there exists a unique $w_\nr$ such that $({\bf x}^\pa_\na(w_\na),\hat{\bf y}^\pa_\nr(w_\nr))\in A^\pa(X_\na {\hat Y}_\nr)$ and $({\bf x}^\pb_\nr(w_\nr),{\bf y}_\na^\pb) \in A^\pb(X_\nr Y_\na)$ and declares a decoding error otherwise.


\section{Achievable rate regions for Compress and Forward and Mixed Protocols}

\label{sec:bounds}

In this Section we present three new achievable rate regions in Theorems \ref{theorem:MABC:CFDF}, \ref{theorem:TDBC:CF} and \ref{theorem:TDBC:CFDF}, and a slight improvement of \cite{Schnurr:2007} in Theorem \ref{theorem:MABC:CF}.  Theorems  \ref{theorem:MABC:CF} and \ref{theorem:MABC:CFDF} are for two phase Multiple Access and Broadcast (MABC) protocols. In MABC protocols, nodes $\na$ and $\nb$ transmit simultaneously as in a standard multiple-access channel in phase 1, the relay processes the received signal (either by decoding, amplifying, or compressing the signal as dictated by the protocol), and during phase 2 the relay broadcasts its signal to the two nodes. In  Theorem \ref{theorem:MABC:CF} the relay simply uses a CF operation for both messages/directions, while in Theorem   \ref{theorem:MABC:CFDF} the relay uses DF to transmit the $w_\na$ message while it uses CF to transmit the $w_\nb$ message. The final two theorems
\ref{theorem:TDBC:CF} and \ref{theorem:TDBC:CFDF} employ three phase Time Division Broadcast (TDBC) protocols. During the first phase, node $\na$  transmits while both the relay and node $\nb$ receive its signal. During phase 2, node $\nb$ is the sole transmitter while node $\na$ and the relay receive its transmission. After these phases,  the relay processes the received signals and is the sole transmitter during phase 3, during which it can aid nodes $\na$ and $\nb$ to recover each others' messages. We now proceed to describe each protocol more precisely and present their respective achievable rate regions.  Again, in Theorem \ref{theorem:TDBC:CF} the relay uses CF when re-transmitting both messages, while in Theorem \ref{theorem:TDBC:CFDF} the relay uses DF to transmit message $w_\na$ and CF to transmit message $w_\nb$. We now proceed to the main technical results of this work.

\subsection{MABC Protocol}

\begin{figure}[t]
 \begin{center}
  \epsfig{figure=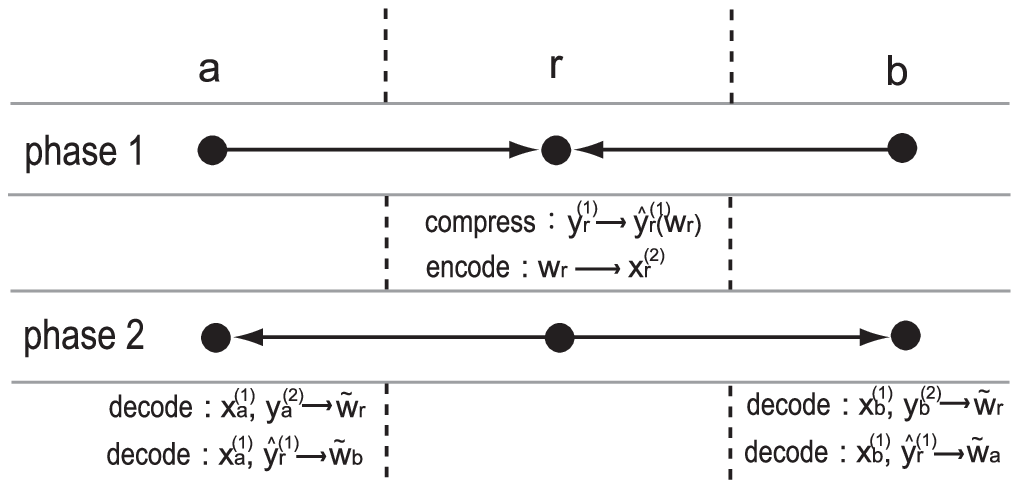, width=8cm}
  \caption{The two-phase MABC protocol with a relay using a CF scheme.}
  \label{fig:MABC:CF}
 \end{center}
\end{figure}

\begin{figure}[t]
 \begin{center}
  \epsfig{figure=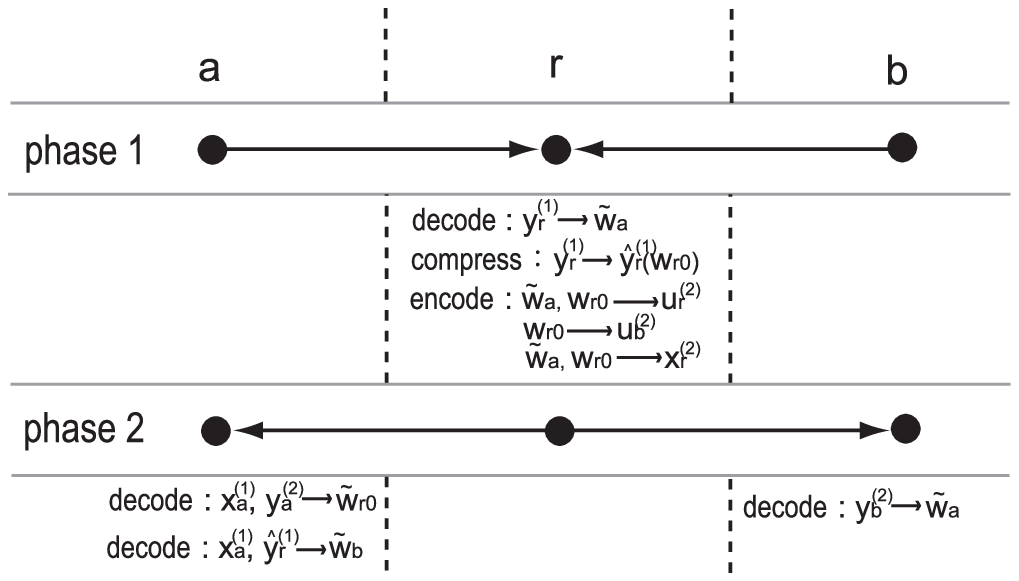, width=8cm}
  \caption{The two-phase MABC protocol with a relay using a mixed scheme.}
  \label{fig:MABC:CFDF}
 \end{center}
\end{figure}

In the MABC Protocol,  message $w_\na$ is communicated from node $\na$ to node $\nb$ and message $w_\nb$ is communicated from node $\nb$ to node $\na$ with the help of the relay in two phases as shown in \Fig \ref{fig:MABC:CF} and \ref{fig:MABC:CFDF}. During phase 1, nodes $\na$ and $\nb$ simultaneously send independent messages $w_\na$ and $w_\nb$ as codewords ${\bf x}_\na^\pa(w_\na)$ and ${\bf x}_\nb^\pa(w_\nb)$ to the relay, forming a classical multiple-access channel. Since we assume half-duplex nodes, neither $\na$ nor $\nb$ can receive the message of the other during phase 1. The relay receives the signal ${\bf y}_\nr^\pa$ according to $p(y_\nr^\pa|x_\na^\pa, x_\nb^\pa)$. Rather than  attempting to decode message $w_\na$ and $w_\nb$ (as in a DF scheme), it compresses the received ${\bf y}_\nr^\pa$ into a  signal $\hat{\bf y}_\nr^\pa(w_\nr)$. The index $w_\nr$ is then mapped in a one-to-one fashion to the codeword ${\bf x}_\nr^\pb(w_\nr)$ which is broadcast in phase 2 back to the relays. The challenge here is to determine the optimal compression strategy such that just enough information is carried back to the nodes to decode the opposite node's message. A key observation is that the nodes may use their own phase 1 transmitted messages as side-information in the decoding of phase 2 signals.

\begin{theorem}
\label{theorem:MABC:CF} An achievable rate region of the half-duplex
bi-directional relay channel with the compress and forward MABC
protocol is the closure of the set of all points $(R_\na, R_\nb)$ satisfying
\begin{align}
R_{\na} &< \Delta_1 I(X_\na^\pa; \hat{Y}_\nr^\pa|X_\nb^\pa, Q)\\
R_{\nb} &< \Delta_1 I(X_\nb^\pa; \hat{Y}_\nr^\pa|X_\na^\pa, Q)
\end{align}
subject to
\begin{align}
\Delta_1 I(Y_\nr^\pa ; \hat{Y}_\nr^\pa | X_\nb^\pa,  Q) &< \Delta_2 I(X_\nr^\pb;Y_\nb^\pb)\label{eq:cf:bound:1}\\
\Delta_1 I(Y_\nr^\pa; \hat{Y}_\nr^\pa | X_\na^\pa,  Q) &<
\Delta_2 I(X_\nr^\pb;Y_\na^\pb)\label{eq:cf:bound:2}
\end{align}
over all joint distributions,
\begin{align}
p(q,x_\na,x_\nb,x_\nr,y_\na,y_\nb,y_\nr,\hat{y}_\nr) =
p^\pa(q,x_\na,x_\nb,y_\nr,\hat{y}_\nr|q)p^\pb(x_\nr,y_\na,y_\nb)
\end{align}
where
\begin{align}
 p^\pa(q,x_\na,x_\nb,y_\nr,\hat{y}_\nr) &=  p^\pa(q)p^\pa(x_\na|q)p^\pa(x_\nb|q)p^\pa(y_\nr|x_\na,x_\nb)p^\pa({\hat y}_\nr|y_\nr,q) \label{eq:dist:1}\\
 p^\pb(x_\nr,y_\na,y_\nb) &=  p^\pb(x_\nr)p^\pb(y_\na, y_\nb|x_\nr)
\end{align}
with $|{\cal Q}| \leq 4$ over the alphabet ${\cal X}_\na \times {\cal X}_\nb \times {\cal X}_\nr
\times {\cal Y}_\na \times {\cal Y}_\nb \times {\cal Y}_\nr \times \hat{{\cal Y}}_\nr$. \thmend
\end{theorem}

\begin{remark}
The bound of Theorem \ref{theorem:MABC:CF} is essentially derived in \cite{Schnurr:2007}; equation \eqref{eq:dist:1} is a slight extension, as we use $p^\pa({\hat y}_\nr|y_\nr,q)$ instead of $p^\pa({\hat y}_\nr|y_\nr)$, i.e., in \cite{Schnurr:2007} the codewords ${\hat {\bf y}}_\nr^\pa$ are generated according to $p^\pa({\hat y}_\nr) = \sum p^\pa(y_\nr)p^\pa({\hat y}_\nr|y_\nr)$, while in \eqref{eq:dist:1} the distribution space $p^\pa({\hat y}_\nr |q) = \sum p^\pa(y_\nr)p^\pa({\hat y}_\nr|y_\nr,q)$ is larger. By conditioning on $q$, one can ``fine-tune'' the distribution of ${\hat y}_\nr^\pa$ for each given $q$ and the left side of \eqref{eq:cf:bound:1} and
\eqref{eq:cf:bound:2} can be reduced. This is because the distributions of $X_\na^\pa$ and $X_\nb^\pa$, and hence $Y_\nr^\pa$, depend on $q$. For example, let $p^\pa(q=1) = \alpha_n$ and $p^\pa(q=2) = 1-\alpha_n$, where $0<\alpha_n <1$. For $q=1$ we optimize $p^\pa({\hat y}_\nr|1)$ and generate $(\alpha_n \Delta_{1,n}\cdot n)$-length sequence ${\hat {\bf y}}_\nr^{\pa,1}(w_{\nr 1})$, $w_{\nr 1} \in \{0,1,\cdots \lfloor 2^{nR_{\nr1}}\rfloor\}$, where $R_{\nr1} = \alpha_n \Delta_{1,n} (I(Y_\nr^\pa;{\hat Y}_\nr^\pa|q=1) + \epsilon)$. Likewise, we generate ${\hat {\bf y}}_\nr^{\pa,2}(w_{\nr 2})$ for $q=2$. To compress ${\bf y}_\nr^\pa$ to ${\hat {\bf y}}_\nr^\pa$, we construct ${\bf y}_\nr^\pa = ({\bf y}_\nr^{\pa,1},{\bf y}_\nr^{\pa,2})$ and ${\hat {\bf y}}_\nr^\pa = ({\hat {\bf y}}_\nr^{\pa,1},{\hat {\bf y}}_\nr^{\pa,2})$ and choose $w_\nr = (w_{\nr 1},w_{\nr 2})$ if both $({\bf y}_\nr^{\pa,1},{\hat {\bf y}}_\nr^{\pa,1}(w_{\nr 1}))$ and $({\bf y}_\nr^{\pa,2},{\hat {\bf y}}_\nr^{\pa,2}(w_{\nr 2}))$ are jointly typical. Then the rate $R_\nr = R_{\nr 1} + R_{\nr 2} = \Delta_1 I(Y_\nr^\pa;{\hat Y}_\nr^\pa|Q)$. However, if one generated ${\hat{\bf y}}^\pa_\nr$ from the distribution $p^\pa({\hat y}_\nr)$ then $R_\nr = \Delta_1 I(Y_\nr^\pa;{\hat Y}_\nr^\pa) \geq\Delta_1 I(Y_\nr^\pa;{\hat Y}_\nr^\pa|Q)$ with strict inequality except in degenerate cases.
\end{remark}

\begin{remark}
Strong typicality is required for the proof of Theorem \ref{theorem:MABC:CF} in order to apply the Markov lemma to $(X_\na^\pa, X_\nb^\pa) \rightarrow Y_\nr^\pa \rightarrow {\hat Y}_\nr^\pa$ for each given $q$. Since strong typicality is defined for discrete alphabets, Theorem \ref{theorem:MABC:CF} cannot be directly extended to continuous alphabets. However, the extended Markov lemma (see Remark 30 of \cite{Kramer:2005} as well as Lemma 3 of \cite{Oohama:1997}) shows that for Gaussian distributions, the Markov lemma still applies.
\end{remark}

The previous theorem assumed the relay used a CF scheme for both messages $w_\na$ and $w_\nb$.
However, in the event of asymmetric channel gains between the two nodes at the relay, it may be beneficial to have the stronger channel use a DF scheme while the weaker channel uses a CF scheme since decoding may not be possible. We next consider a mixed MABC strategy in which phase 1 is still a multiple access channel and phase 2 is still a broadcast channel. However, the relay uses a novel strategy in determining its phase 2 codeword ${\bf x}_\nr^{(2)}$. That is, the relay operates such that the $\na \rightarrow  \nr \rightarrow \nb$ link uses decode and forward while $\na \leftarrow \nr \leftarrow \nb$ link uses compress and forward. Furthermore, the relay applies a Gelfand-Pinsker coding scheme to protect $w_\na$ in the $\nr \rightarrow \nb$ link. In this case, an achievable rate region is given by Theorem \ref{theorem:MABC:CFDF}.

\begin{theorem}
\label{theorem:MABC:CFDF} An achievable rate region of the half-duplex
bi-directional relay channel with the mixed forward MABC protocol,
where $\na \rightarrow \nb$ link uses decode and forward and $\nb
\rightarrow \na$ link uses compress and forward, is the closure of the set of all points $(R_\na, R_\nb)$ satisfying
\begin{align}
R_{\na} &< \min \left\{\Delta_1 I(X_\na^\pa; Y_\nr^\pa|Q), \Delta_2I(U_\nr^\pb;Y_\nb^\pb|Q) - \Delta_2 I(U_\nr^\pb;U_\nb^\pb|Q)\right\}\label{eq:mabc:cfdf:b:1}\\
R_{\nb} &< \Delta_1 I(X_\nb^\pa; \hat{Y}_\nr^\pa|X_\na^\pa, Q)\label{eq:mabc:cfdf:b:3}
\end{align}
subject to
\begin{align}
\Delta_1 I(Y_\nr^\pa;\hat{Y}_\nr^\pa|X_\na^\pa,Q) &< \min\{\Delta_2
I(U_\nr^\pb,U_\nb^\pb;Y_\na^\pb|Q), \Delta_2I(U_\nb^\pb;U_\nr^\pb,Y_\na^\pb|Q)\} \label{eq:mabc:cfdf:b:2}
\end{align}
over all joint distributions,
\begin{align}
p(q,x_\na,x_\nb,x_\nr,u_\na,u_\nb,u_\nr,y_\na,y_\nb,y_\nr,\hat{y}_\nr) =p(q)
p^\pa(x_\na,x_\nb,y_\nr,\hat{y}_\nr|q)p^\pb(u_\nb,u_\nr,x_\nr,y_\na,y_\nb|q)
\end{align}
where
\begin{align}
 p^\pa(x_\na,x_\nb,y_\nr,\hat{y}_\nr|q) &=  p^\pa(x_\na|q)p^\pa(x_\nb|q)p^\pa(y_\nr|x_\na,x_\nb)p^\pa({\hat y}_\nr|y_\nr,q)\\
 p^\pb(u_\nb,u_\nr,x_\nr,y_\na,y_\nb|q) &=  p^\pb(u_\nb,u_\nr|q) p^\pb(x_\nr|u_\nb,u_\nr,q)p^\pb(y_\na,y_\nb|x_\nr)
\end{align}
with $|{\cal Q}| \leq 7$ over the alphabet ${\cal X}_\na \times {\cal X}_\nb \times {\cal X}_\nr \times {\cal U}_\nb \times {\cal U}_\nr
\times {\cal Y}_\na \times {\cal Y}_\nb \times {\cal Y}_\nr \times \hat{{\cal Y}}_\nr$. \thmend
\end{theorem}

\begin{remark}
In the second phase, the relay broadcasts the received signals from the first phase. In contrast to the DF and CF schemes, one of the terminal nodes (in this case, node $\na$) has perfect information of interference at node $\na$, while there remains unknown interference at the other side (at node $\nb$). We use a Gel'fand-Pinsker coding scheme \cite{Gelfand:1980},\cite{Gamal:1981} for the link $\nr \rightarrow \nb$ which yields  the second term of \eqref{eq:mabc:cfdf:b:1}. From the side information $w_\na$ available at node $\na$, node $\na$ is able to reduce the interference, yielding \eqref{eq:mabc:cfdf:b:2}.
\end{remark}

\begin{remark}
If we apply the achievable bound to the Gaussian noise channel without fading with Costa's setup in \cite{Costa:1983} with $|Q|=1$ we have in phase two:
\begin{align}
U_{\nr}^\pb &= V_\nr^\pb + \alpha U_\nb^\pb\\
Y_{\na}^\pb &= V_\nr^\pb + U_\nb^\pb + Z_\na^\pb \label{cfdf:ex:1}\\
Y_{\nb}^\pb &= V_\nr^\pb + U_\nb^\pb + Z_\nb^\pb \label{cfdf:ex:2}
\end{align}
where $Z_\na^\pb\sim {\cal C N}(0,N_\na)$, $Z_\nb^\pb\sim {\cal C N}(0,N_\nb)$, $U_\nb^\pb\sim {\cal C N}(0,P_{U_\nr})$, $V_\nr^\pb\sim {\cal C N}(0,P_{V_\nr})$, and $V_\nr^\pb$, $U_\nb^\pb$ are independent and $P_{U_\nr}+P_{V_\nr} = P_\nr$. $V_\nr^\pb$ is an intermediate random variable generated according to ${\cal C N}(0,P_\nr)$ which contains the information to be transmitted from $\nr$ to $\na$. Then from \eqref{eq:mabc:cfdf:b:1} and \eqref{eq:mabc:cfdf:b:2} the achievable rate of link $\nr\rightarrow\na$, $R_{\nr\na}$ (resp. $R_{\nr\nb}$ for $\nr\rightarrow\nb$) is :
\begin{align}
R_{\nr\na} &= \min \left\{\log_2\left(1+\frac{P_\nr}{N_\na}\right),\log_2\left(\frac{P_{V_\nr} P_{U_\nb}(1-\alpha)^2 + N_\na(P_{V_\nr} + \alpha^2 P_{U_\nb})}{N_\na P_{V_\nr} }\right)\right\}\label{eq:cfdf:g:1}\\
R_{\nr\nb} &= \log_2\left(\frac{P_{V_\nr}(P_{V_\nr} + P_{U_\nb}+ N_\nb)}{P_{V_\nr} P_{U_\nb}(1-\alpha)^2 + N_\nb(P_{V_\nr} + \alpha^2 P_{U_\nb}) }\right)\label{eq:cfdf:g:2}
\end{align}
\end{remark}

\begin{proof}
{\em Random code generation: } For simplicity of exposition, we take $|{\cal Q}| =1 $.
\begin{enumerate}
\item Phase 1: Generate random $(n\cdot\Delta_{1,n})$-length sequences
\begin{itemize}
  \item ${\bf x}^\pa_\na(w_\na)$ i.i.d. with $p^\pa(x_\na)$, $w_\na \in {\cal S}_\na = \{0,1,\cdots,\lfloor 2^{nR_{\na}} \rfloor-1\}$
  \item ${\bf x}^\pa_\nb(w_\nb)$ i.i.d. with $p^\pa(x_\nb)$, $w_\nb \in {\cal S}_\nb = \{0,1,\cdots,\lfloor 2^{nR_{\nb}} \rfloor-1\}$
  \item $\hat{\bf y}_\nr^\pa(w_{\nr0})$ i.i.d. with $p^\pa(\hat{y}_\nr) = \sum_{y_\nr} p^\pa(y_\nr) p^\pa(\hat{y}_\nr|y_\nr)$ , $w_{\nr0} \in \{0,1,\cdots,\lfloor 2^{nR_{\nr0}} \rfloor-1\} \eqdef {\cal S}_{\nr0}$
\end{itemize}
\item Phase 2: Generate random $(n\cdot\Delta_{2,n})$-length sequences
\begin{itemize}
  \item ${\bf u}^\pb_\nr(w_\nr)$ i.i.d. with $p^\pb(u_\nr)$, $w_\nr  \in \{0,1,\cdots,\lfloor 2^{nR_{\nr}} \rfloor-1\} \eqdef  {\cal S}_\nr$
  \item ${\bf u}^\pb_\nb(w_{\nr0})$ i.i.d. with $p^\pb(u_\nb)$, $w_{\nr0} \in {\cal S}_{\nr0}$
\end{itemize}
and define bin $B_i \eqdef \{w_\nr | w_\nr \in [(i-1)\cdot\lfloor2^{n(R_\nr-R_\na)}\rfloor +1, i\cdot\lfloor2^{n(R_\nr-R_\na)}\rfloor]\}$ for $i\in {\cal S}_\na$.
\end{enumerate}

{\em Encoding: } During phase 1, the encoders of node $\na$ and $\nb$ send the codewords $\mathbf{x}^{(1)}_\na(w_\na)$ and $\mathbf{x}^{(1)}_\nb(w_\nb)$ respectively. At the end of phase 1, relay $\nr$ decodes $\tilde{w}_\na$ and maps ${\bf y}_\nr^\pa$ to a message index $w_{\nr0}$ if there exists a $w_{\nr0}$ such that $({\bf y}_\nr^\pa , \hat{\bf y}_\nr^\pa(w_{\nr0}))\in A^\pa(Y_\nr {\hat Y}_\nr)$. Such a $w_{\nr0}$ exists with high probability if
\begin{align}
R_{\nr0} = \Delta_{1,n} I(Y_\nr^\pa;\hat{Y}_\nr^\pa) + \epsilon \label{eq:mabc:cfdf:7}
\end{align}
and $n$ is sufficiently large. We choose
\begin{align}
R_\nr = \Delta_{2,n} I(U_\nr^\pb;Y_\nb^\pb) - 4 \epsilon. \label{eq:mabc:cfdf:0}
\end{align}
To choose $w_\nr$, the relay first selects the bin $B_{{\tilde w}_\na}$ and then it searches for the minimum $w_\nr \in B_{{\tilde w}_\na}$ such that $({\bf u}_\nr^\pb(w_\nr) , {\bf u}_\nb^\pb(w_{\nr0}))\in A^\pb(U_\nr U_\nb)$. This ensures uniqueness of $w_\nr$ if such a $w_\nr$ exists, i.e., $w_\nr$ is a function of $(w_\na,w_{\nr0})$. Such a $w_{\nr}$ exists with high probability if
\begin{align}
|B_{{\tilde w}_\na}| \geq 2^{n(\Delta_{2,n} I(U_\nr^\pb;U_\nb^\pb) + \epsilon)}. \label{eq:mabc:cfdf:5}
\end{align}
Since $|B_i| = 2^{n(R_\nr-R_\na)}$, $\forall i \in {\cal S}_\na$, this condition is equivalent to
\begin{align}
R_\na < \Delta_{2,n} I(U_\nr^\pb;Y_\nb^\pb) - \Delta_{2,n} I(U_\nr^\pb;U_\nb^\pb)- 5 \epsilon. \label{eq:mabc:cfdf:8}
\end{align}
The relay then sends ${\bf x}^\pb_\nr$ randomly generated i.i.d. according to $p^\pb(x_\nr|u_\nr,u_\nb)$ with ${\bf u}_\nr^\pb(w_\nr)$ and ${\bf u}_\nb^\pb(w_{\nr0})$ during phase 2.

{\em Decoding: } Node $\na$ estimates $\tilde{w}_{\nr0}$ after phase 2 using jointly typical decoding. First, since $\na$ knows $w_\na$, it can reduce the cardinality of $w_\nr$ to $\lfloor 2^{n(R_\nr -R_\na)}\rfloor$. Furthermore, it forms two sets of $\tilde{w}_{\nr0}$ based on typical sequences, $\{\tilde{w}_{\nr0}|({\bf x}_\na^\pa(w_\na),\hat{{\bf y}}_\nr^\pa(\tilde{w}_{\nr0}))\in A^\pa(X_\na {\hat Y}_\nr)\}$ and $\{\tilde{w}_{\nr0}|({\bf u}^\pb_\nr({\tilde w}_\nr),{\bf u}^\pb_\nb({\tilde w}_{\nr0}),{\bf y}_\na^\pb) \in A^\pb(U_\nr U_\nb Y_\na), {\tilde w}_\nr \in B_{w_\na}\}$. After decoding $\tilde{w}_{\nr0}$ (which is a success if there is a single common element in both of the previous sets), $\na$ decodes $\tilde{w}_\nb$ using jointly typical decoding of the sequence $({\bf x}_\na^\pa,{\bf x}_\nb^\pa, \hat{{\bf y}}_\nr^\pa)$. Node $\nb$ decodes $\tilde{w}_\nr$ after phase 2 and from the bin index of ${\tilde w}_\nr$ it estimates ${\tilde w}_\na$.

{\em Error analysis: } By the union bound,
\begin{align}
  P[E_{\na,\nb}] & \leq P[E_{\na,\nr}^\pa \cup E_{\nr,\nb}^{\pb}]\\
  & \leq P[E_{\na,\nr}^\pa] + P[E_{\nr,\nb}^{\pb} | \bar{E}_{\na,\nr}^\pa]\\
  P[E_{\nb,\na}] & \leq P[E_{\nr,\na}^\pb \cup E_{\nb,\na}^\pb]\\
  & \leq P[E_{\nr,\na}^\pb] + P[E_{\nb,\na}^\pb | \bar{E}_{\nr,\na}^\pb]
\end{align}
We define error events in each phase as follows:
\begin{enumerate}
\item $E_{\na,\nr}^\pa = E_{\na,\nr}^{\pa,1} \cup E_{\na,\nr}^{\pa,2}$.
\begin{description}
\item[$E_{\na,\nr}^{\pa,1}$ ]: $({\bf x}_\na^\pa(w_\na),{\bf y}_\nr^\pa) \not \in A^\pa(X_\na Y_\nr)$.
\item[$E_{\na,\nr}^{\pa,2}$ ]: there exists ${\tilde w}_\na \neq w_\na$ such that $({\bf x}_\na^\pa({\tilde w}_\na),{\bf y}_\nr^\pa)  \in A^\pa(X_\na Y_\nr)$.
\end{description}
\item $E_{\nr,\nb}^\pb = E_{\nr,\nb}^{\pb,1} \cup E_{\nr,\nb}^{\pb,2}\cup E_{\nr,\nb}^{\pb,3}\cup E_{\nr,\nb}^{\pb,4}$.
\begin{description}
\item[$E_{\nr,\nb}^{\pb,1}$ ]: there does not exist a $w_{\nr 0}$ such that $({\bf y}_\nr^\pa,{\hat {\bf y}}_\nr^\pa(w_{\nr0})) \in A^\pa(Y_\nr {\hat Y}_\nr)$.
\item[$E_{\nr,\nb}^{\pb,2}$ ]: there does not exist a $w_{\nr}\in B_{w_\na}$ such that $({\bf u}_\nr^\pb(w_\nr),{\bf u}_\nb^\pb(w_{\nr0})) \in A^\pb(U_\nr U_\nb)$.
\item[$E_{\nr,\nb}^{\pb,3}$ ]: $({\bf u}_\nr^\pb(w_\nr),{\bf y}_\nb^\pb) \not \in A^\pb(U_\nr Y_\nb)$.
\item[$E_{\nr,\nb}^{\pb,4}$ ]: there exists ${\tilde w}_\nr \neq w_\nr$ such that $({\bf u}_\nr^\pb({\tilde w}_\nr),{\bf y}_\nb^\pb)  \in A^\pb(U_\nr Y_\nb)$.
\end{description}
\item $E_{\nr,\na}^\pb = E_{\nr,\na}^{\pb,1} \cup E_{\nr,\na}^{\pb,2}\cup E_{\nr,\na}^{\pb,3}\cup E_{\nr,\na}^{\pb,4}\cup E_{\nr,\na}^{\pb,5}\cup E_{\nr,\na}^{\pb,6}$.
\begin{description}
\item[$E_{\nr,\na}^{\pb,1}$ ]: there does not exist a $w_{\nr 0}$ such that $({\bf y}_\nr^\pa,{\hat {\bf y}}_\nr^\pa(w_{\nr0})) \in A^\pa(Y_\nr {\hat Y}_\nr)$.
\item[$E_{\nr,\na}^{\pb,2}$ ]: there does not exist a $w_{\nr} \in B_{w_\na}$ such that $({\bf u}_\nr^\pb(w_\nr),{\bf u}_\nb^\pb(w_{\nr0})) \in A^\pb(U_\nr U_\nb)$.
\item[$E_{\nr,\na}^{\pb,3}$ ]: $({\bf x}_\na^\pa(w_{\na}),{\hat{\bf y}}_\nr^\pa(w_{\nr0})) \not \in A^\pa(X_\na {\hat Y}_\nr)$.
\item[$E_{\nr,\na}^{\pb,4}$ ]: $({\bf u}_\nr^\pb(w_\nr),{\bf u}_\nb^\pb(w_{\nr0}),{\bf y}_\na^\pb) \not \in A^\pb(U_\nr U_\nb Y_\na)$.
\item[$E_{\nr,\na}^{\pb,5}$ ]: there exists $({\tilde w}_\nr ,{\tilde w}_{\nr0})$ where ${\tilde w}_\nr \neq w_\nr$ and ${\tilde w}_{\nr0} \neq w_{\nr0}$ such that $({\bf u}_\nr^\pb({\tilde w}_\nr),{\bf u}_\nb^\pb({\tilde w}_{\nr0}),{\bf y}_\na^\pb)  \in A^\pb(U_\nr U_\nb Y_\na)$ and $({\bf x}_\na^\pa(w_\na),{\hat{\bf y}}_\nr^\pa({\tilde w}_{\nr0}))  \in A^\pa(X_\na {\hat Y}_\nr)$. Recall, $w_\nr$ is uniquely specified by $(w_\na,w_{\nr0})$. Hence for a given $w_\na$, there are at most $2^{nR_{\nr0}}$ such $({\tilde w}_\nr,{\tilde w}_{\nr0})$ pairs.
\item[$E_{\nr,\na}^{\pb,6}$ ]: there exists ${\tilde w}_{\nr0} \neq w_{\nr0}$  such that $({\bf u}_\nr^\pb(w_\nr),{\bf u}_\nb^\pb({\tilde w}_{\nr0}),{\bf y}_\na^\pb)  \in A^\pb(U_\nr U_\nb Y_\na)$,  \\$({\bf x}_\na^\pa(w_\na),{\hat{\bf y}}_\nr^\pa({\tilde w}_{\nr0}))  \in A^\pa(X_\na {\hat Y}_\nr)$.
\end{description}
\item $E_{\nb,\na}^\pb = E_{\nb,\na}^{\pb,1} \cup E_{\nb,\na}^{\pb,2}$.
\begin{description}
\item[$E_{\nb,\na}^{\pb,1}$ ]: $({\bf x}_\na^\pa(w_\na),{\bf x}_\nb^\pa(w_\nb),{\hat{\bf y}}_\nr^\pa(w_{\nr0})) \not \in A^\pa(X_\na X_\nb {\hat Y}_\nr)$.
\item[$E_{\nb,\na}^{\pb,2}$ ]: there exists ${\tilde w}_\nb \neq w_\nb$ such that $({\bf x}_\na^\pa(w_\na),{\bf x}_\nb^\pa({\tilde w}_\nb),{\hat{\bf y}}_\nr^\pa(w_{\nr0}))  \in A^\pa(X_\na X_\nb {\hat Y}_\nr)$.
\end{description}
\end{enumerate}
Then:
\begin{align}
P[E_{\na,\nr}^\pa] \leq & P[E_{\na,\nr}^{\pa,1}] + P[E_{\na,\nr}^{\pa,2}]\\
=&P[\bar{D}^\pa({\bf x}_\na(w_\na),{\bf y}_\nr)] + P[\cup_{\tilde{w}_\na \neq w_\na} D^\pa({\bf x}_\na({\tilde w}_\na),{\bf y}_\nr)]\\
\leq &\epsilon + 2^{n(R_\na - \Delta_{1,n}I(X_\na^\pa;Y_\nr^\pa)+ 3\epsilon)}\label{eq:mabc:cfdf:1}\\
P[E_{\nr,\nb}^{\pb} | \bar{E}_{\na,\nr}^\pa] \leq &P[E_{\nr,\nb}^{\pb,1} | \bar{E}_{\na,\nr}^\pa] + P[E_{\nr,\nb}^{\pb,2} | \bar{E}_{\na,\nr}^\pa]+P[E_{\nr,\nb}^{\pb,3} | \bar{E}_{\na,\nr}^\pa]+P[E_{\nr,\nb}^{\pb,4} | \bar{E}_{\na,\nr}^\pa]
\\
\leq& 2\epsilon + P[\bar{D}^\pb({\bf u}_\nr(w_\nr),{\bf y}_\nb)] +
P[\cup_{\tilde{w}_{\nr} \neq w_{\nr}} D^\pb({\bf u}_\nr({\tilde w}_\nr),{\bf y}_\nb)]\label{eq:mabc:cfdf:6}\\
\leq & 3\epsilon + 2^{n(R_\nr - \Delta_{2,n}I(U_\nr^\pb;Y_\nb^\pb) +3\epsilon)} \label{eq:mabc:cfdf:2}
\end{align}
 In \eqref{eq:mabc:cfdf:6}, $P[E_{\nr,\nb}^{\pb,1} | \bar{E}_{\na,\nr}^\pa]$ and $P[E_{\nr,\nb}^{\pb,2} | \bar{E}_{\na,\nr}^\pa]$ are less than $\epsilon$ due to \eqref{eq:mabc:cfdf:7} and \eqref{eq:mabc:cfdf:8}, respectively. Furthermore,
\begin{align}
P[E_{\nr,\na}^\pb] \leq& P[E_{\nr,\na}^{\pb,1}]+P[E_{\nr,\na}^{\pb,2}]+P[E_{\nr,\na}^{\pb,3}]+P[E_{\nr,\na}^{\pb,4}]+P[E_{\nr,\na}^{\pb,5}]+P[E_{\nr,\na}^{\pb,6}]\\
\leq & 2\epsilon + P[\bar{D}^\pa({\bf x}_\na (w_\na),{\hat{\bf y}}_\nr(w_{\nr0}))] +P[\bar{D}^\pb({\bf u}_\nr(w_\nr),{\bf u}_\nb(w_{\nr0}),{\bf y}_\na)]   + \nonumber\\
&P[\cup_{\twolines{{\tilde w}_{\nr} \neq
w_{\nr}}{{\tilde w}_{\nr0} \neq
w_{\nr0}}} D^\pb({\bf u}_\nr({\tilde w}_{\nr}),{\bf u}_\nb({\tilde w}_{\nr0}),{\bf y}_\na),D^\pa({\bf x}_\na (w_\na),{\hat{\bf y}}_\nr({\tilde w}_{\nr0})))] + \nonumber\\
&P[\cup_{\twolines{{\tilde w}_{\nr} =
w_{\nr}}{{\tilde w}_{\nr0} \neq
w_{\nr0}}} D^\pb({\bf u}_\nr(w_{\nr}),{\bf u}_\nb({\tilde w}_{\nr0}),{\bf y}_\na),D^\pa({\bf x}_\na (w_\na),{\hat{\bf y}}_\nr({\tilde w}_{\nr0})))] \label{eq:mabc:cfdf:9}\\
\leq& 4\epsilon +  2^{n(R_{\nr0} - \Delta_{2,n} I(U_\nr^\pb,U_\nb^\pb;Y_\na^\pb)-\Delta_{1,n}I(\hat{Y}_\nr^\pa;X_\na^\pa) + 7\epsilon)}+ \nonumber \\
&2^{n(R_{\nr0} -\Delta_{2,n} I(U_\nb^\pb;U_\nr^\pb,Y_\na^\pb)-\Delta_{1,n}I(\hat{Y}_\nr^\pa;X_\na^\pa) + 7\epsilon)}\label{eq:mabc:cfdf:3}\\
P[E_{\nb,\na}^\pb | \bar{E}_{\nr,\na}^\pb] \leq& P[E_{\nb,\na}^{\pb,1} | \bar{E}_{\nr,\na}^\pb]+P[E_{\nb,\na}^{\pb,2} | \bar{E}_{\nr,\na}^\pb]\\
=&P[\bar{D}^\pa({\bf x}_\na(w_\na),{\bf x}_\nb(w_\nb),{\hat{\bf y}}_\nr(w_{\nr0}))] + P[\cup_{\tilde{w}_\nb \neq w_\nb} D^\pa({\bf x}_\na(w_\na),{\bf x}_\nb(\tilde{w}_\nb),{\hat {\bf y}}_\nr(w_{\nr0}))]\\
\leq&  \epsilon + 2^{n(R_\nb - \Delta_{1,n}I(X_\nb^\pa;\hat{Y}_\nr^\pa|X_\na^\pa)+4\epsilon)}\label{eq:mabc:cfdf:4}
\end{align}
 In \eqref{eq:mabc:cfdf:9}, $P[E_{\nr,\na}^{\pb,1}]$ and $P[E_{\nr,\na}^{\pb,2}]$ are less than $\epsilon$ due to \eqref{eq:mabc:cfdf:7} and \eqref{eq:mabc:cfdf:8}, respectively. In \eqref{eq:mabc:cfdf:3}, $P[\bar{D}^\pa({\bf x}_\na (w_\na),{\hat{\bf y}}_\nr(w_{\nr0}))]$ is less than $\epsilon$ by the Markov lemma and the total cardinality of the case (${\tilde w}_{\nr} \neq
w_{\nr}$ and ${\tilde w}_{\nr0} \neq
w_{\nr0}$) is bounded by $2^{nR_{\nr0}}$ since ${\tilde w}_\nr$ is uniquely specified if $(w_\na,{\tilde w}_{\nr0})$ is given.

Since $\epsilon > 0$ is arbitrary, the conditions of Theorem
\ref{theorem:MABC:CFDF}, \eqref{eq:mabc:cfdf:0} and the AEP property guarantee that the
right hand sides of \eqref{eq:mabc:cfdf:1}, \eqref{eq:mabc:cfdf:2},
\eqref{eq:mabc:cfdf:3} and \eqref{eq:mabc:cfdf:4} corresponding to the first term of \eqref{eq:mabc:cfdf:b:1}, \eqref{eq:mabc:cfdf:0}, \eqref{eq:mabc:cfdf:b:2} and \eqref{eq:mabc:cfdf:b:3} respectively vanish as $n
\rightarrow \infty$.
By the Carath\'{e}odory theorem in \cite{Hiriart:2001}, it is
sufficient to restrict $|{\cal Q}| \leq 7$.
\end{proof}

The mixed MABC region in Theorem \ref{theorem:MABC:CFDF} is outer
bounded by the DF MABC region in Theorem \ref{theorem:MABC:DF}. In
the mixed MABC protocol, the relay $\nr$ has to be able to decode
$w_\na$ correctly after phase 1 without any information about
$w_\nb$. If node $\na$ can decode $w_\nb$ from a compressed version of $y_\nr$ and knowledge of
$w_\na$, then by the information processing inequality, node $\nr$ can decode $w_\nb$ from
$y_\nr$ and $w_\na$.

\begin{theorem}
\label{theorem:MABC:comp}
The achievable rate region of the mixed MABC protocol of Theorem \ref{theorem:MABC:CFDF} is outer bounded by the achievable rate region of the DF MABC protocol of Theorem \ref{theorem:MABC:DF}.
\thmend
\end{theorem}

\begin{proof}
If $R_\na$ and $R_\nb$ lies in the region of the mixed MABC protocol of Theorem \ref{theorem:MABC:CFDF} for a given distribution $p^\pa(x_\na)p^\pa(x_\nb)p^\pa(y_\nr|x_\na,x_\nb)p^\pa(\hat{y}_\nr|y_\nr)p^\pb(x_\nr)p^\pb(y_\na,y_\nb|x_\nr)$  then:
\begin{align}
R_\na &< \Delta_1 I(X_\na^\pa;Y_\nr^\pa) \leq \Delta_1 I(X_\na^\pa;Y_\nr^\pa|X_\nb^\pa) \label{eq:mabc:comp:1}\\
R_\na &< \Delta_2I(U_\nr^\pb;Y_\nb^\pb) - \Delta_2 I(U_\nr^\pb;U_\nb^\pb) \leq \Delta_2I(X_\nr^\pb;Y_\nb^\pb)\label{eq:mabc:comp:3}\\
R_\nb &< \Delta_1 I(X_\nb^\pa; \hat{Y}_\nr^\pa|X_\na^\pa) \leq \Delta_1 I(X_\nb^\pa; Y_\nr^\pa|X_\na^\pa)\\
R_\nb &< \Delta_1 I(X_\nb^\pa; \hat{Y}_\nr^\pa|X_\na^\pa) \leq \Delta_1 I(Y_\nr^\pa;\hat{Y}_\nr^\pa|X_\na^\pa) \nonumber\\
&\leq  \Delta_2 I(U_\nr^\pb,U_\nb^\pb;Y_\na^\pb)\leq \Delta_2 I(X_\nr^\pb;Y_\na^\pb)\label{eq:mabc:comp:4}\\
R_\na + R_\nb &< \Delta_1 I(X_\na^\pa;Y_\nr^\pa) + \Delta_1 I(X_\nb^\pa; Y_\nr^\pa|X_\na^\pa) \leq \Delta_1 I(X_\na^\pa, X_\nb^\pa ; Y_\nr^\pa)\label{eq:mabc:comp:2}
\end{align}
\eqref{eq:mabc:comp:3} and \eqref{eq:mabc:comp:4} are from the Markov process $(U_\nr^\pb,  U_\nb^\pb)\rightarrow X_\nr^\pb \rightarrow (Y_\na^\pb, Y_\nb^\pb)$.
From \eqref{eq:mabc:comp:1} -- \eqref{eq:mabc:comp:2}, $R_\na$ and $R_\nb$ are in the region of the DF MABC protocol in the Theorem \ref{theorem:MABC:DF}. Therefore, every point $(R_\na, R_\nb)$ represented by a convex combination of distributions (or using $Q$) in the Mixed MABC protocol is also achievable with the DF MABC protocol.
\end{proof}
From Theorem \ref{theorem:MABC:comp}, the mixed MABC protocol does not achieve any rate pairs which cannot be achieved by the DF MABC protocol. However, its possible benefit lies in that it only requires the relay $\nr$ to possess \emph{one} of the codebooks of $\na$ and $\nb$. Therefore, in the event that relay $\nr$ has one of the codebooks of the terminal nodes, by employing the Mixed MABC protocol one can achieve a rate region which outperforms that of the CF MABC protocol.
In practice, if there are many terminal and relay nodes a relay may have some but not all codebooks of the terminal nodes. If the relay has full codebook knowledge then DF is possible, otherwise CF (or AF) may be more appropriate.
\subsection{TDBC Protocol}

The Time Division Broadcast (TDBC) protocol consists of three phases rather than the two seen in the MABC protocol. The MABC protocol takes advantage of the various gains provided by multiple-access schemes by having both nodes $\na$ and $\nb$ transmit during phase 1. However, the possible direct links between nodes $\na$ and $\nb$ are not exploited. The TDBC protocol aims to exploit the direct link by having the nodes combine the signals received on the direct link and through the relay node, that is, exploit the side-information available at the decoders.

\begin{figure}
\centerline{\epsfig{figure=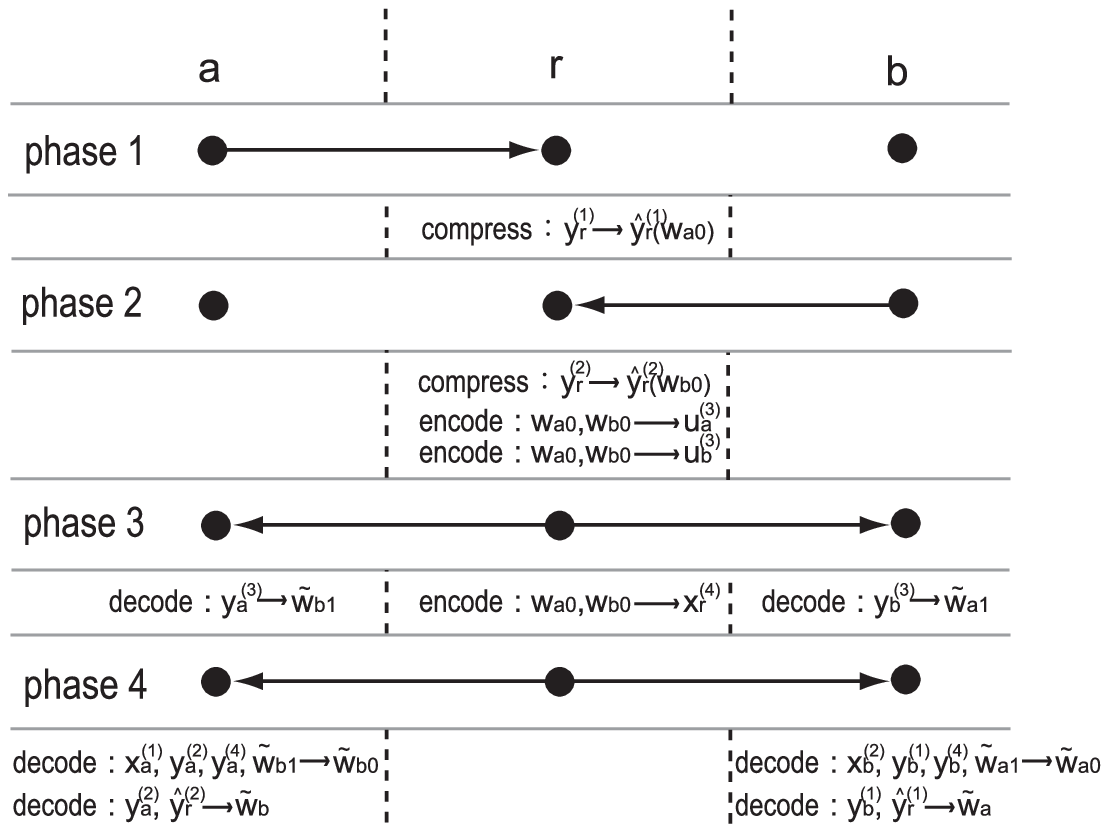, width=8cm}}
\caption{The three-phase TDBC protocol with a relay using a CF scheme.}
\label{fig:TDBC:CF}
\end{figure}

\begin{figure}
\centerline{\epsfig{figure=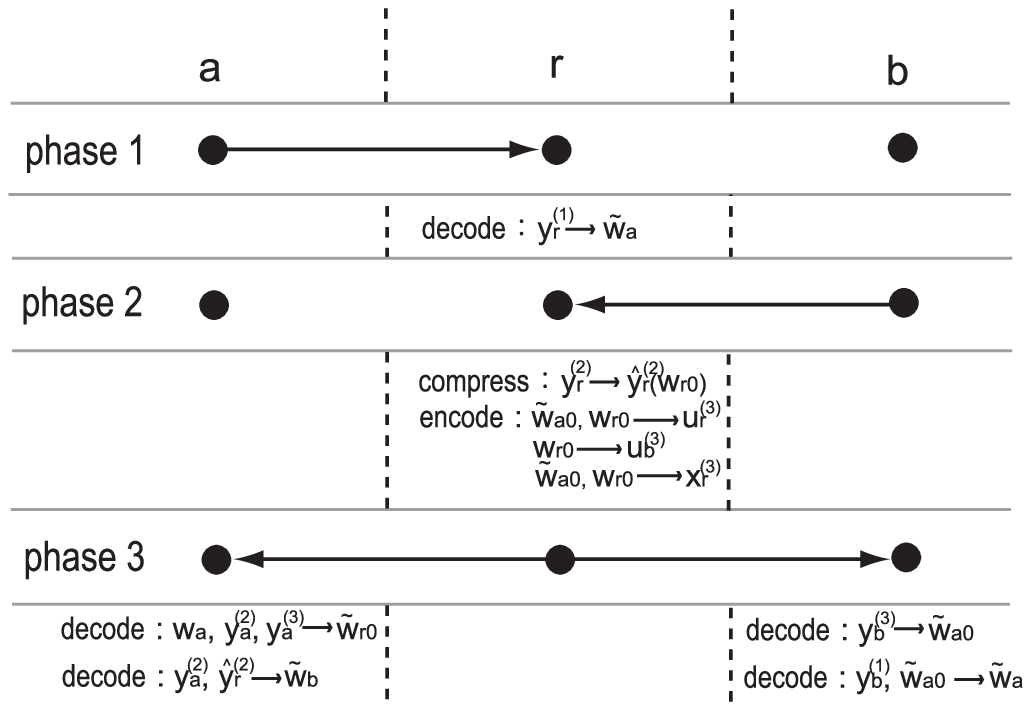, width=8cm}}
\caption{The three-phase TDBC protocol with a relay using a mixed scheme.}
\label{fig:TDBC:CFDF}
\end{figure}

The TDBC protocol consists of three phases, as illustrated in \Fig \ref{fig:TDBC:CF} and \ref{fig:TDBC:CFDF}. During phase 1, node $\na$ is the sole node to transmit, while both the relay and node $\nb$ receive this transmission. During phase 2, node $\nb$ transmits while the relay and node $\na$ receive. After phase 2, the relay processes the signals received during the first 2 phases and proceeds to broadcast to nodes $\na$ and $\nb$ during the third phase.

 In the CF TDBC protocol, we use two different broadcasting schemes in the last phase. For this reason, we divide phase 3 into two sub-phases. In the first relay-broadcast phase, we use Marton's broadcast scheme of \cite{Marton:1979}, in which two different messages are transmitted to the two receivers. In this scheme, neither receiver uses side information ($w_\na$ at node $\na$ and $w_\nb$ at node $\nb$) to decode the messages. In the second relay-broadcasting phase, we assume a compound channel, i.e., a common message is transmitted to the two receivers which have different side information.

For convenience of analysis, we denote the first part of the relay-broadcast phase as phase 3 and the second as phase 4.


\begin{theorem}
\label{theorem:TDBC:CF} An achievable rate region of the half-duplex
bi-directional relay channel with the compress and forward TDBC
protocol is the closure of the set of all points $(R_\na,R_\nb)$ satisfying
\begin{align}
R_{\na} &< \Delta_1 I(X_\na^\pa; \hat{Y}_\nr^\pa, Y_\nb^\pa|Q)\label{thm:TDBC:CF:6}\\
R_{\nb} &< \Delta_2 I(X_\nb^\pb; \hat{Y}_\nr^\pb, Y_\na^\pb|Q)\label{thm:TDBC:CF:7}
\end{align}
subject to
\begin{align}
&\alpha_\na \Delta_1 I(Y_\nr^\pa;{\hat Y}_\nr^\pa|Q) <\Delta_3 I(U_\na^\pc;Y_\nb^\pc|Q)\label{thm:TDBC:CF:3}\\
&\alpha_\nb \Delta_2 I(Y_\nr^\pb;{\hat Y}_\nr^\pb|Q) <\Delta_3 I(U_\nb^\pc;Y_\na^\pc|Q)\label{thm:TDBC:CF:4}\\
&\alpha_\na \Delta_1 I(Y_\nr^\pa;{\hat Y}_\nr^\pa|Q) + \alpha_\nb \Delta_2 I(Y_\nr^\pb;{\hat Y}_\nr^\pb|Q) <\Delta_3 I(U_\na^\pc;Y_\nb^\pc|Q) + \Delta_3 I(U_\nb^\pc;Y_\na^\pc|Q)-\Delta_3 I(U_\na^\pc;U_\nb^\pc|Q)\label{thm:TDBC:CF:5}\\
&(1-\alpha_\na)\Delta_1 I(Y_\nr^\pa;\hat{Y}_\nr^\pa|Q )+\Delta_2 I(Y_\nr^\pb;\hat{Y}_\nr^\pb|X_\nb^\pb,Q) < \Delta_4 I(X_\nr^\pd;Y_\nb^\pd) + \Delta_1I({\hat Y}_\nr^\pa;Y_\nb^\pa|Q)  \label{thm:TDBC:CF:1}\\
&(1-\alpha_\nb)\Delta_2 I(Y_\nr^\pb;\hat{Y}_\nr^\pb|Q )+\Delta_1
I(Y_\nr^\pa;\hat{Y}_\nr^\pa|X_\na^\pa,Q) < \Delta_4
I(X_\nr^\pd;Y_\na^\pd) + \Delta_2 I({\hat Y}_\nr^\pb;Y_\na^\pb|Q) \label{thm:TDBC:CF:2}
\end{align}
where $0< \alpha_\na, \alpha_\nb <1$ over all joint distributions,
\begin{align}
p(q,x_\na,x_\nb,&x_\nr,y_\na,y_\nb,y_\nr,\hat{y}_\nr) \nonumber\\
&=p(q)p^\pa(x_\na,y_\nb,y_\nr,\hat{y}_\nr|q)p^\pb(x_\nb,y_\na,y_\nr,\hat{y}_\nr|q)p^\pc(u_\na,u_\nb,x_\nr,y_\na,y_\nb|q)p^\pd(x_\nr,y_\na,y_\nb)
\end{align}
where
\begin{align}
 p^\pa(x_\na,y_\nb,y_\nr,\hat{y}_\nr|q) &=  p^\pa(x_\na|q)p^\pa(y_\nb, y_\nr|x_\na)p^\pa(\hat{y}_\nr|y_\nr,q)\\
 p^\pb(x_\nb,y_\na,y_\nr,\hat{y}_\nr|q) &=  p^\pb(x_\nb|q)p^\pb(y_\na, y_\nr|x_\nb)p^\pb(\hat{y}_\nr|y_\nr,q)\\
 p^\pc(u_\na,u_\nb,x_\nr,y_\na,y_\nb|q) &=  p^\pc(u_\na,u_\nb|q)p^\pc(x_\nr|u_\na,u_\nb,q)p^\pc(y_\na,y_\nb|x_\nr)\\
 p^\pd(x_\nr,y_\na,y_\nb) &=  p^\pd(x_\nr)p^\pd(y_\na,y_\nb|x_\nr)
\end{align}
with $|{\cal Q}| \leq 13$ over the alphabet ${\cal X}_\na \times {\cal X}_\nb \times {\cal X}_\nr^2
\times {\cal Y}_\na^3 \times {\cal Y}_\nb^3 \times {\cal Y}_\nr^2 \times \hat{{\cal Y}}_\nr^2$. \thmend
\end{theorem}

\begin{remark}
If side information is very limited, then with $\Delta_4 \rightarrow 0$, $\alpha_\na , \alpha_\nb \rightarrow 1$ the relay phase acts as a classical broadcast channel. At the opposite extreme, the side information cancels out all interference, i.e., $I(Y_\nr^\pa;{\hat Y}_\nr^\pa|X_\na^\pa) = I(Y_\nr^\pb;{\hat Y}_\nr^\pb|X_\nb^\pb) = 0$. Then we set $\Delta_3 \rightarrow 0$, $\alpha_\na , \alpha_\nb \rightarrow 0$. \footnote{ This choice of $\Delta_3, \alpha_\na, \alpha_\nb$ is on the boundary of the closure of the achievable rate region.}
\end{remark}

\begin{proof}
{\em Random code generation: } For simplicity of exposition, we take $|{\cal Q}| =1 $.
\begin{enumerate}
\item Phase 1: Generate random $(n\cdot\Delta_{1,n})$-length sequences
\begin{itemize}
  \item ${\bf x}^\pa_\na(w_\na)$ i.i.d. with $p^\pa(x_\na)$, $w_\na \in {\cal S}_\na =\{0,1,\cdots,\lfloor 2^{nR_{\na}} \rfloor-1\} $
  \item $\hat{\bf y}_\nr^\pa(w_{\na0})$ i.i.d. with $p^\pa(\hat{y}_\nr) = \sum_{y_\nr} p^\pa(y_\nr) p^\pa(\hat{y}_\nr|y_\nr)$ , $w_{\na0} \in \{0,1,\cdots,\lfloor 2^{nR_{\na0}} \rfloor-1\} \eqdef {\cal S}_{\na0}$
\end{itemize}
 and generate a partition of ${\cal S}_{\na0}$  randomly by independently assigning every index $w_{\na0} \in {\cal S}_{\na0}$ to a set ${\cal S}_{\na0,i}$, with a uniform distribution over the indices $i \in
\{0, \ldots, \lfloor 2^{nR_{\na1}} \rfloor - 1\} \eqdef \mathcal{S}_{\na1}$. We denote by $s_{\na0}(w_{\na0})$
the index $i$ of ${\cal S}_{\na0,i}$ to which $w_{\na0}$  belongs.
\item Phase 2: Generate random $(n\cdot\Delta_{2,n})$-length sequences
\begin{itemize}
  \item ${\bf x}^\pb_\nb(w_\nb)$ i.i.d. with $p^\pb(x_\nb)$, $w_\nb \in {\cal S}_\nb =\{0,1,\cdots,\lfloor 2^{nR_{\nb}} \rfloor-1\} $
  \item $\hat{\bf y}_\nr^\pb(w_{\nb0})$ i.i.d. with $p^\pb(\hat{y}_\nr) = \sum_{y_\nr} p^\pb(y_\nr) p^\pb(\hat{y}_\nr|y_\nr)$ , $w_{\nb0} \in \{0,1,\cdots,\lfloor 2^{nR_{\nb0}} \rfloor-1\} \eqdef {\cal S}_{\nb0}$
\end{itemize}
 and generate a partition of ${\cal S}_{\nb0}$  randomly by independently assigning every index $w_{\nb0} \in {\cal S}_{\nb0}$ to a set ${\cal S}_{\nb0,i}$, with a uniform distribution over the indices $i \in
\{0, \ldots, \lfloor 2^{nR_{\nb1}} \rfloor - 1\} \eqdef \mathcal{S}_{\nb1}$. We denote by $s_{\nb0}(w_{\nb0})$
the index $i$ of ${\cal S}_{\nb0,i}$ to which $w_{\nb0}$  belongs.
\item Phase 3: Generate random $(n\cdot\Delta_{3,n})$-length sequences
\begin{itemize}
  \item ${\bf u}^\pc_\na(w_{\na2})$ i.i.d with $p^\pc(u_\na)$, $w_{\na2}  \in \{0,1,\cdots,\lfloor 2^{nR_{\na2}} \rfloor-1\} \eqdef  {\cal S}_{\na2}$
  \item ${\bf u}^\pc_\nb(w_{\nb2})$ i.i.d with $p^\pc(u_\nb)$, $w_{\nb2}  \in \{0,1,\cdots,\lfloor 2^{nR_{\nb2}} \rfloor-1\} \eqdef  {\cal S}_{\nb2}$
\end{itemize}
 and define bin $B_j \eqdef \{w_{\na2} | w_{\na2} \in [(j-1)\cdot\lfloor2^{n(R_{\na2}-R_{\na1})}\rfloor +1, j\cdot\lfloor2^{n(R_{\na2}-R_{\na1})}\rfloor]\}$ for $j\in {\cal S}_{\na1}$. Likewise, $C_k \eqdef \{w_{\nb2} | w_{\nb2} \in [(k-1)\cdot\lfloor2^{n(R_{\nb2}-R_{\nb1})}\rfloor +1, k\cdot\lfloor2^{n(R_{\nb2}-R_{\nb1})}\rfloor]\}$ for $k\in {\cal S}_{\nb1}$.
\item Phase 4: Generate random $(n\cdot\Delta_{4,n})$-length sequences
\begin{itemize}
  \item ${\bf x}^\pd_\nr(w_{\na0},w_{\nb0})$ i.i.d with $p^\pd(x_\nr)$, $w_{\na0}\in{\cal S}_{\na0}$ and $w_{\nb0}\in{\cal S}_{\nb0}$.
\end{itemize}
\end{enumerate}

{\em Encoding: }
During phase 1 (resp. phase 2), the encoder of node $\na$ (resp. $\nb$) sends the codeword ${\bf x}^\pa_\na(w_\na)$ (resp. ${\bf x}^\pb_\nb(w_\nb)$). At the end of phase 1, relay $\nr$ compresses the received signal ${\bf y}_\nr^\pa$ into the message $w_{\na0}$ if there exists a $w_{\na0}$ such that $({\bf y}_\nr^\pa , \hat{\bf y}_\nr^\pa(w_{\na0}))\in A^\pa(Y_\nr {\hat Y}_\nr)$. Similarly, $\nr$ compresses ${\bf y}_\nr^\pb$ into the message $w_{\nb0}$ at the end of phase 2. There exist such $w_{\na0}$ and $w_{\nb0}$ with high probability if
\begin{align}
R_{\na0} & = \Delta_{1,n} I(Y_\nr^\pa;\hat{Y}_\nr^\pa) + \epsilon \label{eq:tdbc:cf:6}\\
R_{\nb0} & = \Delta_{2,n} I(Y_\nr^\pb;\hat{Y}_\nr^\pb) + \epsilon \label{eq:tdbc:cf:7}
\end{align}
and $n$ is sufficiently large. Also we choose $R_{\na1}$, $R_{\nb1}$, $R_{\na2}$ and $R_{\nb2}$ as:
\begin{align}
R_{\na1} &= \alpha_\na R_{\na0} = \alpha_\na(\Delta_{1,n} I(Y_\nr^\pa;\hat{Y}_\nr^\pa) + \epsilon)\\
R_{\nb1} &= \alpha_\nb R_{\nb0} = \alpha_\nb(\Delta_{2,n} I(Y_\nr^\pb;\hat{Y}_\nr^\pb) + \epsilon)
\end{align}
and
\begin{align}
R_{\na1} \leq R_{\na2} &= \Delta_{3,n} I(U_\na^\pc;Y_\nb^\pc) -4\epsilon \label{eq:tdbc:cf:3}\\
R_{\nb1} \leq R_{\nb2} &= \Delta_{3,n} I(U_\nb^\pc;Y_\na^\pc) -4\epsilon \label{eq:tdbc:cf:10}.
\end{align}
 From the code constructions of $w_{\na2}$ and $w_{\nb2}$, $R_{\na1}$ and $R_{\nb1}$ have to be less than $R_{\na2}$ and $R_{\nb2}$, respectively. Then the relay constructs $w_{\na1} = s_{\na0}(w_{\na0})$ and $w_{\nb1} = s_{\nb0}(w_{\nb0})$. To choose $w_{\na2}$ and $w_{\nb2}$, the relay first selects the bins $B_{w_{\na1}}$ and $C_{w_{\nb1}}$ and then it searches for a pair $(w_{\na2},w_{\nb2}) \in B_{w_{\na1}} \times C_{w_{\nb1}}$ such that $({\bf u}_\na^\pc(w_{\na2}),{\bf u}_\nb^\pc(w_{\nb2})) \in A^\pc (U_\na U_\nb)$. Such a $(w_{\na2},w_{\nb2})$ exists with high probability if
\begin{align}
R_{\na1} + R_{\nb1} < R_{\na2} + R_{\nb2} - \Delta_{3,n} I(U_\na^\pc;U_\nb^\pc)- \epsilon' \label{eq:tdbc:cf:8}
\end{align}
from the Lemma in \cite{Gamal:1981}.
The relay then sends ${\bf x}^\pc_\nr$ generated i.i.d. according to $p^\pc(x_\nr|u_\na,u_\nb)$ with ${\bf u}_\na^\pc(w_{\na2})$ and ${\bf u}_\nb^\pc(w_{\nb2})$ during phase 3. Finally, the relay sends ${\bf x}^\pd_\nr(w_{\na0}, w_{\nb0})$ during phase 4.

{\em Decoding: } Node $\na$ decodes $\tilde{w}_{\nb2}$ after phase 3 using jointly typical
decoding. Then $\na$ estimates ${\tilde w}_{\nb1}$ from the bin index of ${\tilde w}_{\nb2}$. Node $\na$ decodes ${\tilde w}_{\nb0}$ if there exists a unique ${\tilde w}_{\nb0}$ such that ${\tilde w}_{\nb0} \in {\cal S}_{\nb0,{\tilde w}_{\nb1}}$, $({\bf x}_\nr^\pd(\tilde{w}_{\na0},{\tilde w}_{\nb0}),{\bf y}_\na^\pd)\in A^\pd(X_\nr Y_\na)$, $({\bf
x}_\na^\pa(w_\na),\hat{{\bf y}}_\nr^\pa(\tilde{w}_{\na0}))\in
A^\pa(X_\na {\hat Y}_\nr)$ and $({\hat{\bf
y}}_\nb^\pb(w_{\nb0}),{\bf y}_\na^\pb)\in
A^\pb({\hat Y}_\nr Y_\na)$ . After decoding
$\tilde{w}_{\nb0}$, node $\na$ decodes $\tilde{w}_\nb$ using jointly
typical decoding of the sequence $({\bf x}_\nb^\pb, \hat{{\bf
y}}_\nr^\pb({\tilde w}_{\nb0}),{\bf y}_\na^\pb)$. Similarly, node $\nb$ decodes ${\tilde w}_\na$.

{\em Error analysis: }
\begin{align}
  P[E_{\nb,\na}] & \leq P[E_{\nr,\na}^{\pc} \cup E_{\nr,\na}^\pd \cup E_{\nb,\na}^\pd]\\
  & \leq P[E_{\nr,\na}^{\pc}] + P[E_{\nr,\na}^\pd | \bar{E}_{\nr,\na}^{\pc}] + P[E_{\nb,\na}^\pd | \bar{E}_{\nr,\na}^{\pc}\cap \bar{E}_{\nr,\na}^{\pd}]
\end{align}
We define error events in each phase as follows:
\begin{enumerate}
\item $E_{\nr,\na}^\pc = E_{\nr,\na}^{\pc,1} \cup E_{\nr,\na}^{\pc,2}\cup E_{\nr,\na}^{\pc,3}\cup E_{\nr,\na}^{\pc,4}\cup E_{\nr,\na}^{\pc,5}$.
\begin{description}
\item[$E_{\nr,\na}^{\pc,1}$ ]: there does not exist a $w_{\na0}$ such that $({\bf y}_\nr^\pa,{\hat {\bf y}}_\nr^\pa(w_{\na0})) \in A^\pa(Y_\nr {\hat Y}_\nr)$.
\item[$E_{\nr,\na}^{\pc,2}$ ]: there does not exist a $w_{\nb0}$ such that $({\bf y}_\nr^\pb,{\hat {\bf y}}_\nr^\pb(w_{\nb0})) \in A^\pb(Y_\nr {\hat Y}_\nr)$.
\item[$E_{\nr,\na}^{\pc,3}$ ]: there does not exist a pair $(w_{\na2},w_{\nb2}) \in B_{s_{\na0}(w_{\na0})}\times C_{s_{\nb0}(w_{\nb0})}$ such that $({\bf u}_\na^\pc(w_{\na2}),{\bf u}_\nb^\pc(w_{\nb2})) \in A^\pc(U_\na U_\nb)$.
\item[$E_{\nr,\na}^{\pc,4}$ ]: $({\bf u}_\nb^\pc(w_{\nb2}),{\bf y}_\na^\pc) \not \in A^\pc(U_\nb Y_\na)$.
\item[$E_{\nr,\na}^{\pb,5}$ ]: there exists ${\tilde w}_{\nb2} \neq w_{\nb2}$ such that $({\bf u}_\nb^\pc({\tilde w}_{\nb2}),{\bf y}_\na^\pc)  \in A^\pc(U_\nb Y_\na)$.
\end{description}
\item $E_{\nr,\na}^\pd = E_{\nr,\na}^{\pd,1} \cup E_{\nr,\na}^{\pd,2}\cup E_{\nr,\na}^{\pd,3} \cup E_{\nr,\na}^{\pd,4}$.
\begin{description}
\item[$E_{\nr,\na}^{\pd,1}$ ]: there does not exist a $w_{\na0}$ such that $({\bf y}_\nr^\pa,{\hat {\bf y}}_\nr^\pa(w_{\na0})) \in A^\pa(Y_\nr {\hat Y}_\nr)$.
\item[$E_{\nr,\na}^{\pd,2}$ ]: there does not exist a $w_{\nb0}$ such that $({\bf y}_\nr^\pb,{\hat {\bf y}}_\nr^\pb(w_{\nb0})) \in A^\pb(Y_\nr {\hat Y}_\nr)$.
\item[$E_{\nr,\na}^{\pd,3}$ ]: $({\bf x}_\nr^\pd(w_{\na0},w_{\nb0}),{\bf y}_\na^\pd) \not \in A^\pd(X_\nr Y_\na)$.
\item[$E_{\nr,\na}^{\pd,4}$ ]: there exists $({\tilde w}_{\na0} ,{\tilde w}_{\nb0})$ where ${\tilde w}_{\na0}\neq w_{\na0}$ and ${\tilde w}_{\nb0} \neq w_{\nb0}$ such that $({\bf x}_\nr^\pd({\tilde w}_{\na0},{\tilde w}_{\nb0}),{\bf y}_\na^\pd)  \in A^\pd(X_\nr Y_\na)$, $({\bf x}_\na^\pa(w_\na),{\hat{\bf y}}_\nr^\pa({\tilde w}_{\na0}))  \in A^\pa(X_\na {\hat Y}_\nr)$, $({\hat{\bf y}}_\nr^\pb({\tilde w}_{\nb0}),{\bf y}_\na^\pb)  \in A^\pb({\hat Y}_\nr Y_\na)$, and ${\tilde w}_{\nb0} \in S_{\nb0,s_{\nb0}(w_{\nb0})}$.
\item[$E_{\nr,\na}^{\pd,5}$ ]:  there exists ${\tilde w}_{\nb0} \neq w_{\nb0}$ such that $({\bf x}_\nr^\pd(w_{\na0},{\tilde w}_{\nb0}),{\bf y}_\na^\pd)  \in A^\pd(X_\nr Y_\na)$, $({\hat{\bf y}}_\nr^\pb({\tilde w}_{\nb0}),{\bf y}_\na^\pb)  \in A^\pb({\hat Y}_\nr Y_\na)$, and ${\tilde w}_{\nb0} \in S_{\nb0,s_{\nb0}(w_{\nb0})}$.

\end{description}
\item $E_{\nb,\na}^\pd = E_{\nb,\na}^{\pd,1} \cup E_{\nb,\na}^{\pd,2}$.
\begin{description}
\item[$E_{\nb,\na}^{\pd,1}$ ]: $({\bf x}_\nb^\pb(w_\nb),{\bf y}_\na^\pb,{\hat{\bf y}}_\nr^\pb(w_{\nb0})) \not \in A^\pb(X_\nb Y_\na {\hat Y}_\nr)$.
\item[$E_{\nb,\na}^{\pd,2}$ ]: there exists ${\tilde w}_\nb \neq w_\nb$ such that $({\bf x}_\nb^\pb({\tilde w}_\nb),{\bf y}_\na^\pb,{\hat{\bf y}}_\nr^\pb(w_{\nb0}))  \in A^\pb(X_\nb Y_\na {\hat Y}_\nr)$.
\end{description}
\end{enumerate}
Then,
\begin{align}
P[E_{\nr,\na}^{\pc}] \leq & P[E_{\nr,\na}^{\pc,1}]+P[E_{\nr,\na}^{\pc,2}]+P[E_{\nr,\na}^{\pc,3}]+P[E_{\nr,\na}^{\pc,4}]+P[E_{\nr,\na}^{\pc,5}]\\
\leq & 3\epsilon + P[\bar{D}^\pc({\bf u}_\nb(w_{\nb2}),{\bf y}_\na)] + P[\cup_{\tilde{w}_{\nb2} \neq w_{\nb2}}
D^\pc({\bf u}_\nb(\tilde{w}_{\nb2}),{\bf y}_\na)]\label{eq:tdbc:cf:5}\\
\leq &4\epsilon +2^{n(R_{\nb2}-\Delta_{3,n}I(U_\nb^\pc;Y_\na^\pc) + 3\epsilon)} \label{eq:tdbc:cf:4}
\end{align}
 In \eqref{eq:tdbc:cf:5}, $P[E_{\nr,\na}^{\pc,1}]$, $P[E_{\nr,\na}^{\pc,2}]$ and $P[E_{\nr,\na}^{\pc,3}]$ are less than $\epsilon$ due to \eqref{eq:tdbc:cf:6}, \eqref{eq:tdbc:cf:7} and \eqref{eq:tdbc:cf:8}, respectively.
\begin{align}
P[E_{\nr,\na}^\pd | \bar{E}_{\nr,\na}^{\pc}] \leq & P[E_{\nr,\na}^{\pd,1} | \bar{E}_{\nr,\na}^{\pc}]+P[E_{\nr,\na}^{\pd,2} | \bar{E}_{\nr,\na}^{\pc}] + P[E_{\nr,\na}^{\pd,3} | \bar{E}_{\nr,\na}^{\pc}]+ \nonumber\\
&P[E_{\nr,\na}^{\pd,4} | \bar{E}_{\nr,\na}^{\pc}] +P[E_{\nr,\na}^{\pd,5} | \bar{E}_{\nr,\na}^{\pc}]\\
\leq& P[\bar{D}^\pd({\bf x}_\nr(w_{\na0},w_{\nb0}),{\bf y}_\na)] + \nonumber\\
&P\left[\cup_{\twolines{{\tilde w}_{\na0} \neq
w_{\na0}}{{\tilde w}_{\nb0} \neq
w_{\nb0}}} D^\pd({\bf x}_\nr({\tilde w}_{\na0},{\tilde w}_{\nb0}),{\bf y}_\na),
D^\pa({\bf x}_\na(w_\na),{\hat{\bf y}}_\nr(\tilde{w}_{\na0})),\right.\nonumber\\
&~~~~~~~~~~~~~~\left.D^\pb({\hat{\bf y}}_\nr(\tilde{w}_{\nb0}),{\bf y}_\na) ,s_{\nb0}(\tilde{w}_{\nb0}) = w_{\nb1}\right] +\nonumber\\
&P[\cup_{{\tilde w}_{\nb0} \neq
w_{\nb0}} D^\pd({\bf x}_\nr(w_{\na0},{\tilde w}_{\nb0}),{\bf y}_\na),
D^\pb({\hat{\bf y}}_\nr(\tilde{w}_{\nb0}),{\bf y}_\na) ,s_{\nb0}(\tilde{w}_{\nb0}) = w_{\nb1}]\label{eq:tdbc:cf:9}\\
\leq & \epsilon + 2^{n(
R_{\na0} + R_{\nb0} -\Delta_{4,n}I(X_\nr^\pd;Y_\na^\pd)-\Delta_{1,n}I(\hat{Y}_\nr^\pa;X_\na^\pa) - \Delta_{2,n} I({\hat Y}_\nr^\pb;Y_\na^\pb)- \alpha_\nb R_{\nb0} +\epsilon'')} + \nonumber\\
&2^{n(R_{\nb0} -\Delta_{4,n}I(X_\nr^\pd;Y_\na^\pd) - \Delta_{2,n} I({\hat Y}_\nr^\pb;Y_\nr^\pb)- \alpha_\nb R_{\nb0} +\epsilon''')} \label{eq:tdbc:cf:1}
\end{align}
 In \eqref{eq:tdbc:cf:9}, $P[E_{\nr,\na}^{\pd,1} | \bar{E}_{\nr,\na}^{\pc}]$ and $P[E_{\nr,\na}^{\pd,2} | \bar{E}_{\nr,\na}^{\pc}]$ are zero since $E_{\nr,\na}^{\pd,1}$ and $E_{\nr,\na}^{\pd,2}$ are the same as $E_{\nr,\na}^{\pc,1}$ and $E_{\nr,\na}^{\pc,2}$, respectively. In \eqref{eq:tdbc:cf:1}, the bound for $R_{\nb0}$ in the second term is implied by that in the third term since $R_{\na0}-\Delta_{1,n}I(\hat{Y}_\nr^\pa;X_\na^\pa) = \Delta_{1,n}I(Y_\nr^\pa;{\hat Y}_\nr^\pa|X_\na^\pa) +\epsilon \geq 0$.
\begin{align}
P[E_{\nb,\na}^\pd | \bar{E}_{\nr,\na}^{\pc}\cap \bar{E}_{\nr,\na}^{\pd}] \leq& P[E_{\nb,\na}^{\pd,1} | \bar{E}_{\nr,\na}^{\pc}\cap \bar{E}_{\nr,\na}^{\pd}] + P[E_{\nb,\na}^{\pd,2} | \bar{E}_{\nr,\na}^{\pc}\cap \bar{E}_{\nr,\na}^{\pd}]\\
=&P[\bar{D}^\pb({\bf x}_\nb(w_\nb),{\bf y}_\nb,{\hat{\bf y}}_\nr(w_{\nb0}))] + P[\cup_{\tilde{w}_\nb \neq w_\nb} D^\pb({\bf x}_\nb(\tilde{w}_\nb),{\bf y}_\nb,{\hat{\bf y}}_\nr( w_{\nb0}))]\\
\leq &\epsilon + 2^{n(R_\nb - \Delta_{2,n}I(X_\nb^\pb;\hat{Y}_\nr^\pb,Y_\na^\pb)+3\epsilon)}\label{eq:tdbc:cf:2}
\end{align}

Since $\epsilon > 0$ is arbitrary, a proper choice of $\alpha_\nb$, the conditions of Theorem \ref{theorem:TDBC:CF}, \eqref{eq:tdbc:cf:10}, and the AEP property guarantee that the right hand sides of \eqref{eq:tdbc:cf:4}, \eqref{eq:tdbc:cf:1} and \eqref{eq:tdbc:cf:2} corresponding to \eqref{eq:tdbc:cf:10}, \eqref{thm:TDBC:CF:2} and \eqref{thm:TDBC:CF:7} vanish as $n \rightarrow \infty$. Similarly, $P[E_{\na,\nb}] \rightarrow 0$ as $n \rightarrow \infty$. By the Carath\'{e}odory theorem in \cite{Hiriart:2001}, it is
sufficient to restrict $|{\cal Q}| \leq 13$.
\end{proof}

When the $h_\na$ and $h_\nb$ links are of different strength, a scheme in which one link uses CF and the other uses DF may provide a larger rate region than if both links use CF.  In the next theorem, we provide a rate region for a TDBC scenario in which the forward link uses DF and the reverse link uses CF.

\begin{theorem}
\label{theorem:TDBC:CFDF} An achievable rate region for the half-duplex
bi-directional relay channel with a mixed  TDBC protocol, where the
$\na \rightarrow \nr \rightarrow \nb$ link uses decode and forward
and the  $\nb \rightarrow \nr \rightarrow \na$ link uses compress
and forward, is the closure of the set of all points $(R_\na,R_\nb)$ satisfying
\begin{align}
R_{\na} &< \min \left\{\Delta_1 I(X_\na^\pa; Y_\nr^\pa),\Delta_1 I(X_\na^\pa;Y_\nb^\pa) + \Delta_3I(U_\nr^\pc;Y_\nb^\pc|Q) - \Delta_3 I(U_\nr^\pc;U_\nb^\pc|Q)\right\} \label{thm:TDBC:CFDF:2}\\
R_{\nb} &< \Delta_2 I(X_\nb^\pb; \hat{Y}_\nr^\pb, Y_\na^\pb|Q)\label{thm:TDBC:CFDF:3}
\end{align}
subject to
\begin{align}
\Delta_2 I(Y_\nr^\pb;\hat{Y}_\nr^\pb|Y_\na^\pb,Q)&< \min\{\Delta_3
I(U_\nr^\pc,U_\nb^\pc;Y_\na^\pc|Q),\Delta_3 I(U_\nb^\pc;U_\nr^\pc,Y_\na^\pc|Q)\} \label{thm:TDBC:CFDF:1}
\end{align}
over all joint distributions,
\begin{align}
p(q,x_\na,x_\nb,x_\nr,u_\nb,u_\nr,y_\na,y_\nb,y_\nr,\hat{y}_\nr) =
p(q)p^\pa(x_\na,y_\nb,y_\nr)p^\pb(x_\nb,y_\na,y_\nr,\hat{y}_\nr|q)p^\pc(u_\nb,u_\nr,x_\nr,y_\na,y_\nb|q)
\end{align}
where
\begin{align}
 p^\pa(x_\na,y_\nb,y_\nr) &=  p^\pa(x_\na)p^\pa(y_\nb,y_\nr|x_\na)\\
 p^\pb(x_\nb,y_\na,y_\nr,\hat{y}_\nr|q) &=  p^\pb(x_\nb|q)p^\pb(y_\na, y_\nr|x_\nb)p^\pb(\hat{y}_\nr|y_\nr,q)\\
 p^\pc(u_\na,u_\nb,u_\nr,x_\nr,y_\na,y_\nb|q) &=  p^\pc(u_\nb,u_\nr|q)p^\pc(x_\nr|u_\nb,u_\nr,q)p^\pc(y_\na,y_\nb|x_\nr)
\end{align}
with $|{\cal Q}| \leq 6$ over the alphabet ${\cal X}_\na \times {\cal X}_\nb \times {\cal X}_\nr
\times {\cal U}_\nb \times {\cal U}_\nr
\times {\cal Y}_\na^2 \times {\cal Y}_\nb^2 \times {\cal Y}_\nr^2 \times \hat{{\cal Y}}_\nr$.
\thmend
\end{theorem}
\begin{remark}
We use random binning and a Gel'fand-Pinsker coding scheme in Theorem \ref{theorem:TDBC:CFDF}. The detailed proof is provided in Appendix \ref{app:tdbc:cfdf}.
\end{remark}


In contrast to the MABC protocols, in the TDBC protocols, the mixed TDBC protocol is not outer bounded by the
DF TDBC protocol. In the TDBC protocol, each terminal node obtains side
information, used during decoding, when the opposite node transmits (phase 1 for $\nb$ and
phase 2 for $\na$). The data rate as well as the phase
durations $\Delta_1$ and $\Delta_2$ of the TDBC protocol may vary, and we note that one can
easily find cases in which the mixed TDBC protocol outperforms the DF
TDBC protocol. For example, suppose the channel of the link $\nr
\leftrightarrow \na$ is good enough such that
$I(X_\nr^\pc;Y_\na^\pc) \geq \min\{I(U_\nr^\pc,U_\nb^\pc;Y_\na^\pc),I(U_\nb^\pc;U_\nr^\pc,Y_\na^\pc)\} \gg 0$ and the link mutual informations between
$\nb \leftrightarrow \nr$ and $\nb \leftrightarrow \na$ are the same
such that $I(X_\nb^\pb;Y_\nr^\pb) = I(X_\nb^\pb;Y_\na^\pb)$. Furthermore,
take $\Delta_1 = \epsilon_1$, $\Delta_3 = \epsilon_2$, where
$\epsilon_1$ and $\epsilon_2$ are small positive numbers. We assume the input
distributions are fixed, i.e. $|Q| = 1$. Then, in the DF TDBC
protocol, from Theorem \ref{theorem:TDBC:DF}, an
achievable rate region is given by:
\begin{align}
R_\na &< \epsilon' \label{eq:comp:tdbc:1}\\
R_\nb &< (1-\epsilon_1 -\epsilon_2) I(X_\nb^\pb;Y_\na^\pb) \approx I(X_\nb^\pb;Y_\na^\pb), \label{eq:comp:tdbc:2}
\end{align}
for some $\epsilon'$.
In the mixed TDBC protocol, we find a choice for $\min\{I(U_\nr^\pc,U_\nb^\pc;Y_\na^\pc),I(U_\nb^\pc;U_\nr^\pc,Y_\na^\pc)\}$ which satisfies \eqref{thm:TDBC:CFDF:1} for the same $\Delta_i$'s (this is possible as the $\nr \leftrightarrow \na$ channel is very strong). Then we obtain the following achievable rate region from Theorem \ref{theorem:TDBC:CFDF} :
\begin{align}
R_\na &< \epsilon'' \label{eq:comp:tdbc:3}\\
R_\nb &< (1-\epsilon_1 -\epsilon_2) I(X_\nb^\pb;Y_\na^\pb,\hat{Y}_\nr^\pb) \approx I(X_\nb^\pb;Y_\na^\pb) + I(X_\nb^\pb;\hat{Y}_\nr^\pb|Y_\na^\pb), \label{eq:comp:tdbc:4}
\end{align}
for some $\epsilon''$.
By properly choosing $\epsilon_1$ and $\epsilon_2$,  the regions bounded by \eqref{eq:comp:tdbc:1}, \eqref{eq:comp:tdbc:2}, \eqref{eq:comp:tdbc:3} and \eqref{eq:comp:tdbc:4} demonstrate the case in which a larger rate region is achieved through the mixed TDBC protocol.

\section{Gaussian Case}

\label{sec:gaussian}

We now assume all links in the bi-directional relay channel are subject to independent, identically distributed white Gaussian noise. The commonly considered Gaussian channel will allow us to visually compare different achievable rate regions for the bi-directional relaying channel. Definitions of codes, rate, and achievability in the memoryless Gaussian channels are analogous to those of the discrete memoryless channels.

We apply the previous results to the Gaussian channel. Since strong typicality does not apply to continuous random variables, the achievable rate regions from the theorems in the previous section do not directly apply to continuous domains. However, for the Gaussian input distributions and additive Gaussian noise which we will assume in the following, the Markov lemma of \cite{Oohama:1997}, which generalizes the Markov lemma to the continuous domains, ensures that the achievable rate regions in the previous section hold in the Gaussian case. We use Gaussian input distributions since we assume the average power constraint.

The corresponding  Gaussian channel model is:
\begin{align}
Y_\na[m] &= h_{\nr\na}X_\nr[m] + h_{\nb\na}X_\nb[m] + Z_\na[m]\\
Y_\nb[m] &= h_{\nr\nb}X_\nr[m] + h_{\na\nb}X_\na[m] + Z_\nb[m]\\
Y_\nr[m] &= h_{\na\nr}X_\na[m] + h_{\nb\nr}X_\nb[m] + Z_\nr[m]
\end{align}
where $X_\na[m]$, $X_\nb[m]$ and $X_\nr[m]$ follow the input distributions
$X_\na^{(\ell)}\sim{\cal C N}(0,P_\na)$, $X_\nb^{(\ell)}\sim{\cal C
N}(0,P_\nb)$ and $X_\nr^{(\ell)}\sim{\cal C N}(0,P_\nr)$ respectively during transmitting, where $m \in [n \sum_{j=0}^{\ell-1} \Delta_{j,n} +
1 , n \sum_{j=0}^{\ell} \Delta_{j,n}]$ and ${\cal C N}(\mu, \sigma^2)$
denotes a complex Gaussian random variable with mean $\mu$ and variance $\sigma^2$,
and $\ell$ corresponds to the appropriate phase. If node $i$ is in transmitting mode, the transmit power is bounded by $P_i$, i.e., $E[X_i^2] \leq P_i$.
If node $i$ is in
receiving mode, the input symbol does not exist in the above mathematical
channel model. For example, in the first phase of the TDBC protocol,
the corresponding channel
model is :
\begin{align}
Y_\nb[m] &= h_{\na\nb}X_\na[m] + Z_\nb[m]\\
Y_\nr[m] &= h_{\na\nr}X_\na[m] + Z_\nr[m].
\end{align}
In the above $h_{ij}$ is the effective channel gain between transmitter $i$ and
receiver $j$, which is modeled as a complex number. We assume that
the channel is reciprocal such that $h_{ij} = h_{ji}$ and each node
is fully aware of $h_{\na\nr}$, $h_{\nb\nr}$ and $h_{\na\nb}$ (i.e.
we have full CSI).
 The noise at all receivers $Z_\na,Z_\nb,Z_\nr$ is of unit power, additive, white Gaussian, complex and circularly symmetric. For convenience of analysis, we also define the function $C(x) \eqdef \log_2(1 + x)$.

For the analysis of the Compress and Forward scheme, we assume $\hat{Y}^{(\ell)}_{\nr}$ are zero mean Gaussians and define $P_y^{(\ell)} \eqdef E[(Y_\nr^{(\ell)})^2]$ , $P_{\hat{y}}^{(\ell)} \eqdef E[(\hat{Y}_\nr^{(\ell)})^2]$ and $\sigma_y^{(\ell)} \eqdef E[\hat{Y}_\nr^{(\ell)}Y_\nr^{(\ell)}]$. Then the relation between the received $Y_\nr[m]$ and the compressed $\hat{Y}_\nr[m]$ are given by the following equivalent channel model:
\begin{align}
\hat{Y}_\nr [m] = h_{\nr \hat{\nr}}[m] Y_\nr[m] + Z_{\hat{\nr}}[m]
\end{align}
 where $Y_\nr[m]$, ${\hat Y}_\nr[m]$ and $Z_{\hat{\nr}}[m]$ follow the distributions $Y_\nr^{(\ell)} \sim {\cal CN}(0,P_y^{(\ell)})$, ${\hat Y}_\nr^{(\ell)} \sim {\cal CN}(0,P_{\hat y}^{(\ell)})$ and $Z_{\hat{\nr}}^{(\ell)} \sim {\cal C N}(0,P_{\hat{y}}^{(\ell)} - \frac{(\sigma_y^{(\ell)})^2}{P_y^{(\ell)}})$ and $h_{\nr \hat{\nr}}[m] = \frac{\sigma_y^{(\ell)}}{P_y^{(\ell)}}$, where  $m \in [n \sum_{j=0}^{\ell-1} \Delta_{j,n} + 1 , n \sum_{j=0}^{\ell} \Delta_{j,n}]$. We note that in the following, $P_{\hat{y}}^{(\ell)}$ and $\sigma_y^{(\ell)}$ are unknown variables corresponding to the quantization which we numerically optimize.

We consider four different relaying schemes (i.e. ways in which the relay processes and forwards the received signal) for each MABC and TDBC bi-directional protocol:
{\it Amplify and Forward (AF)}, {\it Decode and Forward (DF)}, {\it Compress and Forward (CF)}, and {\it Mixed Forward (Mixed)}. In addition to achievable rate regions, we apply outer bounds of the MABC and TDBC protocols to the Gaussian channel.

\begin{table}
\caption{Input and output distributions}
\label{table:distribution}
\centering
\begin{tabular}{c||c|c}
  \hline
  Dist. of & during TX & during RX \\
  \hline
  $X_\na^{(\ell)}$ & ${\cal C N}(0,P_\na)$ & N/A \\
  $X_\nb^{(\ell)}$ & ${\cal C N}(0,P_\nb)$ & N/A \\
  $X_\nr^{(\ell)}$ & ${\cal C N}(0,P_\nr)$ & N/A \\
  $Z_\na^{(\ell)}= Z_\nb^{(\ell)} =Z_\nr^{(\ell)}$ & N/A & ${\cal C N}(0,1)$  \\
  $\hat{Y}_\nr^{(\ell)}$ & N/A & ${\cal C N}\left(0,P_{\hat{y}}^{(\ell)}\right)$  \\
  $Z_{\hat{\nr}}^{(\ell)}$ & N/A & ${\cal C N}\left(0,P_{\hat{y}}^{(\ell)} - \frac{(\sigma_y^{(\ell)})^2}{P_y^{(\ell)}}\right)$\\
  \hline
\end{tabular}
\end{table}

\subsection{Amplify and Forward}
In the amplify and forward scheme, all phase durations are equal, since relaying is performed on a  symbol by symbol basis. Therefore, $\Delta_1  = \Delta_2 = \frac12$ for the MABC protocol and $\Delta_1 = \Delta_2 = \Delta_3 = \frac13$ for the TDBD protocol. Furthermore, relay $\nr$ scales the received symbol $y_\nr$ by $\sqrt{\frac{P_\nr}{P_y}}$ to meet the transmit power constraint of $P_\nr$. The following are achievable rate regions for the amplify and forward relaying:
\begin{itemize}
\item MABC Protocol
\begin{align}
R_\na &< \frac12 C\left(\frac{|h_{\na\nr}|^2|h_{\nb\nr}|^2 P_\na P_\nr}{|h_{\na\nr}|^2 P_\na + |h_{\nb\nr}|^2 P_\nb + |h_{\nb\nr}|^2 P_\nr +1}\right) \label{gaussian:mabc:af:1}\\
R_\nb &< \frac12 C\left(\frac{|h_{\na\nr}|^2|h_{\nb\nr}|^2 P_\nb P_\nr}{|h_{\na\nr}|^2 P_\na + |h_{\nb\nr}|^2 P_\nb + |h_{\na\nr}|^2 P_\nr +1}\right)\label{gaussian:mabc:af:2}
\end{align}

\item TDBC Protocol
\begin{align}
R_\na &< \frac13 C\left(|h_{\na\nb}|^2P_\na + \frac{|h_{\na\nr}|^2|h_{\nb\nr}|^2 P_\na P_\nr}{|h_{\na\nr}|^2 P_\na + |h_{\nb\nr}|^2 P_\nb + 2|h_{\nb\nr}|^2 P_\nr +2}\right) \\
R_\nb &< \frac13 C\left(|h_{\na\nb}|^2P_\nb + \frac{|h_{\na\nr}|^2|h_{\nb\nr}|^2 P_\nb P_\nr}{|h_{\na\nr}|^2 P_\na + |h_{\nb\nr}|^2 P_\nb + 2|h_{\na\nr}|^2 P_\nr  +2}\right)
\end{align}
\end{itemize}

\subsection{Decode and Forward}
Applying Theorems \ref{theorem:MABC:DF} and \ref{theorem:TDBC:DF} to the Gaussian case, we obtain the following achievable rate regions:
\begin{itemize}
\item MABC Protocol
\begin{align}
R_\na &< \min \{\Delta_1 C(|h_{\na\nr}|^2 P_\na) , \Delta_2 C(|h_{\nb\nr}|^2 P_\nr)\}\label{gaussian:mabc:df:1}\\
R_\nb &< \min \{\Delta_1 C(|h_{\nb\nr}|^2 P_\nb) , \Delta_2 C(|h_{\na\nr}|^2 P_\nr)\}\label{gaussian:mabc:df:2}\\
R_\na + R_\nb &< \Delta_1 C(|h_{\na\nr}|^2 P_\na +|h_{\nb\nr}|^2 P_\nb )\label{gaussian:mabc:df:3}
\end{align}
\item TDBC Protocol
\begin{align}
R_\na &< \min \{\Delta_1 C(|h_{\na\nr}|^2P_\na), \Delta_1 C(|h_{\na\nb}|^2P_\na) + \Delta_3 C(|h_{\nb\nr}|^2P_\nr)\}\\
R_\nb &< \min \{\Delta_2 C(|h_{\nb\nr}|^2P_\nb), \Delta_2 C(|h_{\na\nb}|^2P_\nb) + \Delta_3 C(|h_{\na\nr}|^2P_\nr)\}
\end{align}
\end{itemize}
When obtaining the regions numerically, we optimize $\Delta_{\ell}$'s for the given channel mutual informations to maximize the achievable rate regions.
\subsection{Compress and Forward}
 Applying Theorem \ref{theorem:MABC:CF} and \ref{theorem:TDBC:CF} to the Gaussian case, we obtain the following achievable rate regions:
\begin{itemize}
\item MABC Protocol
\begin{align}
R_\na &< \Delta_1 C\left(\frac{(\sigma_y^\pa)^2 |h_{\na\nr}|^2P_\na}{P_{\hat{y}}^\pa (P_y^\pa)^2 - (\sigma_y^\pa)^2(P_y^\pa - 1)} \right)\\
R_\nb &< \Delta_1 C\left(\frac{(\sigma_y^\pa)^2 |h_{\nb\nr}|^2P_\nb}{P_{\hat{y}}^\pa (P_y^\pa)^2 - (\sigma_y^\pa)^2(P_y^\pa - 1)} \right)
\end{align}
where,
\begin{align}
\Delta_1 &= \min\left\{\frac{C(|h_{\nb\nr}|^2 P_\nr)}{C\left(\frac{(\sigma_y^\pa)^2(|h_{\na\nr}|^2 P_\na+ 1)}{P_{\hat{y}}^\pa(P_y^\pa)^2 -(\sigma_y^\pa)^2 P_y^\pa} \right) +C(|h_{\nb\nr}|^2 P_\nr)} ,\frac{C(|h_{\na\nr}|^2 P_\nr)}{C\left(\frac{(\sigma_y^\pa)^2(|h_{\nb\nr}|^2 P_\nb+ 1)}{P_{\hat{y}}^\pa (P_y^\pa)^2 -(\sigma_y^\pa)^2 P_y^\pa} \right) +C(|h_{\na\nr}|^2 P_\nr)} \right\}\\
P_y^\pa & = |h_{\na\nr}|^2 P_\na + |h_{\nb\nr}|^2 P_\nb +1
\end{align}

\item TDBC Protocol

 One can show that \eqref{thm:TDBC:CF:3} -- \eqref{thm:TDBC:CF:5} Marton's bound is equivalent to the capacity region of the Gaussian broadcast channel with Costa's setup as follows: let $|h_{\nr\na}|>|h_{\nr\nb}|$ and we set
\begin{align}
\text{In phase 3} \left\{
  \begin{array}{l}
    U_{\nb}[m] = V_\nr[m] + \alpha U_\na[m]\\
   Y_{\na}[m] = h_{\nr\na}(V_\nr[m] + U_\na[m]) + Z_\na[m] \\
    Y_{\nb}[m] = h_{\nr\nb}(V_\nr[m] + U_\na[m]) + Z_\nb[m]
  \end{array}
\right.
\end{align}
where $V_\nr[m]$ and $U_\na[m]$ follow the distributions $V_\nr^\pc\sim {\cal C N}(0,\beta P_\nr)$, $U_\na^\pc\sim {\cal C N}(0,(1-\beta)P_\nr)$ respectively during phase 3, $m \in [n(\Delta_{1,n} + \Delta_{2,n})+1,n]$, where $(0\leq \beta\leq 1)$ and $E[V_\nr^\pc U_\nb^\pc] = 0$, i.e., $V_\nr^\pc$, $U_\na^\pc$ are independent. Also we take $\alpha  = \frac{|h_{\nr\na}|^2 \beta P_\nr}{|h_{\nr\na}|^2 \beta P_\nr + 1}$. Then
\begin{align}
\left\{
  \begin{array}{l}
   I(U_\na^\pc;Y_\nb^\pc)  = C\left(\frac{|h_{\nr\nb}|^2(1-\beta)P_\nr}{|h_{\nr\nb}|^2\beta P_\nr+1}\right)\\
I(U_\nb^\pc;Y_\na^\pc) - I(U_\na^\pc;U_\nb^\pc)= C\left(|h_{\nr\na}|^2 \beta P_\nr\right)
  \end{array}
\right.
\end{align}
Similarly, we obtain the bounds in the case $|h_{\nr\na}|\leq|h_{\nr\nb}|$.
These are the same as the capacity region of the Gaussian broadcast channel ((15.11) and (15.12) in \cite{Cover:2006}).

\begin{align}
R_\na &< \Delta_1 C\left(|h_{\na\nb}|^2 P_\na + \frac{(\sigma_y^\pa)^2|h_{\na\nr}|^2 P_\na}{P_{\hat{y}}^\pa (P_y^\pa)^2 - (\sigma_y^\pa)^2 (P_y^\pa -1)}\right)\\
R_\nb &< \Delta_2 C\left(|h_{\na\nb}|^2 P_\nb + \frac{(\sigma_y^\pb)^2|h_{\nb\nr}|^2 P_\nb}{P_{\hat{y}}^\pb (P_y^\pb)^2 - (\sigma_y^\pb)^2 (P_y^\pb -1) }\right)
\end{align}
where,
\begin{align}
\text{If}~~ |h_{\nr\na}| < |h_{\nr\nb}| :~~
\alpha_\na \Delta_1 C\left(\frac{(\sigma_y^\pa)^2}{P_{\hat{y}}^\pa P_y^\pa -(\sigma_y^\pa)^2}\right) &< \Delta_3 C\left(\beta |h_{\nr\nb}|^2 P_\nr\right)\\
\alpha_\nb \Delta_2 C\left(\frac{(\sigma_y^\pb)^2}{P_{\hat{y}}^\pb P_y^\pb -(\sigma_y^\pb)^2}\right) &< \Delta_3 C\left(\frac{(1-\beta) |h_{\nr\na}|^2 P_\nr}{\beta |h_{\nr\na}|^2 P_\nr + 1}\right),\\
\text{otherwise}~~ :~~
\alpha_\na \Delta_1 C\left(\frac{(\sigma_y^\pa)^2}{P_{\hat{y}}^\pa P_y^\pa -(\sigma_y^\pa)^2}\right) &< \Delta_3 C\left(\frac{(1-\beta) |h_{\nr\nb}|^2 P_\nr}{\beta |h_{\nr\nb}|^2 P_\nr + 1}\right)\\
\alpha_\nb \Delta_2 C\left(\frac{(\sigma_y^\pb)^2}{P_{\hat{y}}^\pb P_y^\pb -(\sigma_y^\pb)^2}\right) &< \Delta_3 C\left(\beta |h_{\nr\na}|^2 P_\nr\right),
\end{align}
and
\begin{align}
(1-\alpha_\na)&\Delta_1 C\left(\frac{(\sigma_y^\pa)^2}{P_{\hat{y}}^\pa P_y^\pa -(\sigma_y^\pa)^2}\right) + \Delta_2 C\left(\frac{(\sigma_y^\pb)^2}{P_{\hat{y}}^\pb (P_y^\pb)^2 - (\sigma_y^\pb)^2 P_y^\pb}\right) \nonumber \\
&< \Delta_3 C(|h_{\nb\nr}|^2P_\nr) + \Delta_1 C\left(\frac{(\sigma_y^\pa)^2 |h_{\na\nb}|^2 |h_{\nr\na}|^2P_\na}{(P_y^\pa)^2 P_{\hat y}^\pa (|h_{\na\nb}|^2P_\na +1) - (\sigma_y^\pa)^2 |h_{\na\nb}|^2|h_{\nr\na}|^2P_\na}\right)\\
(1-\alpha_\nb)&\Delta_2 C\left(\frac{(\sigma_y^\pb)^2}{P_{\hat{y}}^\pb P_y^\pb -(\sigma_y^\pb)^2}\right) + \Delta_1 C\left(\frac{(\sigma_y^\pa)^2}{P_{\hat{y}}^\pa (P_y^\pa)^2 - (\sigma_y^\pa)^2 P_y^\pa}\right) \nonumber \\
&< \Delta_3 C(|h_{\na\nr}|^2P_\nr) + \Delta_2 C\left(\frac{(\sigma_y^\pb)^2 |h_{\na\nb}|^2 |h_{\nr\nb}|^2P_\nb}{(P_y^\pb)^2 P_{\hat y}^\pb (|h_{\na\nb}|^2P_\nb +1) - (\sigma_y^\pb)^2 |h_{\na\nb}|^2|h_{\nr\nb}|^2P_\nb}\right)\\
&P_y^\pa  = |h_{\na\nr}|^2 P_\na +1\\
&P_y^\pb  = |h_{\nb\nr}|^2 P_\nb +1\\
0&<\alpha_\na ,\alpha_\nb, \beta <1
\end{align}
\end{itemize}
Again, when numerically obtaining the regions, we optimize $P_{\hat{y}}^{(\ell)}$, $\sigma_y^{(\ell)}$,  $\Delta_{\ell}$, $\alpha_\na$, $\alpha_\nb$ and $\beta$ to maximize the region boundary.
\subsection{Mixed Forward}
Applying Theorem \ref{theorem:MABC:CFDF} to the Gaussian case with Costa's setup in \cite{Costa:1983} we have the channel in the relay broadcasting phase in the Mixed MABC protocol as:
\begin{align}
\text{In phase 2}\left\{
  \begin{array}{l}
    U_{\nr}[m] = V_\nr[m] + \alpha U_\nb[m]\\
Y_{\na}[m] = h_{\nr\na}(V_\nr[m] + U_\nb[m]) + Z_\na[m] \\
Y_{\nb}[m] = h_{\nr\nb}(V_\nr[m] + U_\nb[m]) + Z_\nb[m]
  \end{array}
\right.
\end{align}
where $V_\nr[m]$ and $U_\nb[m]$ follow the distributions $V_\nr^\pb\sim {\cal C N}(0,\beta P_\nr)$, $U_\nb^\pb\sim {\cal C N}(0,(1-\beta)P_\nr)$ during phase 2, $m\in [\Delta_{1,n}\cdot n +1, n]$, where $(0\leq \beta\leq 1)$, and $E[V_\nr^\pb U_\nb^\pb] = 0$, i.e., $V_\nr^\pb$, $U_\nb^\pb$ are independent. Similarly, we construct the channel in the relay broadcasting phase in the Mixed TDBC protocol.
 Then we obtain the following achievable rate regions, where we numerically optimize $\alpha$, $\beta$, $P_{\hat{y}}^{(\ell)}$, $\sigma_y^{(\ell)}$ and $\Delta_{\ell}$ to maximize their boundary.
\begin{itemize}
\item MABC Protocol
\begin{align}
R_\na &< \min \left\{\Delta_1 C\left(\frac{|h_{\na\nr}|^2 P_\na} {|h_{\nb\nr}|^2 P_\nb + 1} \right), \Delta_2 \log_2 \left(\frac{\beta P_\nr (|h_{\nr\nb}|^2 P_\nr +1)}{|h_{\nr\nb}|^2(1-\alpha)^2\beta(1-\beta)P_\nr^2 + \beta P_\nr + \alpha^2(1-\beta)P_\nr  }\right) \right\}\\
R_\nb &< \Delta_1 C\left(\frac{(\sigma_y^\pa)^2 |h_{\nb\nr}|^2P_\nb}{P_{\hat{y}}^\pa (P_y^\pa)^2 - (\sigma_y^\pa)^2(P_y^\pa - 1)} \right)
\end{align}
where,
\begin{align}
&\Delta_1 C\left(\frac{(\sigma_y^\pa)^2(|h_{\nb\nr}|^2 P_\nb+ 1)}{P_{\hat{y}}^\pa (P_y^\pa)^2 -(\sigma_y^\pa)^2 P_y^\pa} \right)  < \min\left\{\Delta_2 C(|h_{\nr\na}|^2 P_\nr), \Delta_2 C\left(|h_{\nr\na}|^2(1-\alpha)^2(1-\beta)P_\nr + \frac{\alpha^2(1-\beta)}{\beta} \right)\right\}\\
&P_y^\pa  = |h_{\na\nr}|^2 P_\na + |h_{\nb\nr}|^2 P_\nb +1
\end{align}

\item TDBC Protocol
\begin{align}
R_\na &< \min \left\{\Delta_1 C(|h_{\na\nr}|^2 P_\na) , \right.\nonumber\\
&~~~~~~~~~~~\left.\Delta_1 C(|h_{\na\nb}|^2 P_\na) + \Delta_3 \log_2 \left(\frac{\beta P_\nr (|h_{\nr\nb}|^2 P_\nr +1)}{|h_{\nr\nb}|^2(1-\alpha)^2\beta(1-\beta)P_\nr^2 + \beta P_\nr + \alpha^2(1-\beta)P_\nr}\right)\right\} \\
R_\nb &< \Delta_2 C\left(|h_{\na\nb}|^2 P_\nb + \frac{(\sigma_y^\pb)^2|h_{\nb\nr}|^2 P_\nb}{P_{\hat{y}}^\pb (P_y^\pb)^2 - (\sigma_y^\pb)^2 P_y^\pb + (\sigma_y^\pb)^2 }\right)
\end{align}
where,
\begin{align}
\Delta_2 C\left(\frac{(\sigma_y^\pb)^2(1-P^*)}{P_{\hat{y}}^\pb P_y^\pb -(\sigma_y^\pb)^2}\right) &< \min\left\{\Delta_3 C(|h_{\nr\na}|^2 P_\nr), \Delta_3 C\left(|h_{\nr\na}|^2(1-\alpha)^2(1-\beta)P_\nr + \frac{\alpha^2(1-\beta)}{\beta} \right)\right\}\\\
P_y^\pb & = |h_{\nb\nr}|^2 P_\nb +1\\
P^* & = \frac{|h_{\na\nb}|^2 P_\nb}{|h_{\na\nb}|^2 P_\nb +1}\cdot \frac{|h_{\nr\nb}|^2 P_\nb}{|h_{\nr\nb}|^2 P_\nb +1}
\end{align}

\end{itemize}

\subsection{Outer Bound}
Applying Theorems \ref{theorem:MABC:out} and \ref{theorem:TDBC:out} to the Gaussian case, we obtain the following outer bounds. We optimize $\Delta_{\ell}$'s for given channel mutual informations to maximize these outer bounds.
\begin{itemize}
\item MABC Protocol
\begin{align}
R_\na &\leq \min \{\Delta_1 C(|h_{\na\nr}|^2 P_\na) , \Delta_2 C(|h_{\nb\nr}|^2 P_\nr)\}\\
R_\nb &\leq \min \{\Delta_1 C(|h_{\nb\nr}|^2 P_\nb) , \Delta_2 C(|h_{\na\nr}|^2 P_\nr)\}
\end{align}

\item TDBC Protocol
\begin{align}
R_\na &\leq \min \{\Delta_1 C(|h_{\na\nr}|^2P_\na+ |h_{\na\nb}|^2 P_\na), \Delta_1 C(|h_{\na\nb}|^2P_\na) + \Delta_3 C(|h_{\nb\nr}|^2P_\nr)\}\\
R_\nb &\leq \min \{\Delta_2 C(|h_{\nb\nr}|^2P_\nb + |h_{\na\nb}|^2 P_\nb), \Delta_2 C(|h_{\na\nb}|^2P_\nb) + \Delta_3 C(|h_{\na\nr}|^2P_\nr)\}\\
R_\na + R_\nb &\leq \Delta_1 C(|h_{\na\nr}|^2 P_\na) + \Delta_2 C(|h_{\nb\nr}|^2 P_\nb)
\end{align}
\end{itemize}


\section{achievable rate regions in the Gaussian channel}

\label{sec:regions}

In order to obtain an intuitive feel for the regions and to illustrate that the regions are not subsets of one another, the bounds described in Section \ref{sec:gaussian} are plotted in this section for a number of different channel configurations. We first compare the rate regions obtained by the bi-directional protocols and outer bounds in cases in which the links are symmetric ($h_{\na \nr} = h_{\nb \nr} = 1$, $h_{\na \nb} = 0.2$) as well as asymmetric ($h_{\na \nr} = 0.6,\, h_{\nb \nr} = 20, h_{\na \nb} = 0.5$ and $h_{\na\nr} = 20, h_{\nb \nr} = 0.6, h_{\na \nb} = 0.5$) for two different transmit SNRs of $0$ and $20$dB. The protocols considered are:
\begin{enumerate}
\item AF MABC: Amplify and Forward Multiple Access Broadcast.
\item AF TDBC: Amplify and Forward Time Division Broadcast.
\item DF MABC: Decode and Forward Multiple Access Broadcast.
\item DF TDBC: Decode and Forward Time Division Broadcast.
\item CF MABC: Compress and Forward Multiple Access Broadcast.
\item CF TDBC: Compress and Forward Time Division Broadcast.
\item Mixed MABC: Mixed scheme with Multiple Access Broadcast.
\item Mixed TDBC: Mixed scheme with Time Division Broadcast.
\item Outer MABC: Outer bound when using MABC protocol.
\item Outer TDBC: Outer bound when using TDBC protocol.
\end{enumerate}
We then proceed to examine the maximal sum-rate $R_\na +R_\nb$ of the ten schemes as a function of the transmit SNR. Finally, we evaluate the maximal sum-rate and maximal constrained sum-rates (that is we require the rate $R_\na = R_\nb$ as well as $R_\na = 2R_\nb$) of the schemes as a function of the relay position. The main conclusions to be drawn are that different schemes are optimal under different channel conditions. We provide further discussions in the following subsections.

\subsection{Achievable rate region comparisons}

We compare the achievable rate regions and outer bounds of the 10 aforementioned protocols for both symmetric and asymmetric source to relay channel gains at transmit SNRs of $0$ and $20$dB.

\subsubsection{Symmetric Case}

\begin{figure*}
  \hfill
  \begin{minipage}[t]{.47\textwidth}
    \begin{center}
      \epsfig{figure=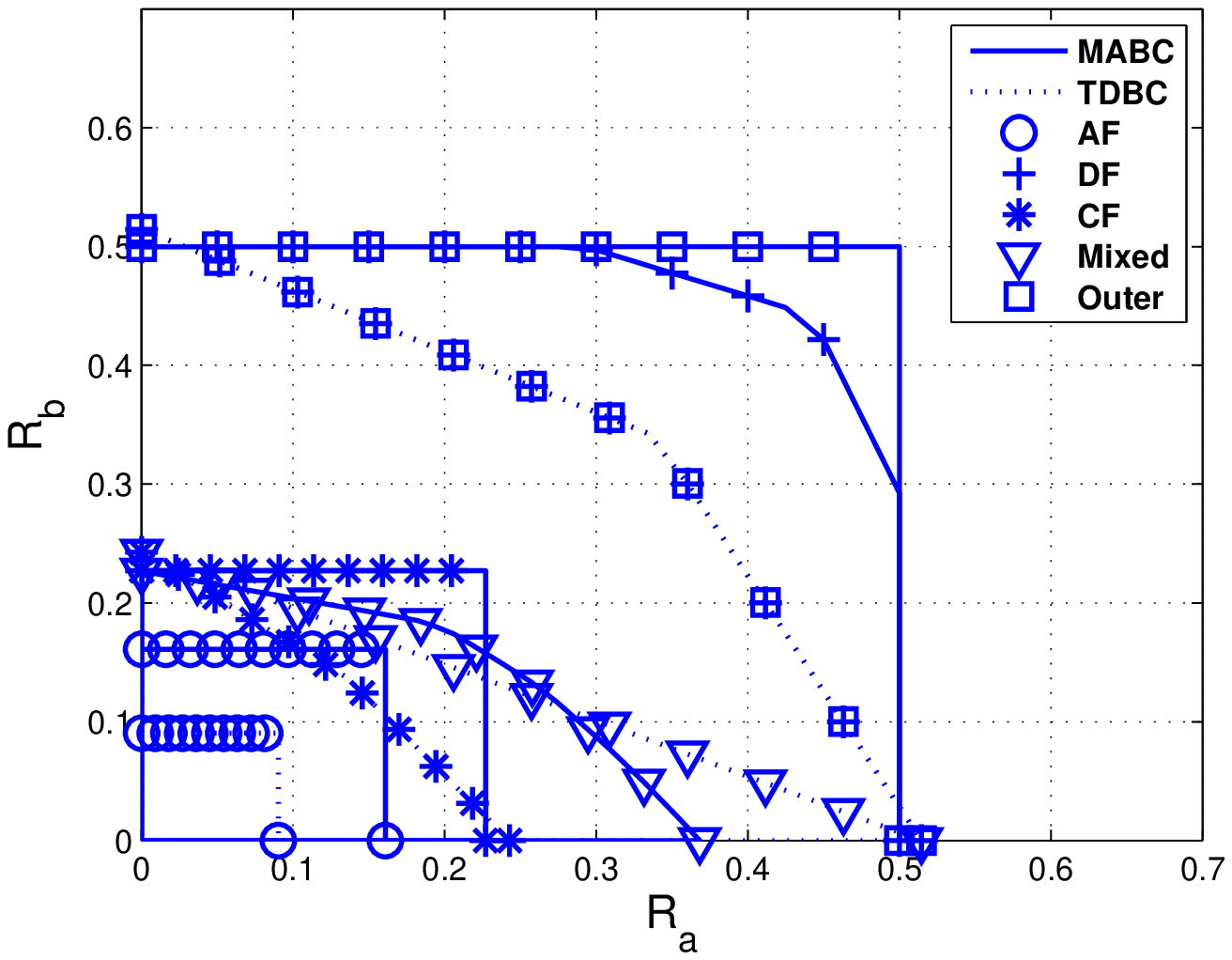, width=8.5cm}
      \caption{\baselineskip=10pt Comparison of bi-directional regions with $h_{\na\nr}=h_{\nb\nr}=1$, $h_{\na\nb}=0.2$, $P_\na=P_\nb=P_\nr=0$ dB and $N_\na=N_\nb=N_\nr=1$.}
            \label{fig:sym_low}
    \end{center}
  \end{minipage}
  \hfill
  \begin{minipage}[t]{.47\textwidth}
    \begin{center}
      \epsfig{figure=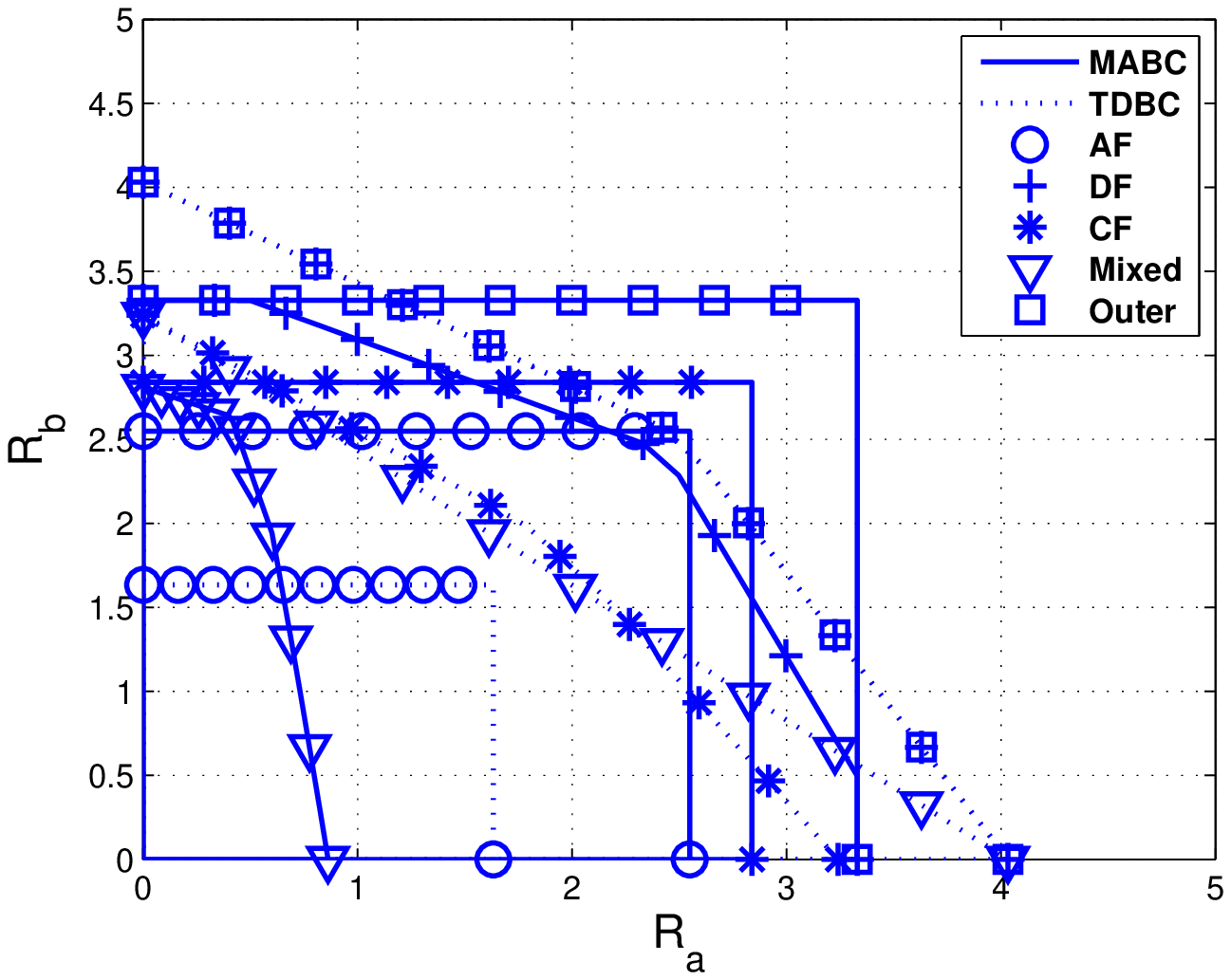, width=8.5cm}
      \caption{\baselineskip=10pt Comparison of bi-directional regions with $h_{\na\nr}=h_{\nb\nr}=1$, $h_{\na\nb}=0.2$, $P_\na=P_\nb=P_\nr=20$ dB and $N_\na=N_\nb=N_\nr=1$.}
      \label{fig:sym_med}
    \end{center}
  \end{minipage}
  \hfill
\end{figure*}

In this case $h_{\na \nr} = h_{\nb \nr} = 1$ (\Figs \ref{fig:sym_low}, \ref{fig:sym_med}). In the low SNR regime, the DF MABC protocol dominates the other protocols. The MABC protocol in general outperforms the TDBC protocol as the benefits of side information and reduced interference are relatively small in this regime. The DF scheme outperforms the other schemes since the relatively large amount of noise in the first phase (and the second phase in the TDBC protocol) can be eliminated in the DF scheme, which cannot be done using the other schemes. In contrast, the DF TDBC protocol dominates the other protocols at high SNR since the direct link is strong enough to convey information in this regime.

In the high SNR regime (when $P_\na = P_\nb = P_\nr = P$ is sufficiently large), the AF MABC protocol outperforms the DF MABC protocol. From \eqref{gaussian:mabc:af:1}, \eqref{gaussian:mabc:af:2}, the achievable rate region of the AF MABC protocol is:
\begin{align}
R_\na &< \frac12 \log \left(1 + \frac{P^2}{3P+1}\right) \thickapprox \frac12 \log P\\
R_\nb &< \frac12 \log \left(1 + \frac{P^2}{3P+1}\right) \thickapprox \frac12 \log P
\end{align}
also from  \eqref{gaussian:mabc:df:1}, \eqref{gaussian:mabc:df:2} and  \eqref{gaussian:mabc:df:3}, the achievable rate region of the DF MABC protocol is:
 \begin{align}
R_\na &< \min\{\Delta_1,1-\Delta_1\}\cdot \log(1+P) \thickapprox \min\{\Delta_1,1-\Delta_1\}\cdot \log P < \frac12 \log P\\
R_\nb &< \min\{\Delta_1,1-\Delta_1\}\cdot \log(1+P) \thickapprox \min\{\Delta_1,1-\Delta_1\}\cdot \log P < \frac12 \log P\\
R_\na + R_\nb &< \Delta_1 \log(1 + 2P) \thickapprox \Delta_1 \log P  \label{gaussian:mabc:df:4}
\end{align}
From \eqref{gaussian:mabc:df:4}, we can conclude that the achievable rate region of the DF MABC protocol is outer-bounded by the AF MABC protocol.

In the TDBC protocol, the CF scheme does not
outperform the DF scheme 
since the DF uses two parallel channels in phase one and three, while the CF uses one channel in phase one with two receivers. In other words, $R_\na^{DF} < \Delta_1 C(\cdot) + \Delta_3 C(\cdot)$ for the DF as opposed to $R_\na^{CF}<\Delta_1 C(\sum\cdot)$ for the CF scheme.
 However, under the MABC protocol, the CF scheme
outperforms the DF scheme in the high SNR regime. This is because
the interference of the transmission of two terminal nodes affects
the DF MABC scheme due to the multiple-access nature but not the CF
scheme (as it must not decode the signals).

The achievable rate region of the Mixed TDBC protocol lies between
the CF TDBC protocol and the DF TDBC protocol. In the TDBC protocol,
$\max_{R_\nb} R_\na^{MIX} = \max_{R_\nb} R_\na^{DF}$, where
$R_\na^{MIX}$ is the data rate of node $\na$ in the Mixed scheme. Here the $\max$ is taken over all
rates in the achievable rate regions.  The $\max R_\na^{MIX}$ is achieved by taking $\Delta_2 = 0$. The rate  $R_\na^{DF}$ is
similarly defined and  $\max  R_\na^{DF}$ is
achieved in an analogous manner. Therefore, the point $(\max
R_\na^{DF},0)$ lies in both the Mixed scheme and the DF scheme.  While in the TDBC scheme a particular rate may be
set to 0 by indirectly setting the appropriate interval $\Delta_i$
to 0, in the MABC protocol this is not possible. In the MABC
protocol, even when $R_\na = 0$ or $R_\nb = 0$,  the transmit power
($P_\na$ or $P_\nb$) remains constant in the first phase and does
not decrease to 0, acting as the additional noise for the opposite
transmission. Therefore, $\max_{R_\nb} R_\na^{MIX} \leq \max_{R_\nb}
R_\na^{DF}$. This interference seen in MABC protocols is especially
pronounced in the high SNR regime, where the gap between the
intercept points of the Mixed scheme and the DF scheme is seen to
grow as the interference increases (with SNR).  We note that this
effect is due to our assumption that both transmitters transmit at full
power regardless of the rate. If we were to allow
for optimization of the transmission power, larger achievable rate
regions for the Mixed MABC protocol could result.

In \Figs \ref{fig:sym_low} and \ref{fig:sym_med}, the AF scheme is always outer bounded by the CF
scheme. We thus expect that when relay $\nr$ does not know the
codebooks of $\na$ and $\nb$ (and hence cannot decode as in the DF
scheme), that the CF scheme is a better strategy than the AF scheme.

In the low SNR regime, the achievable rate region of the DF MABC protocol
and the outer bound of the MABC protocol are tight, while in the
high SNR regime, the achievable rate region of the CF MABC protocol is
tight. For the TDBC protocol, there is a very small gap between the
achievable rate region of the DF TDBC protocol and the outer bound of the
TDBC protocol since interference is not an issue for the TDBC
protocol and hence decoding is thus, intuitively, near optimal.\\

\subsubsection{Asymmetric Cases}

\begin{figure*}
  \hfill
  \begin{minipage}[t]{.47\textwidth}
   \begin{center}
      \epsfig{figure=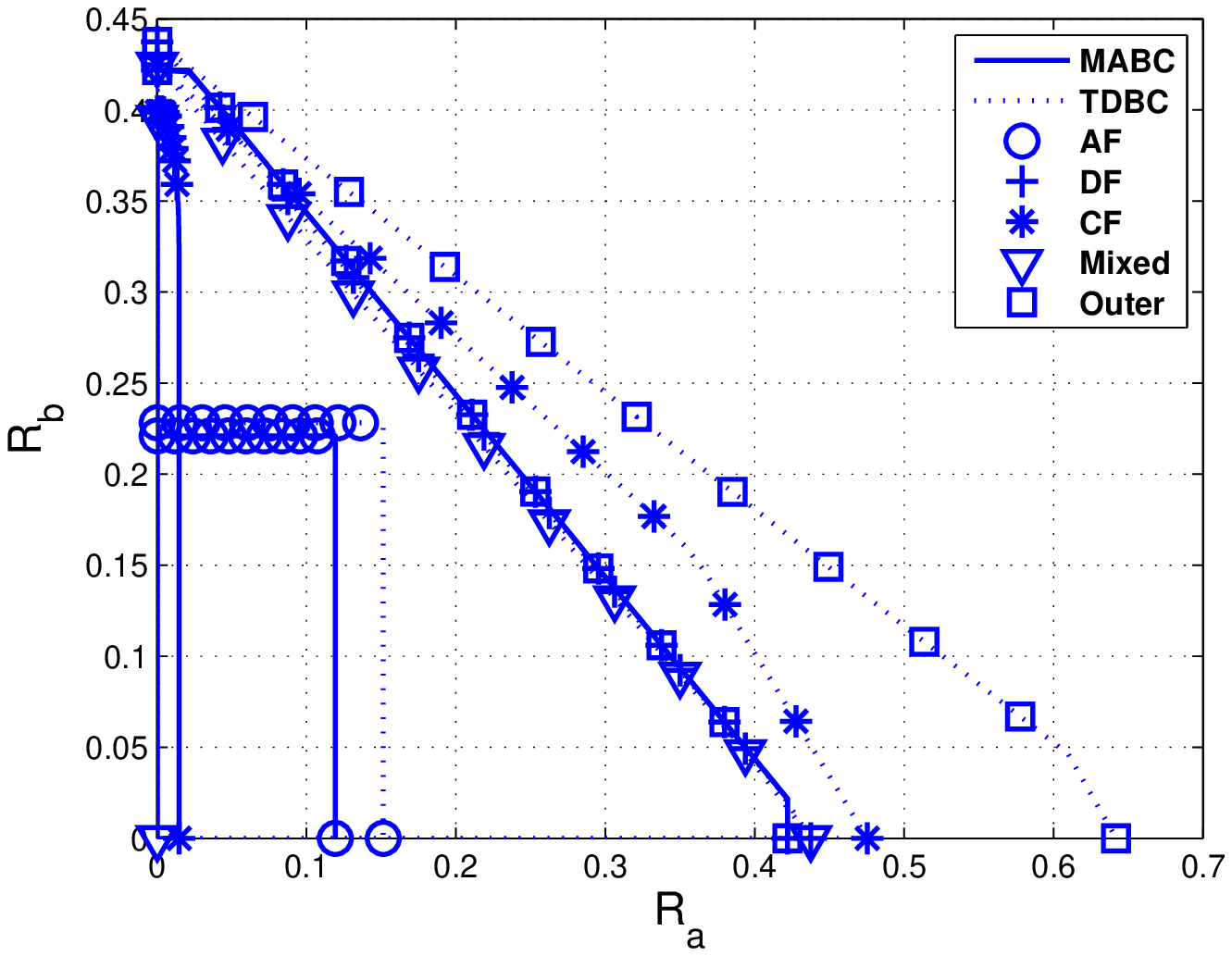, width=8.5cm}
     \caption{\baselineskip=10pt Comparison of bi-directional regions with $h_{\na\nr}=0.6$, $h_{\nb\nr}=20$, $h_{\na\nb}=0.5$, $P_\na=P_\nb=P_\nr=0$ dB and $N_\na=N_\nb=N_\nr=1$.}
             \label{fig:asym1_low}
    \end{center}
  \end{minipage}
  \hfill
  \begin{minipage}[t]{.47\textwidth}
    \begin{center}
     \epsfig{figure=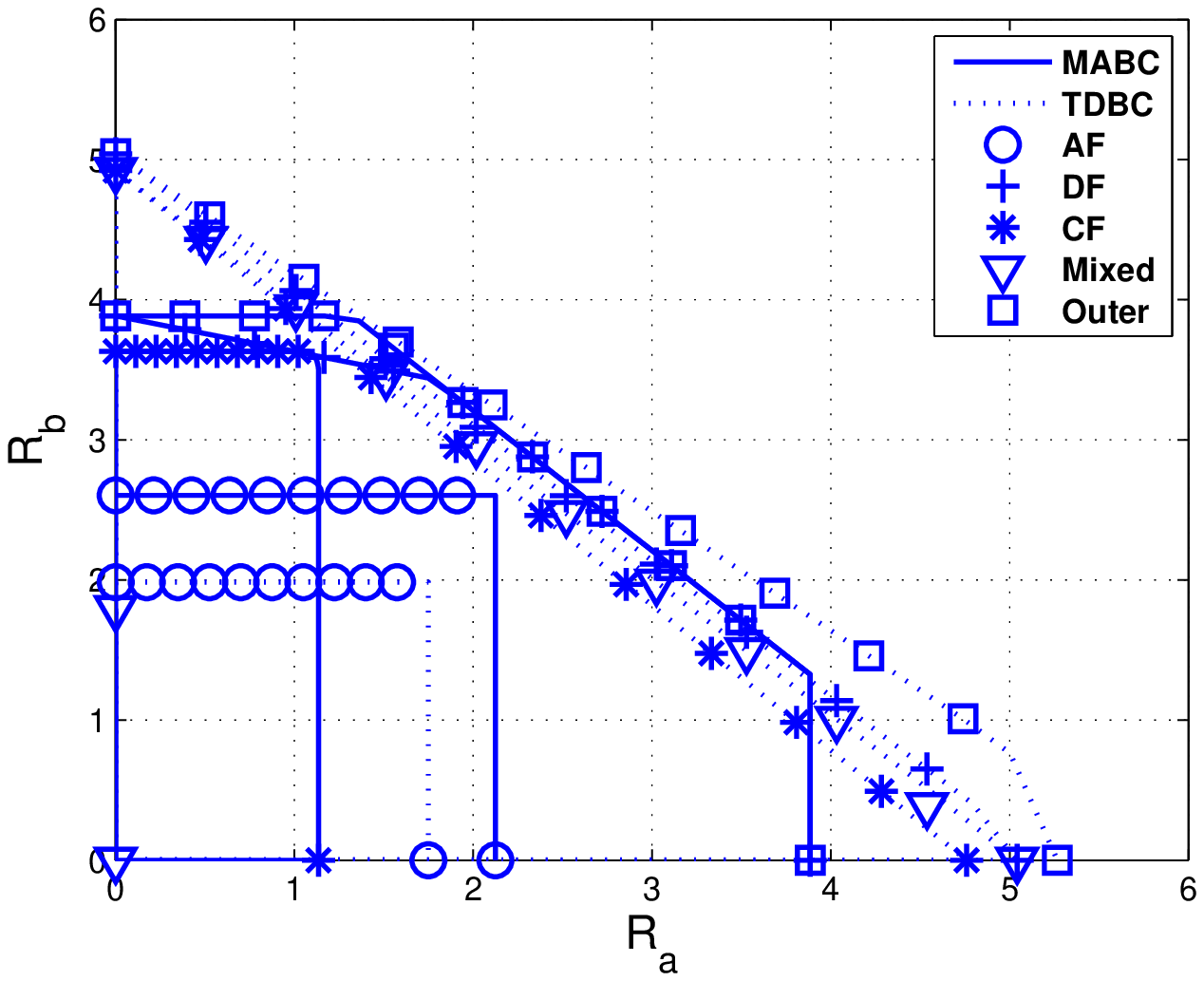, width=8.5cm}
   \caption{\baselineskip=10pt Comparison of bi-directional regions with $h_{\na\nr}=0.6$, $h_{\nb\nr}=20$, $h_{\na\nb}=0.5$, $P_\na=P_\nb=P_\nr=20$ dB and $N_\na=N_\nb=N_\nr=1$.}
      \label{fig:asym1_med}
    \end{center}
  \end{minipage}
  \hfill
\end{figure*}

\begin{figure*}
  \hfill
  \begin{minipage}[t]{.47\textwidth}
    \begin{center}
      \epsfig{figure=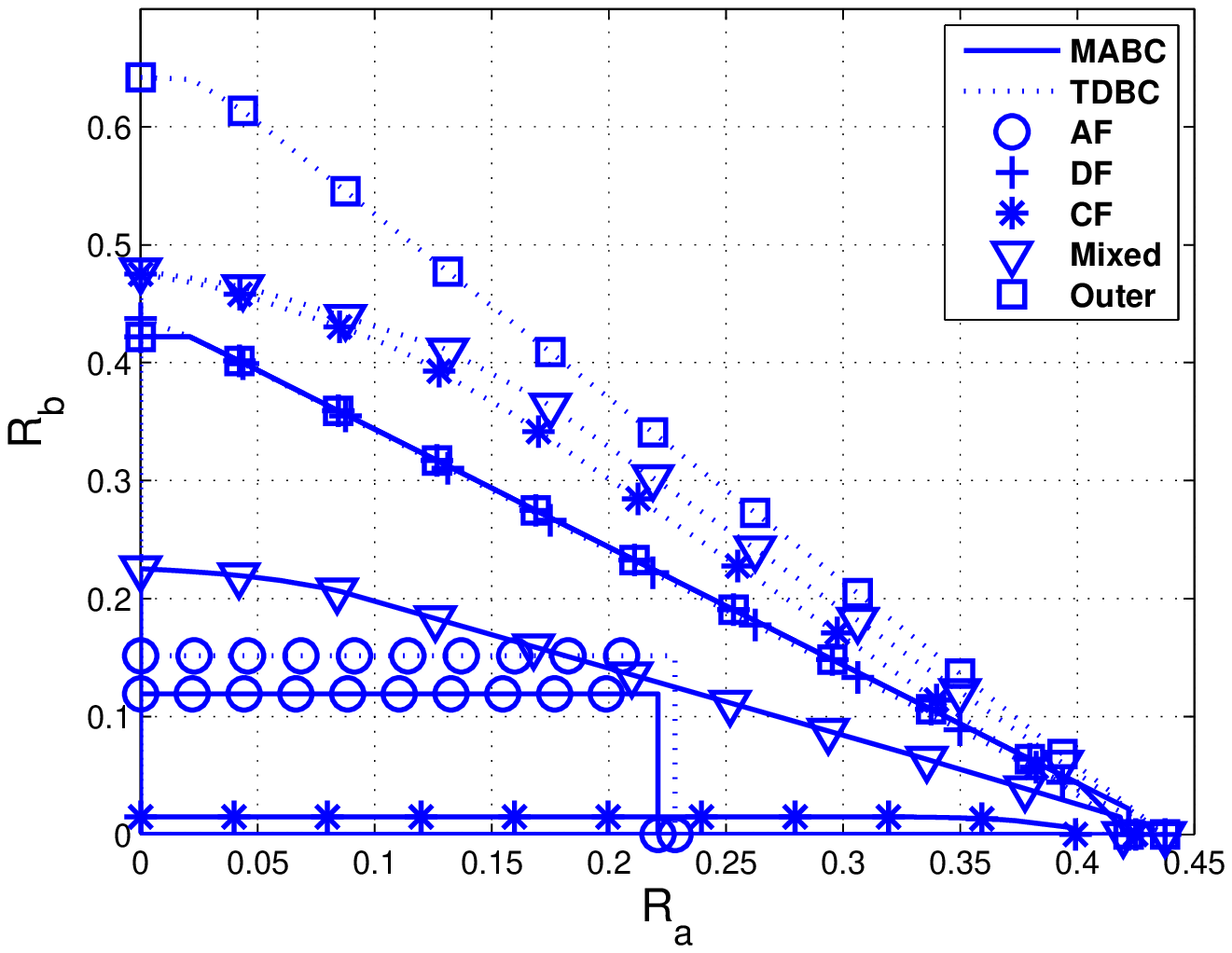, width=8.5cm}
      \caption{\baselineskip=10pt Comparison of bi-directional regions with $h_{\na\nr}=20$, $h_{\nb\nr}=0.6$, $h_{\na\nb}=0.5$, $P_\na=P_\nb=P_\nr=0$ dB and $N_\na=N_\nb=N_\nr=1$.}
           \label{fig:asym2_low}
    \end{center}
  \end{minipage}
  \hfill
  \begin{minipage}[t]{.47\textwidth}
    \begin{center}
       \epsfig{figure=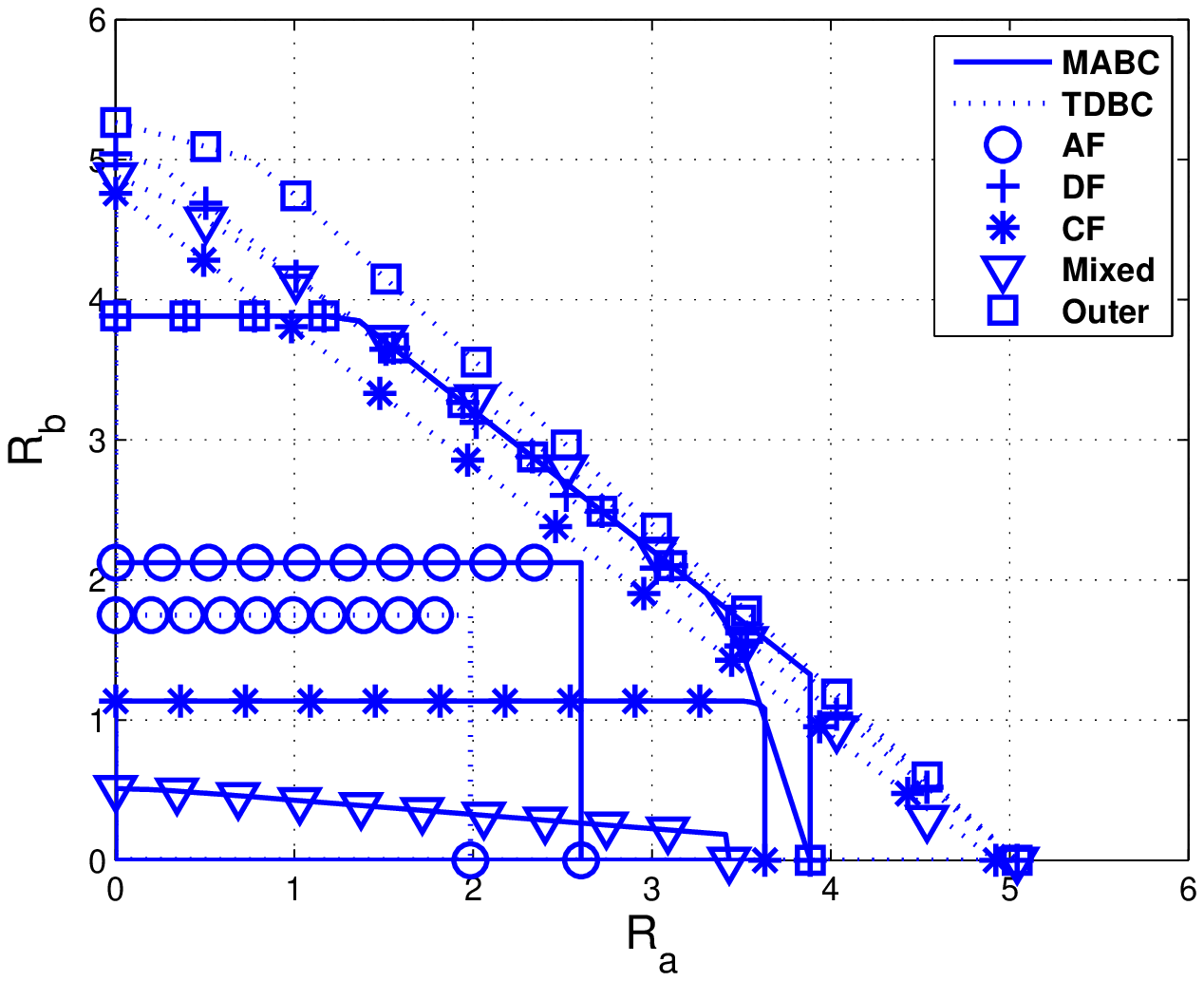, width=8.5cm}
      \caption{\baselineskip=10pt Comparison of bi-directional regions with $h_{\na\nr}=20$, $h_{\nb\nr}=0.6$, $h_{\na\nb}=0.5$, $P_\na=P_\nb=P_\nr=20$ dB and $N_\na=N_\nb=N_\nr=1$.}
      \label{fig:asym2_med}
    \end{center}
  \end{minipage}
  \hfill
\end{figure*}

In these cases $h_{\na \nr} = 0.6,\, h_{\nb \nr} = 20, h_{\na \nb} = 0.5$ (\Figs \ref{fig:asym1_low}, \ref{fig:asym1_med}) and $h_{\na\nr} = 20, \, h_{\nb \nr} = 0.6, h_{\na \nb}=0.5$ (\Figs \ref{fig:asym2_low}, \ref{fig:asym2_med}). Note that these two asymmetric cases are different for the mixed forwarding cases, which assume CF in one direction and DF in the other. In the low SNR regime, the CF TDBC and mixed TDBC protocol achieve the best performance in \Fig \ref{fig:asym1_low} and \Fig \ref{fig:asym2_low}, respectively. However, in the high SNR regime, the DF MABC protocol and the DF TDBC protocol yields larger regions than the other protocols. In contrast to the symmetric case, the AF MABC protocol is not outer bounded by the CF MABC protocol.

The mixed forwarding scheme is the only one which has different performance in the two asymmetric cases. In the mixed MABC protocol, if $h_{\nb\nr} > h_{\na\nr}$ then the noise seen at relay $\nr$ when it decodes $\tilde{w}_\na$ is larger than when  $h_{\nb\nr} <  h_{\na\nr}$. Therefore, the corresponding achievable rate region is also relatively smaller. In particular,  in the high SNR regime (\Fig \ref{fig:asym1_med}), the achievable rate region for the mixed MABC protocol is only able to achieve, under our input assumptions, rate $R_\na$ close to $0$ because of the effective noise from node $\nb$ during the first phase. In the mixed TDBC protocol, if $h_{\na\nr} > h_{\nb\nr}$, then we have a larger achievable rate region since the first link is more critical to the performance of the DF scheme. As the SNR increases, the difference between the  two asymmetric cases decreases.

\subsection{Maximum Sum Data Rate}

\begin{figure}
\begin{center}
 \epsfig{figure=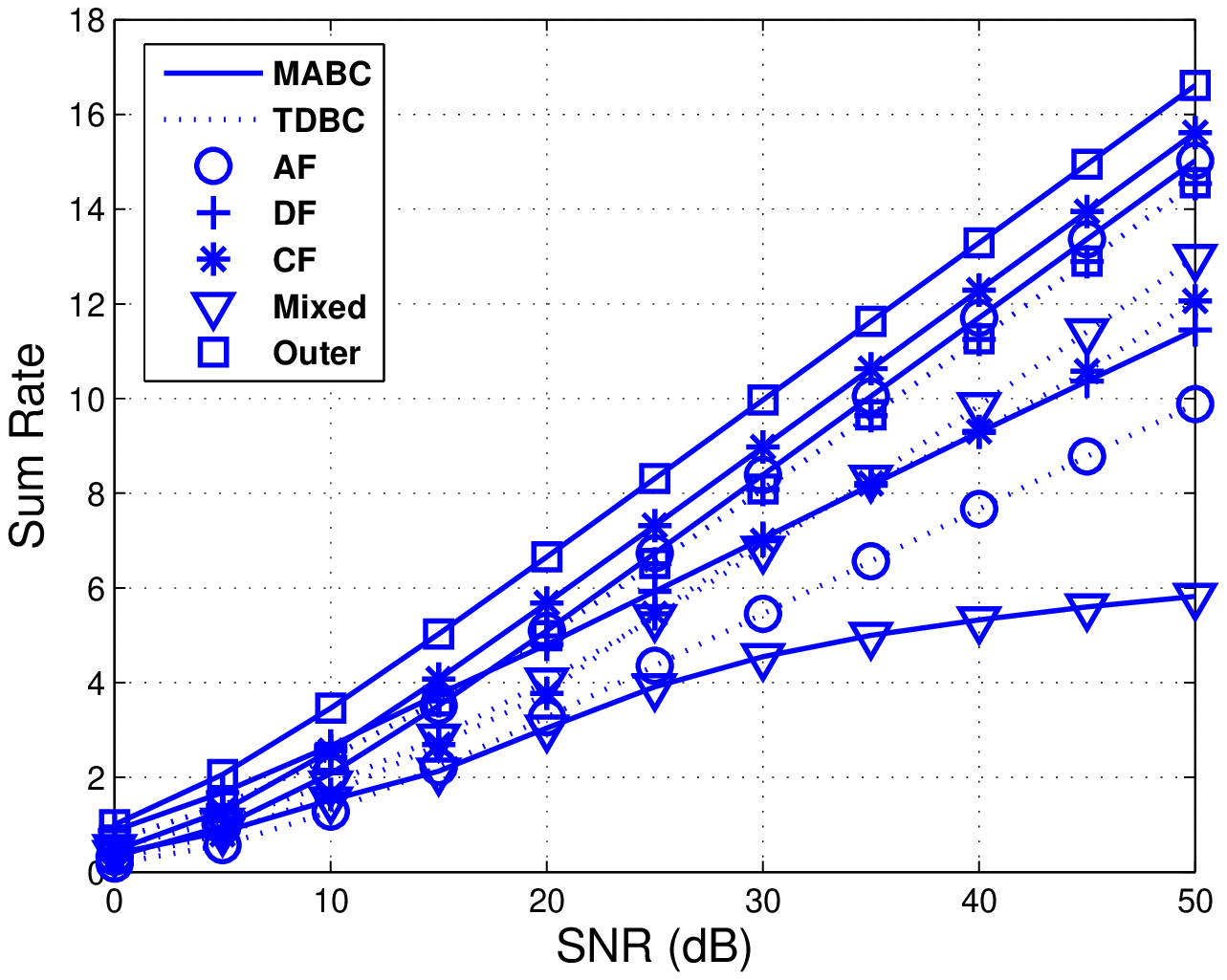, width=8.5cm}
\end{center}
 \caption{\baselineskip=10pt Maximum sum-rate of the 8 bi-diretional protocols and 2 outer bounds at different SNR. Here $h_{\na\nr}=h_{\nb\nr}=1$ and $h_{\na\nb}=0.2$}
            \label{fig:snr_sym}
\end{figure}

 In this subsection we plot the maximum sum-rate $R_\na+R_\nb$ as a function of the transmit SNR for the symmetric and two asymmetric cases of the previous subsection. As expected, different schemes dominate for different SNR values. The sum-rate is basically proportional to the SNR in dB scale since the sum-rate is roughly the logarithm of the SNR. In \Fig \ref{fig:snr_sym} around $12$ dB the relative performance of the CF MABC protocol and the DF MABC protocol changes. At lower SNRs, the DF MABC protocol is better, while at higher SNRs, the CF MABC protocol is better.
 The AF MABC protocol is always worse than the CF MABC protocol in the symmetric case (\Fig \ref{fig:snr_sym}).
 In the TDBC protocol, the sum-rate of the mixed TDBC protocol lies between the DF scheme and the CF scheme in \Fig \ref{fig:snr_sym}.

\begin{figure*}
  \hfill
  \begin{minipage}[t]{.47\textwidth}
    \begin{center}
      \epsfig{figure=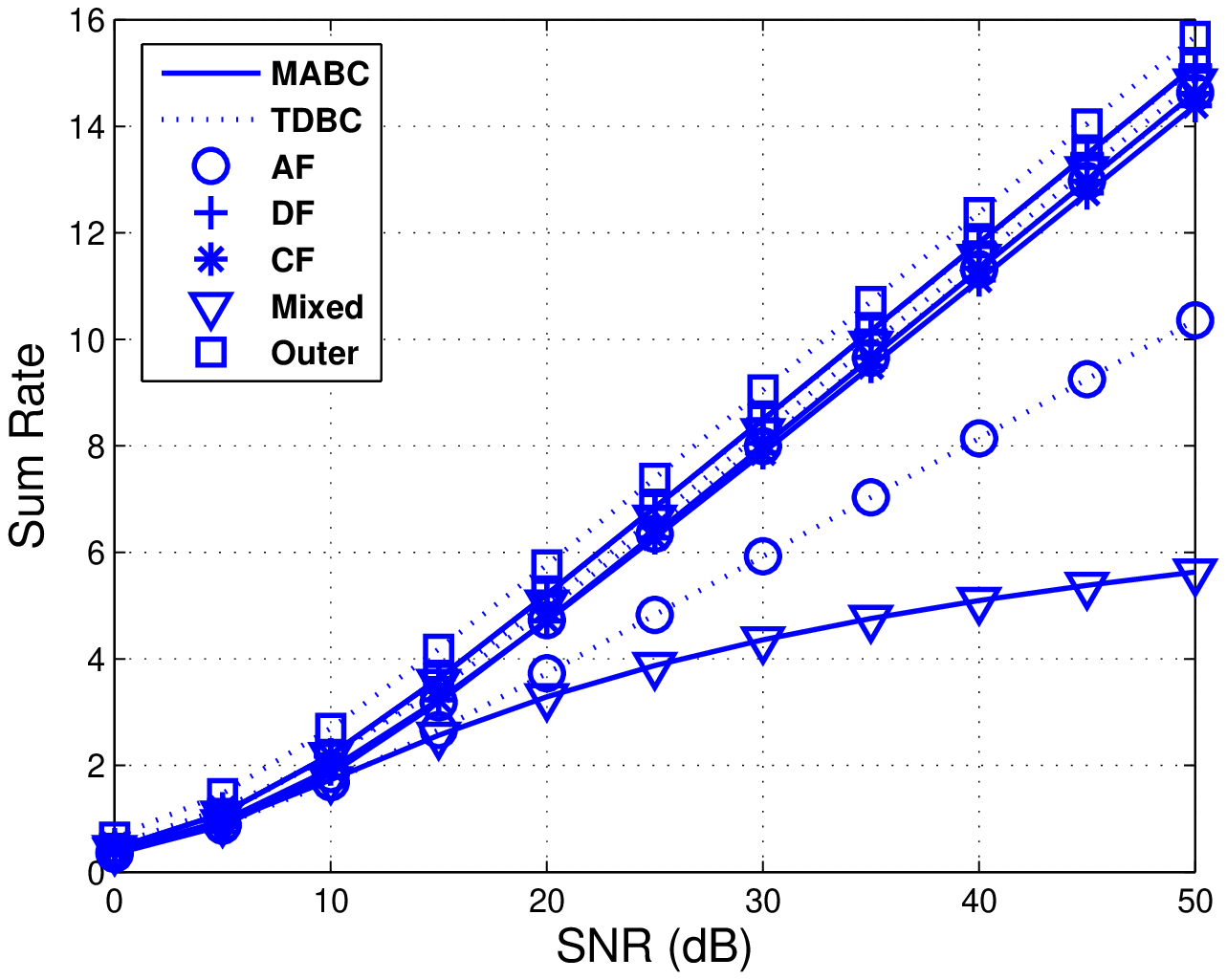, width=8.5cm}
      \caption{\baselineskip=10pt Maximum sum-rate of the 8 bi-diretional protocols and 2 outer bounds at different SNR. Here $h_{\na\nr}=0.6$, $h_{\nb\nr}=20$ and $h_{\na\nb}=0.5$}
      \label{fig:snr_asym1}
    \end{center}
  \end{minipage}
  \hfill
  \begin{minipage}[t]{.47\textwidth}
    \begin{center}
      \epsfig{figure=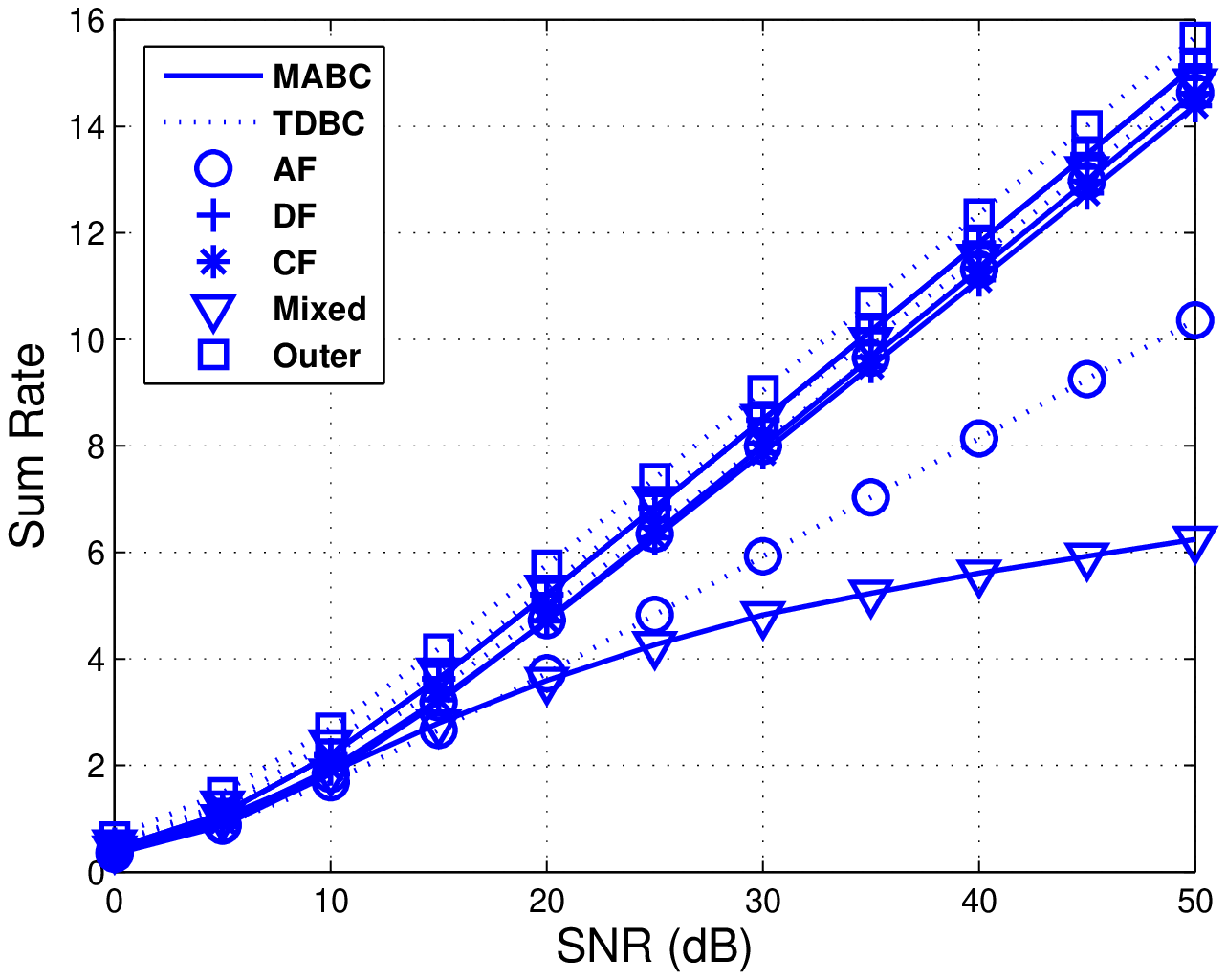, width=8.5cm}
      \caption{\baselineskip=10pt
      Maximum sum-rate of the 8 bi-diretional protocols and 2 outer bounds at different SNR. Here $h_{\na\nr}=20$, $h_{\nb\nr}=0.6$ and $h_{\na\nb}=0.5$}
      \label{fig:snr_asym2}
    \end{center}
  \end{minipage}
  \hfill
\end{figure*}

\subsection{Relay position}

\begin{figure}
\begin{center}
 \epsfig{figure=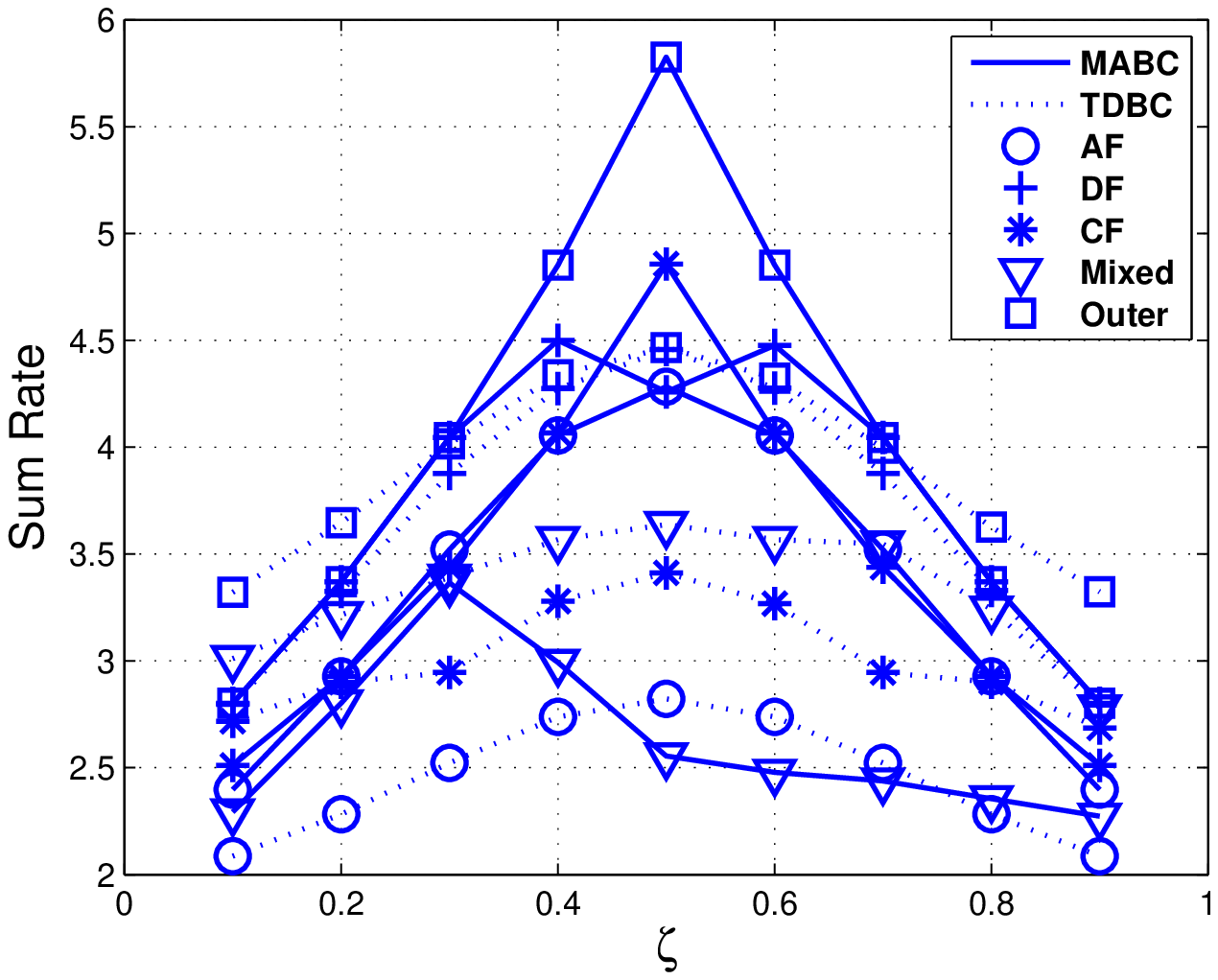, width=8.5cm}
\end{center}
\caption{\baselineskip=10pt Maximum sum-rate of the 8 bi-diretional protocols at different relay position. Here $h_{\na\nb}=0.2$ and no rate constraints. }
            \label{fig:dist_1}
\end{figure}

\begin{figure*}

  \hfill
  \begin{minipage}[t]{.47\textwidth}
    \begin{center}
      \epsfig{figure=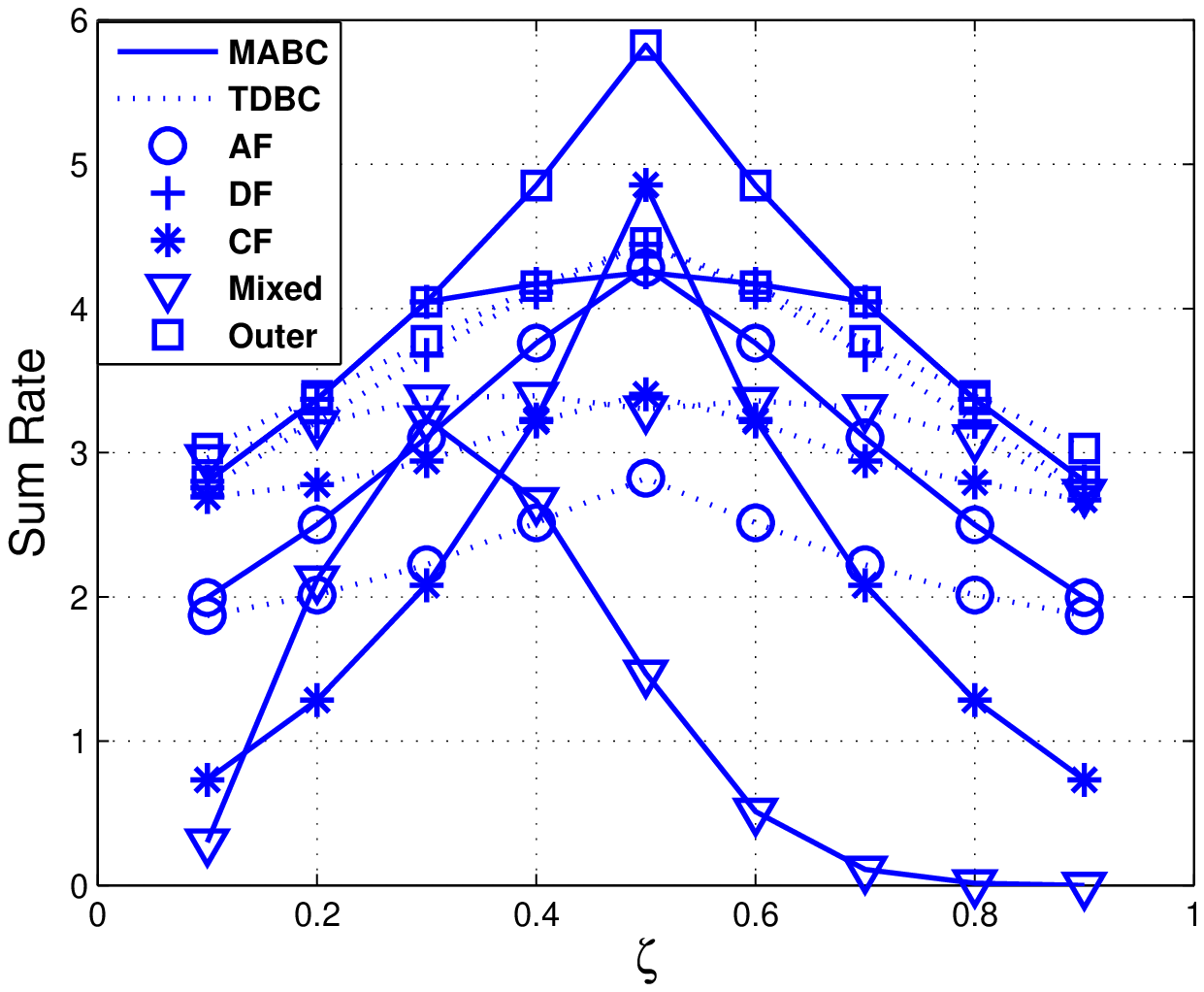, width=8.5cm}
      \caption{\baselineskip=10pt Maximum sum-rate of the 8 bi-diretional protocols at different relay position. Here $h_{\na\nb}=0.2$ and $\sigma = 1$ ($R_\na = R_\nb$).}
      \label{fig:dist_2}
    \end{center}
  \end{minipage}
  \hfill
  \begin{minipage}[t]{.47\textwidth}
    \begin{center}
      \epsfig{figure=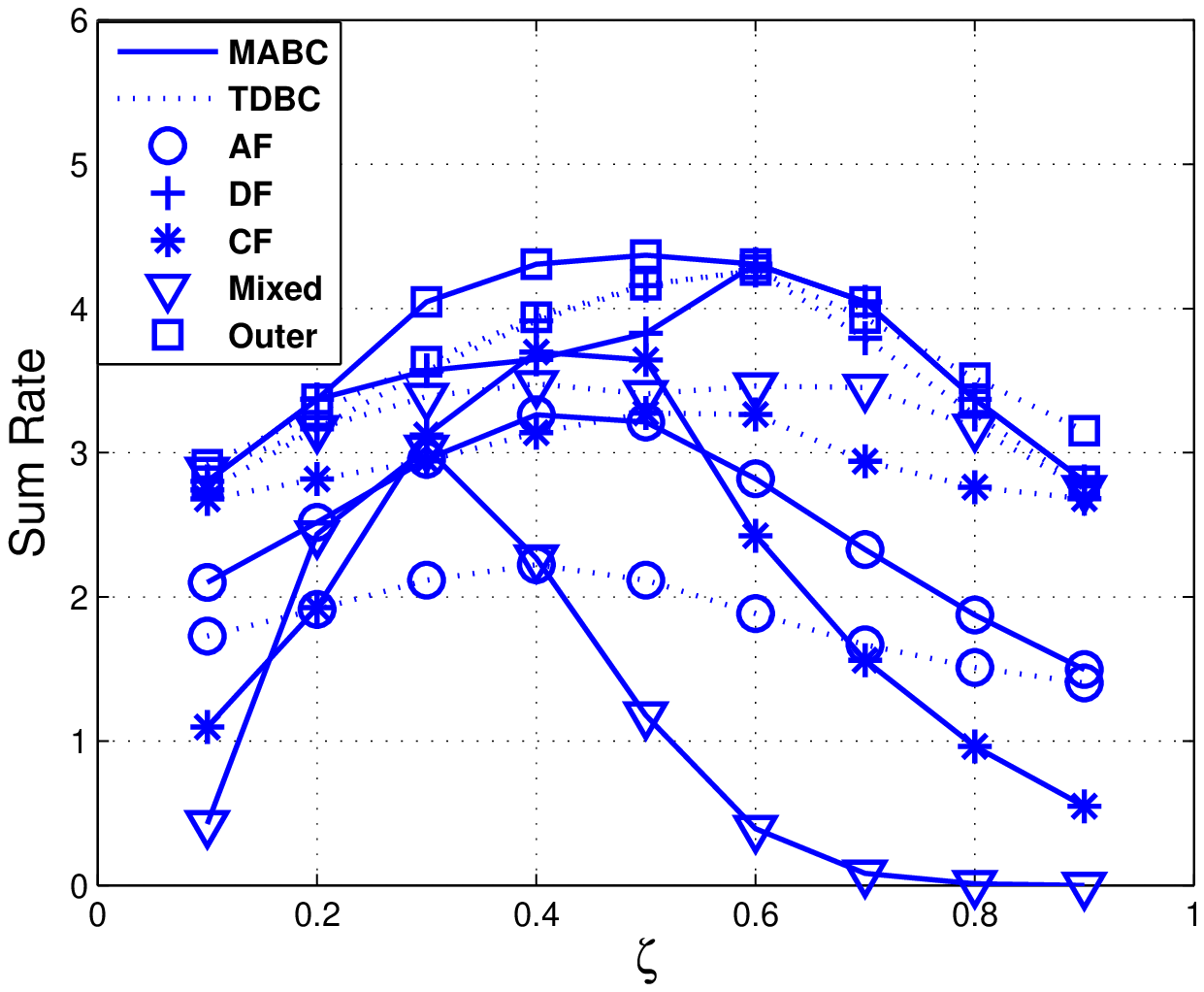, width=8.5cm}
      \caption{\baselineskip=10pt Maximum sum-rate of the 8 bi-diretional protocols at different relay position. Here $h_{\na\nb}=0.2$ and $\sigma = 2$ $(R_\na  = 2R_\nb$).}
      \label{fig:dist_3}
    \end{center}
  \end{minipage}
  \hfill
\end{figure*}

 In this subsection we plot the maximum sum-rate $R_\na+R_\nb$ as a function of the relay position $d_{\na\nr} = \zeta d_{\na\nb}$ ($0<\zeta<1$) when the relay $\nr$ is located on the line between $\na$ and $\nb$. Thus, $d_{\nb\nr} = (1-\zeta) d_{\na\nb}$. We apply $h_{\na\nb} = 0.2$ and $P_\na = P_\nb = P_\nr = 20$ dB and let $|h_{ij}|^2 = k  /d_{ij}^{3.8}$ for $k$ constant and a path-loss exponent of $3.8$. We consider three constraints on the sum-rate in the three \Figs \ref{fig:dist_1}, \ref{fig:dist_2} and \ref{fig:dist_3}. In the first, the sum-rate is maximized without any additional constraints. For the latter two we consider more realistic scenarios in which the sum-rate is constrained. In many communication systems,  uplink and downlink rates are not equal. More specifically, it is not uncommon for the dowlink rate to be 2 to 4 times greater than that of the uplink. In \Figs \ref{fig:dist_2} and \ref{fig:dist_3} we plot the maximal sum-rate of the protocols under a $\sigma = R_\na/R_\nb$ rate ratio restriction. These constrained sum-rates are obtained by optimizing the Gaussian regions of Section \ref{sec:gaussian} with the additional constraint $R_\na = \sigma R_\nb$.

For the MABC protocol, if the relay location is biased (not midway between nodes $\na$ and $\nb$), then the DF MABC protocol outperforms the CF MABC protocol and the AF MABC protocol outperforms the CF MABC protocol. This effect is more explicit in the constrained cases. In contrast, for the TDBC protocol, in order of increasing complexity, (and performance), the
relaying schemes are AF, CF, Mixed and DF in all cases when $(0.1\leq \zeta \leq 0.9)$.
As expected, the sum-rate for the mixed MABC protocol is worse than those of the other protocols.

The sum-rate plot for the mixed protocol is not symmetric since it uses different forwarding schemes for each link.
In addition, the sum-rates in the constrained case where $\sigma=2$ (\Fig \ref{fig:dist_3}) are asymmetric even for non-mixed protocols. The intuitive reason for this is that the rates are constrained in an asymmetric way, and hence a particular, non-midpoint distance will be optimal even for CF and DF forwarding schemes. The performance of the $\sigma=2$ sum-rate for the DF MABC protocol is remarkably asymmetric where it peaks and almost touches the outer bound at $\zeta=0.6$. These plots and region optimizations may be useful when determining the optimal relay position subject to particular rate constraints.

\section{Conclusion}

\label{sec:conclusion}
In this paper, we derived achievable rate regions for 4 new half-duplex bi-directional relaying protocols. We have specialized the 8 different achievable rate regions and the 2 outer bounds to the Gaussian case and numerically evaluated them under various channel conditions. For the MABC protocol,  DF or CF is the optimal scheme, depending on the given channel and SNR regime.
In the TDBC protocol, the relative performance of the forwarding schemes depends on the given channel condition. Notably, we have determined an example of a channel condition in which the mixed TDBC protocol outperforms the other proposed protocols. In general, the MABC protocol outperforms the TDBC protocol in the low SNR regime, while the reverse is true in the high SNR regime.

\appendices


\section{Proof of Theorem \ref{theorem:TDBC:CFDF}}

\label{app:tdbc:cfdf}

  \begin{proof}
  {\em Random code generation: } For simplicity of exposition, we take $|{\cal Q}| =1 $.
\begin{enumerate}
\item Phase 1: Generate random $(n\cdot\Delta_{1,n})$-length sequences
\begin{itemize}
  \item ${\bf x}^\pa_\na(w_\na)$ i.i.d. with $p^\pa(x_\na)$, $w_\na \in {\cal S}_\na = \{0,1,\cdots,\lfloor 2^{nR_{\na}} \rfloor-1\}$
\end{itemize}
and generate a partition of ${\cal S}_\na$ randomly by independently assigning every index $w_\na \in {\cal S}_\na$
to a set ${\cal S}_{\na,i}$, with a uniform distribution over the indices $i \in
\{0, \ldots, \lfloor 2^{nR_{\na0}} \rfloor - 1\} \eqdef \mathcal{S}_{\na0}$. We denote by $s_\na(w_\na)$ the index $i$ of ${\cal S}_{\na,i}$ to which $w_\na$ belongs.

\item Phase 2: Generate random $(n\cdot\Delta_{2,n})$-length sequences
\begin{itemize}
  \item ${\bf x}^\pb_\nb(w_\nb)$ i.i.d. with $p^\pb(x_\nb)$, $w_\nb \in {\cal S}_\nb = \{0,1,\cdots,\lfloor 2^{nR_{\nb}} \rfloor-1\}$
  \item $\hat{\bf y}_\nr^\pb(w_{\nr0})$ i.i.d. with $p^\pb(\hat{y}_\nr) = \sum_{y_\nr} p^\pb(y_\nr) p^\pb(\hat{y}_\nr|y_\nr)$ , $w_{\nr0} \in \{0,1,\cdots,\lfloor 2^{nR_{\nr0}} \rfloor-1\} \eqdef {\cal S}_{\nr0}$
\end{itemize}
\item Phase 3: Generate random $(n\cdot\Delta_{3,n})$-length sequences
\begin{itemize}
  \item ${\bf u}^\pc_\nr(w_\nr)$ i.i.d. with $p^\pc(u_\nr)$, $w_\nr  \in \{0,1,\cdots,\lfloor 2^{nR_{\nr}} \rfloor-1\} \eqdef  {\cal S}_\nr$
  \item ${\bf u}^\pc_\nb(w_{\nr0})$ i.i.d. with $p^\pc(u_\nb)$, $w_{\nr0} \in {\cal S}_{\nr0}$
\end{itemize}
and define bin $B_i \eqdef \{w_\nr | w_\nr \in [(i-1)\cdot\lfloor2^{n(R_\nr-R_{\na0})}\rfloor +1, i\cdot\lfloor2^{n(R_\nr-R_{\na0})}\rfloor]\}$ for $i\in {\cal S}_{\na0}$.
\end{enumerate}

{\em Encoding: } During phase 1 (resp. phase 2), the encoder of node $\na$ (resp. $\nb$) sends the codeword ${\bf x}^\pa_\na(w_\na)$ (resp. ${\bf x}^\pb_\nb(w_\nb)$). At the end of phase 1, relay $\nr$ estimates (or decodes)  $\tilde{w}_\na$. At the end of phase 2, the relay compresses the received ${\bf y}_\nr^\pb$ to a message $w_{\nr0}$ if there exists a $w_{\nr0}$ such that $({\bf y}_\nr^\pb , \hat{\bf y}_\nr^\pb(w_{\nr0}))\in A^\pb(Y_\nr {\hat Y}_\nr) $. Such an $w_{\nr0}$ exists with high probability if
\begin{align}
R_{\nr0} = \Delta_{2,n} I(Y_\nr^\pb;\hat{Y}_\nr^\pb) + \epsilon \label{eq:tdbc:cfdf:8}
\end{align}
and $n$ is sufficiently large. Also we choose
\begin{align}
R_\nr = \Delta_{3,n} I(U_\nr^\pc;Y_\nb^\pc) - 4\epsilon \label{eq:tdbc:cfdf:0}
\end{align}
To choose $w_\nr$, the relay first selects $w_{\na0} = s_\na({\tilde w}_\na)$ and the bin $B_{w_{\na0}}$, then it searches the minimum $w_\nr \in B_{w_{\na0}}$ such that $({\bf u}_\nr^\pc(w_\nr) , {\bf u}_\nb^\pc(w_{\nr0}))\in A^\pc(U_\nr U_\nb)$. This ensures uniqueness of $w_\nr$ if such a $w_\nr$ exists, i.e., $w_\nr$ is a function of $(w_{\na0},w_{\nr0})$. Such a $w_{\nr}$ exists with high probability if
\begin{align}
|B_{s_\na({\tilde w}_\na)}| = 2^{n(\Delta_{3,n} I(U_\nr^\pc;U_\nb^\pc) + \epsilon)}
\end{align}
Since $|B_j| = 2^{n(R_\nr-R_{\na0})}$, $\forall j \in {\cal S}_{\na0}$, the condition is equivalent to
\begin{align}
R_{\na0} < \Delta_{3,n} I(U_\nr^\pc;Y_\nb^\pc) - \Delta_{3,n} I(U_\nr^\pc;U_\nb^\pc)- 5 \epsilon \label{eq:tdbc:cfdf:3}
\end{align}
The relay then sends ${\bf x}^\pc_\nr$ randomly generated i.i.d. according to $p^\pc(x_\nr|u_\nr,u_\nb)$ with ${\bf u}_\nr^\pc(w_\nr)$ and ${\bf u}_\nb^\pc(w_{\nr0})$ during phase 2.

{\em Decoding: } Node $\na$ estimates $\tilde{w}_{\nr0}$ after phase 3 using jointly typical decoding. First, since node $\na$ knows $w_\na$, it can reduce the cardinality of $w_\nr$ to $\lfloor 2^{n(R_\nr -R_{\na0})}\rfloor$. Furthermore, it forms two sets of $\tilde{w}_{\nr0}$ based on typical sequences, $\{\tilde{w}_{\nr0}|({\bf y}_\na^\pb,\hat{{\bf y}}_\nr^\pb(\tilde{w}_{\nr0}))\in A^\pb(Y_\na {\hat Y}_\nr)\}$ and $\{\tilde{w}_{\nr0}|({\bf u}^\pc_\nr({\tilde w}_\nr),{\bf u}^\pc_\nb({\tilde w}_{\nr0}),{\bf y}_\na^\pc) \in A^\pc(U_\nr U_\nb Y_\na), {\tilde w}_\nr \in B_{w_{\na0}}\}$. After decoding $\tilde{w}_{\nr0}$ (which is a success if there is a single common element in both of the previous sets), node $\na$ decodes $\tilde{w}_\nb$ using jointly typical decoding of the sequence $({\bf x}_\nb^\pb, \hat{{\bf y}}_\nr^\pb,{\bf y}_\na^\pb)$. Node $\nb$ decodes $\tilde{w}_\nr$ after phase 3 and from the bin index of ${\tilde w}_\nr$ it estimates ${\tilde s}_\na$. Then node $\nb$ decodes the index as $\tilde{w}_\na$ if there exists a unique $\tilde{w}_\na \in S_{\na,\tilde{s}_\na}$ such that $({\bf x}_\na^\pa({\tilde w}_\na),{\bf y}_\nb^\pa)\in A^\pa(X_\na Y_\nb)$.

{\em Error analysis: }
\begin{align}
  P[E_{\na,\nb}] & \leq P[E_{\na,\nr}^\pa \cup E_{\nr,\nb}^{\pc} \cup E_{\na,\nb}^\pc]\\
  & \leq P[E_{\na,\nr}^\pa] + P[E_{\nr,\nb}^{\pc} | \bar{E}_{\na,\nr}^\pa] + P[E_{\na,\nb}^\pc| \bar{E}_{\na,\nr}^\pa \cap \bar{E}_{\nr,\nb}^{\pc}]\\
  P[E_{\nb,\na}] & \leq P[E_{\nr,\na}^\pc \cup E_{\nb,\na}^\pc]\\
  & \leq P[E_{\nr,\na}^\pc] + P[E_{\nb,\na}^\pc | \bar{E}_{\nr,\na}^\pc]
\end{align}
We define error events in each phase as follows:
\begin{enumerate}
\item $E_{\na,\nr}^\pa = E_{\na,\nr}^{\pa,1} \cup E_{\na,\nr}^{\pa,2}$.
\begin{description}
\item[$E_{\na,\nr}^{\pa,1}$ ]: $({\bf x}_\na^\pa(w_\na),{\bf y}_\nr^\pa) \not \in A^\pa(X_\na Y_\nr)$.
\item[$E_{\na,\nr}^{\pa,2}$ ]: there exists ${\tilde w}_\na \neq w_\na$ such that $({\bf x}_\na^\pa({\tilde w}_\na),{\bf y}_\nr^\pa)  \in A^\pa(X_\na Y_\nr)$.
\end{description}
\item $E_{\nr,\nb}^\pc = E_{\nr,\nb}^{\pc,1} \cup E_{\nr,\nb}^{\pc,2}\cup E_{\nr,\nb}^{\pc,3}\cup E_{\nr,\nb}^{\pc,4}$.
\begin{description}
\item[$E_{\nr,\nb}^{\pc,1}$ ]: there does not exist a $w_{\nr 0}$ such that $({\bf y}_\nr^\pb,{\hat {\bf y}}_\nr^\pb(w_{\nr0})) \in A^\pb(Y_\nr {\hat Y}_\nr)$.
\item[$E_{\nr,\nb}^{\pc,2}$ ]: there does not exist a $w_{\nr}\in B_{w_{\na0}}$ such that $({\bf u}_\nr^\pc(w_\nr),{\bf u}_\nb^\pc(w_{\nr0})) \in A^\pc(U_\nr U_\nb)$.
\item[$E_{\nr,\nb}^{\pc,3}$ ]: $({\bf u}_\nr^\pc(w_\nr),{\bf y}_\nb^\pc) \not \in A^\pc(U_\nr Y_\nb)$.
\item[$E_{\nr,\nb}^{\pc,4}$ ]: there exists ${\tilde w}_\nr \neq w_\nr$ such that $({\bf u}_\nr^\pc({\tilde w}_\nr),{\bf y}_\nb^\pc)  \in A^\pc(U_\nr Y_\nb)$.
\end{description}
\item $E_{\na,\nb}^\pc = E_{\na,\nb}^{\pc,1} \cup E_{\na,\nb}^{\pc,2}$
\begin{description}
\item[$E_{\na,\nb}^{\pc,1}$ ]: $({\bf x}_\na^\pa(w_\na),{\bf y}_\nb^\pa) \not \in A^\pa(X_\na Y_\nb)$.
\item[$E_{\na,\nb}^{\pc,2}$ ]: there exists ${\tilde w}_\na \neq w_\na$ such that $({\bf x}_\na^\pa({\tilde w}_\na),{\bf y}_\nb^\pa)  \in A^\pa(X_\na Y_\nb)$ and ${\tilde w}_\na \in S_{\na, w_{\na0}}$.
\end{description}
\item $E_{\nr,\na}^\pc = E_{\nr,\na}^{\pc,1} \cup E_{\nr,\na}^{\pc,2}\cup E_{\nr,\na}^{\pc,3}\cup E_{\nr,\na}^{\pc,4}\cup E_{\nr,\na}^{\pc,5}\cup E_{\nr,\na}^{\pc,6}$.
\begin{description}
\item[$E_{\nr,\na}^{\pc,1}$ ]: there does not exist a $w_{\nr 0}$ such that $({\bf y}_\nr^\pb,{\hat {\bf y}}_\nr^\pb(w_{\nr0})) \in A^\pb(Y_\nr {\hat Y}_\nr)$.
\item[$E_{\nr,\na}^{\pc,2}$ ]: there does not exist a $w_{\nr} \in B_{w_{\na0}}$ such that $({\bf u}_\nr^\pc(w_\nr),{\bf u}_\nb^\pc(w_{\nr0})) \in A^\pc(U_\nr U_\nb)$.
\item[$E_{\nr,\na}^{\pc,3}$ ]: $({\bf y}_\na^\pb,{\hat{\bf y}}_\nr^\pb(w_{\nr0})) \not \in A^\pb(Y_\na {\hat Y}_\nr)$.
\item[$E_{\nr,\na}^{\pc,4}$ ]: $({\bf u}_\nr^\pc(w_\nr),{\bf u}_\nb^\pc(w_{\nr0}),{\bf y}_\na^\pc) \not \in A^\pc(U_\nr U_\nb Y_\na)$.
\item[$E_{\nr,\na}^{\pc,5}$ ]: there exists $({\tilde w}_\nr ,{\tilde w}_{\nr0})$ where ${\tilde w}_\nr \neq w_\nr$ and ${\tilde w}_{\nr0} \neq w_{\nr0}$ such that $({\bf u}_\nr^\pc({\tilde w}_\nr),{\bf u}_\nb^\pc({\tilde w}_{\nr0}),{\bf y}_\na^\pc)  \in A^\pc(U_\nr U_\nb Y_\na)$ and $({\bf y}_\na^\pb,{\hat{\bf y}}_\nr^\pb({\tilde w}_{\nr0}))  \in A^\pb(Y_\na {\hat Y}_\nr)$. Recall, $w_\nr$ is uniquely specified by $(w_{\na0},w_{\nr0})$. Hence for a given $w_{\na0}$, there are at most $2^{nR_{\nr0}}$ such $({\tilde w}_\nr,{\tilde w}_{\nr0})$ pairs.
\item[$E_{\nr,\na}^{\pc,6}$ ]: there exists ${\tilde w}_{\nr0} \neq w_{\nr0}$  such that $({\bf u}_\nr^\pc(w_\nr),{\bf u}_\nb^\pc({\tilde w}_{\nr0}),{\bf y}_\na^\pc)  \in A^\pc(U_\nr U_\nb Y_\na)$,  \\$({\bf y}_\na^\pb,{\hat{\bf y}}_\nr^\pb({\tilde w}_{\nr0}))  \in A^\pb(Y_\na {\hat Y}_\nr)$.
\end{description}
\item $E_{\nb,\na}^\pc = E_{\nb,\na}^{\pc,1} \cup E_{\nb,\na}^{\pc,2}$.
\begin{description}
\item[$E_{\nb,\na}^{\pc,1}$ ]: $({\bf x}_\nb^\pb(w_\nb),{\bf y}_\na^\pb,{\hat{\bf y}}_\nr^\pb(w_{\nr0})) \not \in A^\pb(X_\nb Y_\na {\hat Y}_\nr)$.
\item[$E_{\nb,\na}^{\pc,2}$ ]: there exists ${\tilde w}_\nb \neq w_\nb$ such that $({\bf x}_\nb^\pb({\tilde w}_\nb),{\bf y}_\na^\pb, {\hat{\bf y}}_\nr^\pb(w_{\nr0}))  \in A^\pb(X_\nb Y_\na {\hat Y}_\nr)$.
\end{description}
\end{enumerate}

Then,
\begin{align}
P[E_{\na,\nr}^\pa] \leq & P[E_{\na,\nr}^{\pa,1}] + P[E_{\na,\nr}^{\pa,2}]\\
\leq&P[\bar{D}^\pa({\bf x}_\na(w_\na),{\bf y}_\nr)] + P[\cup_{\tilde{w}_\na \neq w_\na} D^\pa({\bf x}_\na({\tilde w}_\na),{\bf y}_\nr)]\\
\leq &\epsilon + 2^{n(R_\na -\Delta_{1,n}I(X_\na^\pa;Y_\nr^\pa)+3\epsilon)}\label{eq:tdbc:cfdf:1}\\
P[E_{\nr,\nb}^{\pc} | \bar{E}_{\na,\nr}^\pa] \leq & P[E_{\nr,\nb}^{\pc,1}| \bar{E}_{\na,\nr}^\pa] +  P[E_{\nr,\nb}^{\pc,2}| \bar{E}_{\na,\nr}^\pa]+ P[E_{\nr,\nb}^{\pc,3}| \bar{E}_{\na,\nr}^\pa]+ P[E_{\nr,\nb}^{\pc,4}| \bar{E}_{\na,\nr}^\pa] \\
\leq&2\epsilon+ P[\bar{D}^\pc({\bf u}_\nr(w_\nr),{\bf y}_\nb)] +
P[\cup_{\tilde{w}_\nr \neq w_\nr} D^\pc({\bf u}_\nr({\tilde w}_\nr),{\bf y}_\nb)] \label{eq:tdbc:cfdf:7}\\
\leq& 3\epsilon +
2^{n(R_\nr-\Delta_{3,n}I(U_\nr^\pc;Y_\nb^\pc)+3\epsilon)} \label{eq:tdbc:cfdf:2}
\end{align}
In \eqref{eq:tdbc:cfdf:7}, $P[E_{\nr,\nb}^{\pc,1}| \bar{E}_{\na,\nr}^\pa]$ and $P[E_{\nr,\nb}^{\pc,2}| \bar{E}_{\na,\nr}^\pa]$ are less than $\epsilon$ due to \eqref{eq:tdbc:cfdf:8} and \eqref{eq:tdbc:cfdf:3}, respectively.
\begin{align}
P[E_{\na,\nb}^\pc| \bar{E}_{\na,\nr}^\pa \cap \bar{E}_{\nr,\nb}^{\pc}] \leq &  P[E_{\na,\nb}^{\pc,1}| \bar{E}_{\na,\nr}^\pa \cap \bar{E}_{\nr,\nb}^{\pc}] + P[E_{\na,\nb}^{\pc,2}| \bar{E}_{\na,\nr}^\pa \cap \bar{E}_{\nr,\nb}^{\pc}]\\
\leq& P[\bar{D}^\pa({\bf x}_\na(w_\na),{\bf y}_\nb)] +  P[\cup_{\tilde{w}_\na \neq w_\na} D^\pa({\bf x}_\na(w_\na),{\bf y}_\nb),s_\na(w_\na) = s_\na(\tilde{w}_\na)]\\
\leq & \epsilon + 2^{n (R_\na - \Delta_{1,n}I(X_\na^\pa;Y_\nb^\pa)-R_{\na0} + 3\epsilon)}\label{eq:tdbc:cfdf:4}\\
P[E_{\nr,\na}^\pc] \leq&  P[E_{\nr,\na}^{\pc,1}]+ P[E_{\nr,\na}^{\pc,2}]+ P[E_{\nr,\na}^{\pc,3}]+ P[E_{\nr,\na}^{\pc,4}]+ P[E_{\nr,\na}^{\pc,5}]+  P[E_{\nr,\na}^{\pc,6}]\\
\leq&2\epsilon + P[\bar{D}^\pb({\bf y}_\na,{\hat{\bf y}}_\nr(w_{\nr0}))] +P[\bar{D}^\pc({\bf u}_\nr(w_\nr),{\bf u}_\nb(w_{\nr0}),{\bf y}_\na))] + \nonumber\\
&P[\cup_{\twolines{\tilde{w}_{\nr} \neq w_{\nr}}{\tilde{w}_{\nr0} \neq w_{\nr0}}} D^\pc({\bf u}_\nr({\tilde w}_\nr),{\bf u}_\nb(\tilde{w}_{\nr0}),{\bf y}_\na), D^\pb({\bf y}_\na,{\hat {\bf y}}_\nr({\tilde w}_{\nr0}))] + \nonumber \\
&P[\cup_{\twolines{\tilde{w}_{\nr} = w_{\nr}}{\tilde{w}_{\nr0} \neq w_{\nr0}}} D^\pc({\bf u}_\nr(w_\nr),{\bf u}_\nb(\tilde{w}_{\nr0}),{\bf y}_\na), D^\pb({\bf y}_\na,{\hat {\bf y}}_\nr({\tilde w}_{\nr0}))]\label{eq:tdbc:cfdf:9} \\
\leq & 3\epsilon + 2^{n( R_{\nr0}-\Delta_{3,n}I(U_\nr^\pc,U_\nb^\pc;Y_\na^\pc) - \Delta_{2,n}I(Y_\na^\pb;{\hat Y}_\nr^\pb) +7\epsilon)} + \nonumber \\
&2^{n( R_{\nr0}-\Delta_{3,n}I(U_\nb^\pc;U_\nr^\pc,Y_\na^\pc) - \Delta_{2,n}I(Y_\na^\pb;{\hat Y}_\nr^\pb) +7\epsilon)} \label{eq:tdbc:cfdf:5}
\end{align}
In \eqref{eq:tdbc:cfdf:9}, $ P[E_{\nr,\na}^{\pc,1}]$ and $ P[E_{\nr,\na}^{\pc,2}]$ are less than $\epsilon$ due to \eqref{eq:tdbc:cfdf:8} and \eqref{eq:tdbc:cfdf:3}, respectively. In \eqref{eq:tdbc:cfdf:5},  $P[\bar{D}^\pb({\bf y}_\na,{\hat{\bf y}}_\nr(w_{\nr0}))]$ is less than $\epsilon$ by the Markov lemma.
\begin{align}
P[E_{\nb,\na}^\pc | \bar{E}_{\nr,\na}^\pc] \leq & P[E_{\nb,\na}^{\pc,1}| \bar{E}_{\nr,\na}^\pc] + P[E_{\nb,\na}^{\pc,2}| \bar{E}_{\nr,\na}^\pc]\\
\leq&P[\bar{D}^\pb({\bf x}_\nb(w_\nb),{\bf y}_\na,{\hat{\bf y}}_\nr(w_{\nr0}))] + P[\cup_{\tilde{w}_\nb \neq w_\nb} D^\pb({\bf x}_\nb({\tilde w}_\nb),{\bf y}_\na,{\hat{\bf y}}_\nr(w_{\nr0}))]\\
\leq &\epsilon + 2^{n(R_\nb -
\Delta_{2,n}I(X_\nb^\pb;Y_\na^\pb,\hat{Y}_\nr^\pb)+4\epsilon)}
\label{eq:tdbc:cfdf:6}
\end{align}
Since $\epsilon > 0$ is arbitrary, the proper choice of $R_{\na0}$, the conditions of Theorem \ref{theorem:TDBC:CFDF}, \eqref{eq:tdbc:cfdf:0}, \eqref{eq:tdbc:cfdf:3} and the AEP
guarantee that the right hand sides of \eqref{eq:tdbc:cfdf:1},
\eqref{eq:tdbc:cfdf:2},
\eqref{eq:tdbc:cfdf:4}, \eqref{eq:tdbc:cfdf:5} and \eqref{eq:tdbc:cfdf:6} corresponding to the first term of \eqref{thm:TDBC:CFDF:2}, \eqref{eq:tdbc:cfdf:0}, (the second term of \eqref{thm:TDBC:CFDF:2} and  \eqref{eq:tdbc:cfdf:3}), \eqref{thm:TDBC:CFDF:1} and \eqref{thm:TDBC:CFDF:3} vanish as $n
\rightarrow \infty$. By the Carath\'{e}odory theorem in \cite{Hiriart:2001}, it is
sufficient to restrict $|{\cal Q}| \leq 6$.
\end{proof}
\bibliographystyle{IEEEtranS}
\bibliography{sang}
\end{document}